\documentclass[letterpaper,12pt,reqno]{article}
\usepackage{paper}
\hypersetup{pdftitle={Productivity and Efficiency in Science}}

\newcommand{\bib}{bibliography.bib}

\usepackage[parfill]{parskip}

\usepackage{upgreek,array}

\usepackage[labelsep=newline,justification=centering]{caption}

\usepackage{soul}

\begin{document}

\title{\vspace{-1.5em}Productivity Beliefs and \\ Efficiency in Science}

\author{%
\begin{tabular}{wc{4cm} c wc{4cm} c wc{4cm}}
\large{Fabio Bertolotti} & & \large{Kyle R. Myers} & & \large{Wei Yang Tham} \\
\small{Bank of Italy} & & \small{Harvard \& NBER} & & \small{Univ. of Toronto}\\ 
& \\
\end{tabular}
%
\thanks{Fabio Bertolotti: \href{mailto:fabiobertolotti@hotmail.it}{fabiobertolotti@hotmail.it}. Kyle Myers: \href{mailto:kmyers@hbs.edu}{kmyers@hbs.edu}. Wei Yang Tham: \href{mailto:weiyang.tham@rotman.utoronto.ca}{weiyang.tham@rotman.utoronto.ca}. This project received financial support from The Alfred P. Sloan Foundation and the Harvard Business School. There are no financial relationships or other potential conflicts of interest that apply to the authors. This project would not have been possible without the excellent work of Rachel Mural, Nina Cohodes, and Yilun Xu as well as input from Marie Thursby, Jerry Thursby, and Karim Lakhani. We received helpful comments from Nick Bloom, Judy Chevalier, Matt Clancy, Zoë Cullen, Shane Greenstein, Matt Grennan, Daniel Gross, Jorge Guzman, Bronwyn Hall, Sabrina Howell, Bruce Kogut, Petra Moser, David Rivers, Tim Simcoe, Chad Syverson, Joel Waldfogel, as well as seminar participants at Columbia, JOMT, MIT, the Workshop on the Organisation Economics and Policy of Scientific Research, the Banff Productivity Innovation and Economic Growth Conference, the NBER Productivity Lunch, Boston University TPRI, IIOC, Washington University in St. Louis, and the Conference on the Economics of Innovation in Memory of Zvi Griliches.}
}

\date{September 2025}

\begin{titlepage}
\maketitle

We develop a method to estimate producers' productivity beliefs when output quantities and input prices are unobservable, and we use it to evaluate the market for science. Our model of researchers' labor supply shows how their willingness to pay for inputs reveals their productivity beliefs. We estimate the model's parameters using data from a nationally representative survey of researchers and find the distribution of productivity to be very skewed. Our counterfactuals indicate that a more efficient allocation of the current budget could be worth billions of dollars. There are substantial gains from developing new ways of identifying talented scientists.

\end{titlepage}

\section{Introduction}

Science has continuously proven to be the engine of economic growth.\footnote{For motivating theory, see: \cite{jones2005growth}. For a recent review, see: \cite{bryan2021innovation}.} Academic science, in particular, is now a \$100 billion engine for the United States (\citealt{gibbons2024higher}). How efficiently is this engine operating? How much more knowledge could be produced if we optimized the allocation of resources to better support the most talented scientists? In this paper, we formalize and answer these questions.

A long line of work has documented frictions in science; some researchers are denied resources for reasons plausibly unrelated to their productivity.\footnote{For example: \cite{azoulay2019does,hager2024measuring}.} However, economists have struggled to quantify how much these frictions slow down the production of knowledge overall. Despite growing evidence that the \emph{average} marginal returns to investments in science are significant, the \emph{distribution} of marginal returns across researchers --- estimates of each researcher's productivity --- has remained elusive. And without this distribution of marginal returns, we cannot make statements about efficiency in science.

Estimating researcher-level productivity in science using conventional methods (e.g., via factor shares or control functions) poses some major challenges. First, the output of science is difficult to observe; how does one quantify a unit of knowledge? Bibliometric data is commonly used; however, as noted by \cite{adams1998research}, ``\emph{what constitutes a scientific paper makes for an elastic yardstick of scientific achievement}.'' Second, scientific inputs are often not allocated via price mechanisms and researchers face large adjustment costs (\citealt{myers2020elasticity,baruffaldi2025returns}). This severs the link between producers' productivity and their observed input choices, which is at the core of conventional methods. 

To overcome these challenges, we develop a new method for estimating productivity in teh absence of data on output quantities or inputs prices. The logic of our method is similar to conventional methods --- producers' input demand reveals information about their productivity. However, our approach involves a new step: the direct solicitation of producers' willingness to pay (WTP) for inputs. This new step, combined with additional assumptions, allows us to recover the distribution of researcher-level productivity, which is very skewed, and quantify the value from more efficiently allocating inputs in science, which is very large.

To illustrate our approach clearly, the following simple example shows how soliciting producers' WTP for some quantity of an input can be used to estimate their productivities:
\begin{quote}
 Profit maximizing producers are indexed by $i$. Output ($Y$) is produced using a single variable input ($X$) per the production function: $Y_i=\alpha_i X_i$, where $\alpha_i$ is the focal productivity parameter. Producers are price-takers and face a common output price ($p>0$), which we can either observe or estimate. Input costs are heterogeneous and convex as given by: $X_i^{c_i}$, where $c_i>1$. 

 We cannot observe the productivity or cost parameters ($\alpha_i, c_i$), and we can never observe output ($Y_i$). Our goal is to estimate productivity ($\alpha_i$).\footnote{Note that, even if we could also observe or estimate producers' expected input choice, $X^*_i$, then the first order condition, $X^*_i=(p \alpha_i/c_i)^{1/(c_i-1)}$, still does not separately identify the productivity and cost parameters ($\alpha_i, c_i$).}
 
 Our solution is to solicit producers' willingness to pay ($\text{WTP}_i$) to obtain a fixed quantity of input from outside the market (e.g., from us, the experimenters). Consider an experiment where producers are asked to report their WTP to obtain 1 unit of $X$ in this way. Producers' WTP will equate their profits inside the market whether or not they purchase the 1 unit of $X$ from outside the market. Formally, their $\text{WTP}_i$ should make the following equality hold: $\max_{X_i}\, [p\alpha_i X_{i} - X_{i}^{c_i}] = \max_{X_i}\,[p\alpha_i (X_{i}+1) - X_{i}^{c_i} - \text{WTP}_i]$. 
 
 This yields a simple expression for identifying producers' productivity that does not require output quantities or input prices: $\alpha_i=\text{WTP}_i/p$.
\end{quote}

The example above reflects the canonical setting of manufacturing firms producing physical goods that generate revenue via prices in the product market. But the underlying concept applies to science as well: researchers use their inputs to produce knowledge that generates utility via the incentive structures of science. Just as a firm's WTP for inputs reveals their belief about how productively they can produce goods, a researcher's WTP for scientific inputs reveals their belief about how productively they can produce knowledge.

In general, the mapping between productivity and willingness to pay for inputs is less transparent than in the simple example above. Below, we develop the general methodological framework that leverages this intuition. Our approach is not without limitations, and it requires some unconventional data. But engaging with these challenges can yield estimates of productivity in settings where other methods fail. We tailor our method to estimating productivity in science, but the general approach provides a path forward for productivity estimation in settings where heterogeneous producers (e.g., individual workers) obtain inputs that are not explicitly priced and generate output that is difficult to observe.

In order to apply this method to the market for science, we first present a new model of researchers' production and consumption decisions. Researchers have heterogeneous preferences and derive utility from three sources: their salary, their scientific output (i.e., the quantity of knowledge they produce), and their leisure time. Researchers choose how to allocate their time across fundraising (e.g., writing grants), research, and leisure given their beliefs, budgets, and constraints. Each researcher has a unique production function with constant returns to scale, defined by two researcher-specific parameters: ($i$) a funding-intensity parameter that determines the relative weight of the two key inputs, funding and time, and ($ii$) a total factor productivity (TFP) parameter that describes the efficiency with which researchers produce scientific output using their inputs.

Using the model, we can write a researcher's willingness to trade off their salary for inputs as a function of their productivity beliefs. This trade-off mirrors real decisions researchers make in the job market, and it forms the basis of the experiments that provide the variation necessary to identify the model's parameters. By identifying researchers' input demand via their WTP, we can handle the fact that inputs are not explicitly priced. Moreover, since the survey also solicits researchers' actual input levels, we can still estimate researchers' scientific output despite the fact that we never need to observe the knowledge they expect to produce. Throughout the paper, we highlight several important limitations to our general approach and the specific model. 

In order to generate the data necessary to estimate the model, we make use of a nationally representative survey of research-active professors across all major fields of science at roughly 150 major institutions of higher education in the US; for details, see \cite{myers2023new}.\footnote{Evidence is provided in \cite{myers2023new} suggesting the presence of non-response bias in the sample is very low on observable dimensions such as institutional rank, grant funding, and publication rates; some of this is reprinted in the Appendix. The survey recruitment included randomized incentives and reminders, which we use to test for sample selection per \cite{heckman1979sample} and find little evidence of any sample selection bias.} The survey solicits researchers' salaries, time allocations, and access to inputs. Importantly, the survey also includes a series of hypothetical experiments that solicit researchers' willingness to trade off their salary for more research funding or fewer administrative duties.

Researchers' willingness to pay for inputs are of plausible magnitudes and behavior. The median researcher is willing to pay 10\unskip cents for \$1 of research funding and \$68\unskip for one less hour of administrative duties. The components of the model explain a large fraction of the variation in researchers' responses (82--96\unskip\%) and, for the most part, responses do not appear to reflect sample selection or systematic noise in respondents' willingness to pay for any hypothetical good. Furthermore, researchers' willingness to pay for free time is strongly correlated with their implied hourly wages as expected.\footnote{We also implement \citeauthor{dizon2022improving}'s (\citeyear{dizon2022improving}) approach of using a ``benchmark good'' as a part of our willingness-to-pay elicitation to test whether respondents exhibit systematic noise in their stated preferences.} 

Our estimates of productivity beliefs vary widely across researchers, even after accounting for outliers. Within major fields of study, the ratio of the 90$^{th}$ and 10$^{th}$ percentile of TFP is approximately 52\unskip. When we look in narrower fields of study and attempt to control for other sources of heterogeneity, we estimate 90--10 TFP ratios to be approximately 29\unskip. This represents a high degree of dispersion compared to what is observed in commercial markets at the firm-level (e.g., \citealt{syverson2011determines}); however, this scale of dispersion in productivity across individual workers has been observed in some high-skilled settings (e.g., \citealt{sackman1968exploratory}). Motivated by \cite{gabaix2009power}, we investigate the upper tail of the TFP distribution and find it to exhibit power laws.

As a sign of face validity, our productivity estimates are positively correlated with common metrics of knowledge production (e.g., publications, citations, grant funding) and exhibits a similar degree of dispersion as those metrics. Notably, our estimates indicate that approximately half of the variance in scientific output across individual researchers is due to variance in their productivity.\footnote{We are unable to distinguish the degree to which our productivity estimates reflects an individual's fixed capabilities as a researcher versus any cumulative advantage they have acquired (c.f., \citealt{hall2024explorations}).} The high degree of dispersion suggests that it may be hard for the market to facilitate positive selection on productivity.

To evaluate allocative efficiency, we compare inputs and outputs under actual allocations, to those under alternative allocations. Specifically, we consider two alternative objectives: ($i$) maximize total scientific output, or ($ii$) maximize researchers' private utility. Using the model, we can solve for the allocation of inputs that achieves an objective while allowing for researchers' behavioral responses as they re-optimize their choices.

Overall, we find evidence of a moderate degree of efficiency given our proposed objectives. The correlations between researchers' actual input levels and the optima implied by our model generally span 0.4--0.8\unskip; more productive researchers acquire more inputs on average. However, there are significant gains from alternative allocations. Our counterfactuals suggest that total annual scientific output could be increased by approximately 160\unskip\%. The private value to researchers of this additional output is on the scale of 5\unskip\%. Estimating the social value of this growth requires assumptions about externalities, which we explore. 

To provide another way of characterizing the gains from reallocation, we ask the following: how much would funding levels need to increase under actual allocations to produce the same growth in output as our alternative allocations that hold input levels fixed? We find that aggregate funding levels would need to increase roughly 40\%\unskip to achieve the same growth in output our alternative allocations can achieve. That we can obtain a 160\unskip\% growth in aggregate output from a 40\%\unskip increase in aggregate funding is due to the combination of a mechanical composition effect and a endogenous behavioral response, which we detail further. Conservative approaches to scaling these estimates to the size of the population imply gains from reallocation that are equivalent to multi-billion dollar increases in annual funding.

We also evaluate the degree to which differences in aggregate output across major fields of study are due to differences in the number of researchers, their productivities, their input levels, or the fields' allocative efficiency. Overall, differences in allocative efficiency are the largest determinant of differences in aggregate output. At efficient allocations, the gaps in output between fields shrink by 20--50\%. 

Lastly, we unpack the counterfactuals to explore the following questions: Is the efficient allocation of inputs implied by the model more or less concentrated than the actual allocation? How does the reallocation of each input (i.e., funding and time) independently change the results? How much does the efficient allocation change under different objectives? What is the distribution of input wedges (i.e., the difference between their optimal and actual input levels) across researchers? Are researchers' input wedges predictable given their observable features?

Our estimates come with many caveats due to our method and setting. First, we face a set of challenges common to all existing productivity estimation techniques.\footnote{Specifically: ($i$) noise due to mismeasurement or misspecification; ($ii$) the presence of unmeasured, tradable inputs; and ($iii$) the presence of unmodeled heterogeneity in output prices (or preferences over payoffs).} Second, we face some unique limitations: ($i$) our method provides only ex-ante productivity beliefs and has no way of identifying ex-post differences in production; and ($ii$) the identifying variation in our data is based on stated preferences from hypothetical experiments, which may suffer from a range of biases. Throughout the paper we engage with these limitations by providing robustness tests of our assumptions and face validity tests of the data. Overall, the results from these tests give us confidence in our conclusions. Still, we interpret our results as plausible upper bounds on the productivity dispersion and gains from reallocation in science. Given the dearth of quantitative evidence on these points, we view this as an important step forward.

After a brief review of our connection to the literature, the rest of the paper proceeds as follows: Section \ref{sec:methodology} provides the general framework of and key assumptions underpinning our methodology; Section \ref{sec:data} details the survey data; Section \ref{sec:model} describes our model of researchers' production and consumption; Section \ref{sec:experiment} describes and reports the results of the survey experiments; Section \ref{sec:estimates} provides the model estimates; Section \ref{sec:efficiency} contains our counterfactual allocation exercises; and Section \ref{sec:discuss} concludes with a discussion of our results and the usefulness of our methodology more generally.

\subsection*{Related Literature}

Our paper sits at the intersection of two bodies of literature: empirical studies of science and markets for innovation (i.e., \citealt{merton1973sociology,stephan1996economics,bryan2021innovation}); and economic studies of producers' productivity and factor misallocation (i.e., \citealt{syverson2011determines,restuccia2017causes,de2021industrial}).

The meta-science literature has long been interested in misallocation. Following a long line of sociological work (\citealt{merton1973sociology,zuckerman1988sociology,shapin1995here}) economists have quantified some status-based frictions (\citealt{azoulay2019does}) and have also studied potential frictions such as: political lobbying (\citealt{hegde2015can}), information asymmetries regarding researchers' output (\citealt{hager2024measuring}), and competitive pressures from priority-based credit mechanisms (\citealt{hill2025race}).

Looking towards the productivity literature, our methodology is centered on understanding producers' factor demand. This concept is at the core of prevailing methods for estimating productivity, where identification can depend on input cost shares (e.g., \citealt{hsieh2009misallocation}) or the inversion of input demand into control functions (e.g., \citealt{olley1996dynamics,levinsohn2003estimating, ackerberg2015identification}). Much work has been done to extend these methods to account for features such as unobservable prices (\citealt{de2016prices}), adjustment costs (\citealt{petrin2013estimating,asker2014dynamic}), as well as  measurement error and heterogeneity (\citealt{il2016estimating,gollin2021heterogeneity}). To our knowledge, we are the first to formalize an approach for estimating production functions without data on output quantities and input prices.

We also follow a growing body of work using surveys to study the determinants of productivity in settings with inputs and outputs that are subjective or difficult to measure (e.g., \citealt{bloom2012organization,atkin2019survey}). We are not the first to solicit producers' WTP for inputs (c.f., \citealt{cole2013barriers,wossen2024estimating}); however, those studies tend to have an inherent interest in producers' demand for a specific input. Our work also runs parallel to the development of methods that use (quasi-)experimental variation to estimate misallocation (e.g., \citealt{sraer2023use,carrillo2023misallocation}).

\section{Methodological Framework}\label{sec:methodology}

In this section, we describe a producer's optimization problem and show how their productivity can be estimated by using information about their WTP for a fixed amount of inputs. The setup is general; the producers of interest may be organizations or individuals. Our approach requires four key assumptions that we describe in detail.

\subsection{Setup}
Our goal is to estimate producers' productivity beliefs: their rational expectations about their productivity in a future period of production. Rational expectations implies that producers have known beliefs about relevant variables; for example, a business manager can answer the question ``\emph{how many employees do you plan to have for next year?}'', or an individual researcher can answer the question ``\emph{how many hours per week do you plan to spend on your work next semester?}'' Despite revolving around forecasts, we present the framework as static; there is no dynamic optimization.\footnote{We also omit the expectation operator despite variables being forecasts.} 

There are $N$ producers indexed by $i$. For simplicity, we focus on the case where the producer uses a single, variable input $X$ to produce some quantity $Q$ of output; extensions to multiple inputs and stocks are possible. Producers have heterogeneous production functions with a Hicks-neutral total factor productivity (TFP) term: $Q_i=\alpha_i f(X_i,\bm{\upmu}_i)$, which is monotonically increasing in $X$.\footnote{Thus, $\alpha_i$ is a TFP-Quantity (TFPQ) parameter.} Heterogeneity depends on the vector $\bm{\upmu}_i$, which can include producers' observable characteristics $T_i$ and some common parameters designated $\bm{\upmu}$. Estimating the productivity parameter $\alpha_i$ is our primary goal.

Payoff from production is governed by a benefit function $b(M_i,P_iQ_i,\bm{\upmu}_i)$, which depends on output prices ($P_i$) and output quantity ($Q_i$), some endowment of liquid capital that is valued by the producer and guaranteed regardless of output ($M_i$), and parameters $\bm{\upmu}_i$. For a business, $M_i$ could reflect cash reserves and the benefit function can be viewed as a generalized revenue function that includes value to the business from these cash reserves. For individual researchers, $M_i$ could reflect their guaranteed salary and the benefit function describes their utility from their salary as well as any additional benefits that they expect to receive from producing more output (e.g., prestige, expectations of a promotion).\footnote{For simplicity, we do not allow for financial markets, but they could be incorporated.} Total costs depend on an input cost function $c(X_i,\bm{\upmu}_i)$ and there may also be constraints on input levels denoted by $g(X_i,\bm{\upmu}_i)=0$.

Thus, producers choose input levels that maximize their payoff per:
\begin{equation}\label{eq_obj_actual}
\begin{aligned}
 \max_{X_i} \;\; & b\bigl(M_i,P_i\alpha_i f_i(X_i), \bm{\upmu}_i \bigr) - c(X_i, \bm{\upmu}_i) \\
 & \text{subject to} \;\; g(X_i, \bm{\upmu}_i)=0 \;,
\end{aligned}
\end{equation}
where we will define $X^*_i$ as the argument that maximizes Eq. \eqref{eq_obj_actual}. Next, we walk through the key assumptions of our framework that facilitate identification of the focal productivity parameter $\alpha_i$ based on producers' WTP for additional units of the input from us, the experimenters.

\paragraph{Assumption 1 --- Observable Plans} \emph{Producers' optimal input plans $X^*_i$ are observable}. Variation in planned input levels (absent the WTP experiment) are necessary.\footnote{There are some knife-edge cases where input plans need not be observed, such as the example in the Introduction. However, those cases are likely rare in practice.}

\paragraph{Assumption 2 --- Output Prices} \emph{Producers are price takers and output prices are either observable or depend on observable covariates.} Our framework does not allow for unobservable horizontal differentiation as we would not be able to separately identify producer-specific output prices and productivity. Prices must either be observable (and assumed to reflect all quality differences) or homogeneous conditional on other observables, so that productivity reflects all quality differences (and producers are producing a commodity).

\paragraph{Assumption 3 --- Convex Input Costs} \emph{The direct costs of inputs are convex: $c_X(X_i,\bm{\upmu}_i)>0$ and $c_{XX}(X_i,\bm{\upmu}_i)>0$.} These convexities can be due to any sort of friction or adjustment cost. If this assumption does not hold, then the producer's WTP in the experiment may not depend on their productivity, as we illustrate below.\footnote{This is perhaps the least intuitive of our assumptions. Formally, this assumption ensures the necessary rank condition for estimation. Intuitively, the WTP experiment operates via an implicit linear cost schedule, so if the producer already faces a linear cost schedule in the actual input market, our WTP experiment will simply reflect those linear costs. It is trivial to show that if a producer uses a single input and faces a linear price for that input, then their WTP will not depend on their productivity, because they will simply report the price they face in their input market (i.e., the producer would never pay us, the experimenters, more for an input than what they pay in their own input market).}

\paragraph{Assumption 4 --- Monotonic Payoffs} \emph{The benefit function $b(\cdot)$ is strictly monotonically increasing in $Q_i$ and $M_i$.} This implies that more output or more liquid resources always increases the producers' payoff and, furthermore, that $b(\cdot)$ is invertible.\footnote{Extensions to settings where producers have market power may be possible.}

\subsection{Productivity and First Order Conditions}
In what follows, we assume there are no constraints on the producer and we set the output price to 1 as the numeraire in order to keep our expressions simpler. Solving the first order conditions of the producer's problem (Eq. \ref{eq_obj_actual}), sans constraints, yields:
\begin{equation}\label{eq_foc}
 \alpha_i b_{X} \bigl(M_i, \alpha_i f(X_i,\bm{\upmu}_i),\bm{\upmu}_i\bigr) f_X(X_i,\bm{\upmu}_i) - c_X(X_i,\bm{\upmu}_i)=0 \;,
\end{equation}
which shows that productivity ($\alpha_i$) is an implicit function of observables and the unknown parameters $\bm{\upmu}_i$ that govern the functions $b$,$f$, and $c$. This illustrates the value of Assumptions 1 and 2, but still leaves productivity unidentified due to the unknown parameters in $\bm{\upmu}_i$.

\subsection{WTP Solicitation}
In order to identify the productivity and other parameters in Eq. \ref{eq_foc}, we experimentally solicit producers' WTP for a fixed quantity of the input outside their input market (e.g., directly from us, the experimenters). For example, we offer the producer $\Delta$ units of the input and solicit their WTP (e.g., via stated preferences or incentive-compatible revealed preferences).

To see the value of this experiment, first consider the producer's optimization problem in the scenario where they choose to pay their WTP to purchase $\Delta$ inputs from us:
\begin{equation}\label{eq_obj_exp}
\begin{aligned}
 \max_{\widetilde{X}_i} \;\; & b(M_i-WTP_i,\alpha_i f(\widetilde{X}_i+\Delta,\bm{\upmu}_i),\bm{\upmu}_i) - c(\widetilde{X}_i,\bm{\upmu}_i) \;,
\end{aligned}
\end{equation}
where we define $\widetilde{X}^*_i$ as the argument that maximizes Eq. \eqref{eq_obj_exp}. In this scenario, the producer's $WTP_i$ is subtracted from their $M_i$ since they are both in monetary units and are perfect substitutes in the benefit function. The inputs purchased, $\Delta$, are added to the production function, but the producer's cost function $c$ still only depends on the quantity of inputs they choose to purchase in the market.

Therefore, the producer's WTP for $\Delta$ units of the input equates the following:
\begin{equation}\label{eq_wtpequate}
\begin{aligned}
 \underbrace{b(M_i, \alpha_i f(X^*_i,\bm{\upmu}_i),\bm{\upmu}_i) - c(X^*_i,\bm{\upmu}_i)}_{\substack{\text{expected payoff in} \\ \text{no-purchase scenario}}} = \underbrace{b(M_i-WTP_i,\alpha_i f(\widetilde{X}^*_i+\Delta,\bm{\upmu}_i),\bm{\upmu}_i) - c(\widetilde{X}^*_i,\bm{\upmu}_i)}_{\substack{\text{expected payoff in} \\ \text{scenario paying $WTP$ for $\Delta$}}} \;.
\end{aligned}
\end{equation}
That is, their WTP equates their net expected payoff in both ($i$) the scenario where they don't purchase the inputs from us (left-hand side of Eq. \ref{eq_wtpequate}) and ($ii$) the scenario where they do purchase from us (right-hand side of Eq. \ref{eq_wtpequate}). Assumption 3 --- convex input costs --- guarantees that the $WTP_i$ that solves Eq. \eqref{eq_wtpequate} depends on productivity ($\alpha_i$) and, therefore, on the parameters in $\bm{\mu}_i$ per Eq. \eqref{eq_foc}.\footnote{In the Appendix, we show how linear cost functions yield corner solutions for producers' $WTP$ that do not reflect their productivity and instead reflect the linear input costs the producers face in their market.}

\subsection{Estimation}

As written, identifying the productivity term ($\alpha_i$) from Eq. \eqref{eq_wtpequate} appears challenging because it depends on two unobserved components: ($i$) the producer's optimal input choice in the purchase scenario ($\widetilde{X}^*_i$), and ($ii$) the unknown parameters of $\bm{\upmu}_i$.\footnote{Recall, the vector $\bm{\upmu}_i$ includes unknown common parameters $\bm{\upmu}$ and observable individual-specific attributes $T_i$.} However, we know that $\widetilde{X}^*_i$ --- the solution to Eq. \eqref{eq_obj_exp} --- is itself an implicit function of productivity ($\alpha_i$), the amount of input offered ($\Delta$), and the vector $\bm{\upmu}_i$. 

Per Assumption 4 --- the benefit function ($b$) is invertible --- we can write a producer's $WTP_i$ as:
\begin{equation}\label{eq_wtp_phi}
WTP_i = \Pi(\Delta, M_i, X^*_i, T_i, \bm{\upmu}_i) \equiv W\widehat{T}P (\Delta, M_i, X^*_i, T_i,\bm{\upmu}_i) \;,
\end{equation}
which leaves only the unknown parameters $\bm{\upmu}_i$ to be estimated. This approach makes use of the fact that productivity ($\alpha_i$) and optimal allocations in the purchase scenario ($\tilde{X}^*_i$) are implicit functions of observables and unknown parameters.\footnote{Specifically, Assumptions 3 and 4 generally guarantees that $WTP_i$ is a direct function of $\alpha_i$, which, by Eq. \eqref{eq:alphahat} in the Appendix, depends on $\bm{\upmu}$.}
At this point we have a theoretical prediction of a producer's $WTP_i$ for $\Delta$ units of the input as well as the empirical value solicited in the experiment, call this $WTP^\text{obs}_i$. 

For estimation, we leverage the parameterization of $\bm{\upmu}_i$, which includes some common parameters $\bm{\upmu}$ and may depend on producers' observable characteristics $T_i$. The GMM estimator for $\bm{\upmu}$ solves the minimization problem:
\begin{equation}
 \min_{\bm{\upmu}} \sum_{i=1}^N \Bigl(W\widehat{T}P(\Delta, M_i, X^*_i, T_i, \bm{\upmu}) - WTP^\text{obs}_i \Bigr)^2 \;,
\end{equation}
which identifies $\bm{\upmu}$ given standard GMM identification conditions being met.

Particularly relevant for identification is the rank condition, which requires that the matrix of derivatives of the stacked moment conditions with respect to the vector of parameters has full rank. This condition fails, for example, if the WTP does not directly depend on a specific parameter, or if a parameter is redundant. It can be shown that Assumption 3 (Convex Input Costs) is a sufficient condition for the rank condition being met.

\subsection{Examples}

In Appendix \ref{app:method}, we walk through four additional example settings of different forms of production and cost functions to show the usefulness and limitations of our approach. We provide a variety of examples where the methodology is applicable as well as an example where identification is not achieved. For illustrative purposes, we focus on settings with an analytical characterization, but of course the method extends to cases where the solution is numerical, as in our application to researchers below.

\section{Survey and Data Overview}\label{sec:data}

\subsection{Survey Design}
We use the National Survey of Academic Researchers (\citealt{myers2023new}) and provide a brief overview of the survey methodology here. The population target is U.S. professors who conduct research at major institutions of higher education. To construct the sampling frame, information on professors was collected from the 158 largest institutions in the US by total R\&D funding using the National Science Foundation's 2019 Higher Education R\&D survey (HERD; \citealt{nsf2023herd}). 

The population consisted of 264,036 unique e-mails. A total of 131,672 individuals were e-mailed and 4,388 (3.33\%) completed the survey.\footnote{The IRB approval permitted e-mailing only 50\% of the population. The response rate is more than twice what has been obtained from sourcing academic researcher contacts from the corresponding author data contained within the publication record (e.g., \citealt{myers2020unequal}).} We then restrict the sample to the 4,003 individuals (91.2\% of respondents) who reported being a professor, spending a non-zero amount of time on research, and having a non-zero salary from their primary institution.

During recruitment, incentives and reminders were randomly assigned. The four incentive arms were: ($i$) no incentive, ($ii$) entry into a lottery to win a gift card, ($iii$) the ability to vote for a set of charities to receive a donation, and ($iv$) both the second and third incentives. The reminder arms were zero, one, or two follow-up emails. Each email was randomly assigned to one incentive arm and one reminder arm with equal probability, resulting in twelve possible combinations.

The randomized incentives and reminders provide us with instruments that we can use to implement a sample selection correction (i.e., \citealt{heckman1979sample}). The validity of this approach relies on having variables that cause entry into the sample (i.e., completing the survey) but do not affect the outcomes of interest. This allows us to adjust for unobservable differences between the population and our sample. Appendix Table \ref{tab_responserate_byincent} reports the results from a regression of an indicator for survey completion on the different incentive and reminder arms, showing that all arms had a statistically significant positive effect on researchers' propensity to complete the survey.

In addition to adjusting for unobservable differences, \cite{myers2023new} also checks the representativeness of the respondent sample by comparing it to the invited sample on observable characteristics. First, the authors explore a series of observable characteristics at the researcher level by comparing the grant and publication histories of respondents and non-respondents using the Dimensions database (\citealt{dimension2018data}), which collects and disambiguates scientific metrics for researchers worldwide.\footnote{Using a fuzzy name matching process, \cite{myers2023new} are able to confidently match 87,000 (66\%)\unskip of the researchers to their records in Dimensions.} Appendix Figure \ref{fig_nonresponse_dim} (replicated from \cite{myers2023new}) shows invite-respondent overlap on various measures of scientific inputs and outputs. Overall, there is little difference between the respondents and non-respondents both economically and statistically speaking.

Looking at the institutional level, Appendix Figure \ref{fig_nonresponse_herd} (replicated from \cite{myers2023new}) shows invite-respondent overlap on various measures of institution funding derived from the HERD survey. As in the case of the researcher-level comparison, the distributions overlap substantially. In this case, there are some statistically significant differences; on average, respondents come from institutions with slightly less research funding (4--6\%).

\subsection{Summary Statistics: Researchers and their Inputs}
Table \ref{tab_sumstat_modelvar} reports the summary statistics of the key covariates in our analyses. See \cite{myers2023new} for a more detailed investigation of these summary statistics. 

\paragraph{Fields of Study} Using the name of the professor's department, we assign them to a narrow set of twenty ``minor'' fields of study and aggregate those fields into five broader ``major'' fields: Humanities and related; Engineering, Math, and related; Medicine and Health Sciences; Natural Sciences; and Social Sciences. Unless otherwise noted, our counterfactual analyses constrain the reallocation of inputs to only occur \emph{within} the major fields. 

\begin{table}[h!]\centering \small
\caption{Summary Statistics---Salary, Funding, Time, and Fields}\label{tab_sumstat_modelvar}
{
\def\sym#1{\ifmmode^{#1}\else\(^{#1}\)\fi}
\begin{tabular}{l*{1}{rrrrrr}}
\hline\hline
                    & \emph{mean}& \emph{s.d.}& $p$50\\
\hline
\underline{\emph{Salary and research funds, \$/year}}&            &            &            \\
Total salary        &  159,028.23&   74,516.13&  140,000.00\\
Guaranteed \& existing funding&   52,223.08&   83,420.47&    5,000.00\\
Fundraising expectations&   93,909.19&  144,013.69&   20,000.00\\
[0.5em] \underline{\emph{Work time, hrs./week}}&            &            &            \\
Total work          &       48.55&       10.00&       48.00\\
Research            &       18.51&        9.60&       17.40\\
Fundraising         &        4.47&        5.09&        2.65\\
Administration      &        7.48&        6.17&        5.80\\
Teaching and other work&       18.09&        9.78&       17.20\\
[0.5em] \underline{\emph{Major field}, \{0,1\}}&            &            &            \\
Engineering, math, \& related&        0.17&        0.37&        \\
Humanities \& related&        0.19&        0.39&        \\
Medical \& health sciences&        0.28&        0.45&        \\
Natural sciences    &        0.16&        0.37&        \\
Social sciences     &        0.21&        0.40&        \\
\hline\hline
\end{tabular}
}
\note{\emph{Note}: Reports summary statistics for 4,003 researcher-level observations. Unless otherwise noted, all variables are continuous and bound below by zero. \emph{\{0,1\}} indicates binary variables.
} 
\end{table}

\paragraph{Salaries and Research Inputs} The survey solicits a range of details regarding researchers' salary, their guaranteed funding (e.g., from prior grants or institutional guarantees), expected funds they will raise over the coming five years, and their time allocations. Importantly, most variables are elicited as expectations over the coming five years to ensure the responses span the same time horizon as the thought experiments described below. In the Appendix, we replicate a test of respondents' self-reporting by comparing their self-reported salaries to the publicly-reported salaries we are able to locate for a subset of researchers. Overall, there is a high degree of alignment (see Appendix Figure \ref{fig_salcompare_pub_slf}).

\paragraph{Position and Socio-demographics} Appendix Table \ref{tab_sumstat_othervar1}provides a full summary of all other major features collected regarding researchers' positions (e.g., rank and tenure status) and their socio-demographics (e.g., gender, race/ethnicity, citizenship). 

\begin{table}[h!]\centering \small
\caption{Summary Statistics---Subjective Output}\label{tab_sumstat_qualoutput}
{
\def\sym#1{\ifmmode^{#1}\else\(^{#1}\)\fi}
\begin{tabular}{l*{1}{rrrrrr}}
\hline\hline
                    & \emph{mean}& \emph{s.d.}\\
\hline
\underline{\emph{Intended research outputs}, \{0,1,2\}}&            &            \\
Journal articles    &        1.87&        0.37\\
Books               &        0.52&        0.68\\
Materials or methods&        0.70&        0.68\\
Products            &        0.47&        0.64\\
[1em] \underline{\emph{Intended research audience}, \{0,1,2\}}&            &            \\
Academic peers      &        1.88&        0.36\\
Policymakers        &        0.84&        0.69\\
Businesses          &        0.54&        0.62\\
General public      &        0.83&        0.62\\
[1em] \underline{\emph{Riskiness of own research}, [0,10]}&            &            \\
Own belief          &        4.63&        2.34\\
Belief of peers' beliefs&        4.57&        2.37\\
[1em] \underline{\emph{Theoretical vs. empirical}, [0,10]}&            &            \\
Ask [0] or answer [10] questions&        4.89&        2.54\\
\hline\hline
\end{tabular}
}
\note{\emph{Note}: Reports summary statistics for 4,003 researcher-level observations. The intended research outputs and audience variables are coded responses to questions of the form \emph{How often are the following the intended output / audience of your research}, where responses are coded as follows: \emph{Rarely}=0, \emph{Sometimes}=1, \emph{Very often}=2. The question about risk used a scale where 0 indicated no risk and 10 indicated very high risk. The question about theory versus empirics used a scale where 0 indicated that the researcher focused on asking new questions and 10 indicated that the researcher focsed on answering existing questions.
}
\end{table}

\subsection{Summary Statistics: Subjective Output Measures}
Our methodology for estimating productivity is explicitly designed to avoid the need to quantify the output produced by researchers' efforts. Still, it is useful to have a more qualitative understanding of what researchers are producing. Fortunately, the survey includes a number of subjective questions regarding researchers' output. Table \ref{tab_sumstat_qualoutput} summarizes answers to these questions, which asked the frequency with which researchers intended to produce outputs of certain types (e.g., articles, books, materials, products) for certain types of audiences (e.g., peers, policymakers, businesses, general public) as well as other measures of riskiness and the degree to which the researcher focused on asking new questions or answering existing questions.

The traditional proxy for scientific output is peer-reviewed journal articles, and the data indicate that this is the most common output type that researchers intend to produce. However, there is a considerable amount of variation and a significant amount of attention focused on producing output of types and for audiences that may never be codified in a journal article. More importantly, these measures are often negatively correlated in a way that suggests strategic substitution (see Appendix Table \ref{tab_pwcorr_qualoutput}). For instance, researchers who focus more on publishing articles for their academic peers are less likely to focus on publishing books or developing products intended for non-academic audiences. This highlights the limitation of observable output proxies.

\section{Model of Science}\label{sec:model}
In this section, we model researchers' labor supply, which describes their utility from production and consumption. First, we present the environment and researchers' optimal decisions (Subsection \ref{subsec:theory}). Then we incorporate externalities in production to define social welfare (Subsection \ref{subsec:theory_socval}) and highlight the connection between the model and the survey thought experiments that solicit researchers' willingness to pay (WTP) for different factors (Subsection \ref{subsec:theory_wtp}). Additional details regarding the model and estimation are contained in Appendix \ref{app:moremodel}.


\subsection{Researchers' Labor Supply}\label{subsec:theory}
There are $N$ researchers indexed by $i$ who each choose how much total time to work ($H_i$) and how to allocate their time between research $R_i$ and fundraising $F_i$ over a fixed horizon. Their choices maximize their utility conditional on a contract from their primary institution, which is a triplet of state variables $\text{\textbf{S}}_i = (M_i,G_i,D_i)$: salary $M_i$ (\$), guaranteed funding $G_i$ (\$), and administrative and teaching duties $D_i$ (hours).\footnote{We use the term ``\emph{contract}'' loosely, since, for example, a researcher's guaranteed funding may come in part from future flows of funding from grants obtained outside the institution. In the context of the methodological framework presented in Section \ref{sec:methodology}, the function $u_{1i}(\cdot)+u_{2i}(\cdot)$ serves the role of the ``benefit'' function, while labor disutility $u_{3i}(\cdot)$ represents the cost function, which is parameterized to be convex.} Researchers' indirect utility is given by:
\begin{subequations}\label{eq:utmaxproblem}
\begin{align}
 \mathcal{V}(\text{\textbf{S}}_i\,,\, \bm{\uptheta}_i\,,\, \bm{\upmu}_i) = \max_{R_i\,,\, F_i\,,\, H_i} & \; u_{1i}(M_i) + u_{2i}(Y_i) - u_{3i}(R_i\,,\, F_i\,,\, D_i) \label{eq:utmaxproblem_u} \\
 & \text{subject to} \nonumber \\
 & B_i = B_{\text{min}} + G_i + \phi_i F_i \label{eq:utmaxproblem_B} \\
 & R_i + F_i + D_i = H_i \label{eq:utmaxproblem_H} \\
 & Y_i = \alpha_i B_i^{\gamma_i} R_i^{1-\gamma_i} \label{eq:utmaxproblem_Y} \;,
\end{align}
\end{subequations}
where $\bm{\uptheta}_i=(\alpha_i\,,\, \gamma_i\,,\, \phi_i)$ is the vector of individual-specific attributes related to scientific activity and $\bm{\upmu}_i$ is a vector of parameters that govern the shape of the $u$ functions.

The quantity of knowledge $Y_i$ that researchers produce is a Cobb-Douglas function of total funding $B_i$ and research time $R_i$, which includes researcher-specific productivities ($\alpha_i$) and output elasticities (per $\gamma_i$). Additionally, we assume a minimum funding amount $B_{\text{min}}$, and that all work hours are non-negative and have an upper bound $H_{\text{max}}$.

Utility from salary, output, and effort are given by the following functional forms:
\begin{subequations}\label{eq:Uform}
\begin{align}
 u_{1i}(\cdot) & = \omega \frac{(M_i)^{1-\sigma_i}}{1-\sigma_i} \label{eq:Uform1}\\ 
 u_{2i}(\cdot) & = \frac{(Y_i)^{1-\eta_i}}{1-\eta_i} \label{eq:Uform2} \\ 
 u_{3i}(\cdot) & = \psi \frac{(R_i+F_i+D_i^{\xi_i})^{1+\zeta_i}}{1+\zeta_i} \;. \label{eq:Uform3} 
\end{align}
\end{subequations}
The vector $\bm{\upmu}_i$ collects the parameters $(\omega,\sigma_i,\eta_i,\psi,\xi_i,\zeta_i)$.\footnote{The $\xi_i$ parameter allows for additional disutility from duty-related work (e.g., administration or teaching), which improves the model's fit to the data. We also assume the following: $\eta_i \in (0,1)$, $\psi>0$, $\xi_i >0$, and $\zeta_i>0$.} 

The policy functions $\mathcal{R}(\text{\textbf{S}}_i, \bm{\uptheta}_i, \bm{\upmu}_i)$, $\mathcal{F}(\cdot)$, and $\mathcal{H}(\cdot)$ characterize the solutions to Equation \eqref{eq:utmaxproblem}. For each individual researcher, these policies determine optimal $(R^*_i, F^*_i, H^*_i)$ as a function of states, attributes, and parameters. The derivations of these policy functions are described in Appendix \ref{app:policyfct}.

Ideally, we would have enough variation to estimate all researcher-specific attributes including the production-related parameters ($\alpha_i, \gamma_i, \phi_i$) as well as all of the consumption-related parameters that govern researchers' utility from salary, output, and effort ($\sigma_i, \eta_i, \xi_i, \zeta_i$). However, as shown below, we only have enough structural conditions to uniquely identify the production-related parameters ($\alpha_i, \gamma_i, \phi_i$). Thus, we choose to specify each of the consumption-related parameters as parametric function of researchers' observable features, whose common parameters we then estimate by GMM consistent with the framework of Section \ref{sec:methodology}.

Fortunately, we have a large vector of observable features ($\mathbf{X}_i$) that describe researchers' positions, their backgrounds, and their subjective descriptions of their scientific output (i.e., the features summarized in Table \ref{tab_sumstat_qualoutput} and Appendix Tables \ref{tab_sumstat_othervar1}). This is useful because it allows us to incorporate more heterogeneity into the consumption components of the model and limit the degree to which variation in the data might otherwise cause us to overstate the heterogeneity in the production parameters.

Unfortunately, the computational demands of estimating the model limit the flexibility with which we can incorporate the dozens of features available. Thus, to balance the benefits of allowing for heterogeneity in consumption with the benefits of simpler estimation, we use $k$-means clustering to reduce the full set of observable features ($\mathbf{X}_i$) into a one-dimensional index. We assume there are two clusters of researcher types ($k=2$) and estimate each researcher's distance from these clusters with their Euclidean similarity score. This distance provides a one-dimensional description of all of the different ways researchers responded to questions that could plausibly be driven by heterogeneous preferences. 

We refer to this resulting index as describing researchers' type, $T_i$. Appendix \ref{app:kmeans} reports the results of the $k$-means estimation showing the distribution of researcher type $T$ in the sample as well as a view of the mean differences in the features of researchers per their type.

We then specify each of the consumption-related parameters ($\sigma_i, \eta_i, \xi_i, \zeta_i$) to be simple, univariate functions of a researcher's type. For instance, the parameter that governs the utility from private consumption is parameterized as: $\sigma_i = \exp(\delta_{\sigma,0} + T_i \delta_{\sigma,1})$. We similarly parameterize $\zeta_i$ and $\xi_i$ with exponential functions with intercept and slope, respectively. The curvature of utility in scientific output, $\eta_i$, is modeled with a logistic function bounded in the interval $(0,1)$. Therefore, we express individual-specific parameters $\bm{\upmu}_i(T_i,\bm{\upmu})$ as a function of a researcher's type $T_i$ and of the vector of common parameters to be estimated, denoted by $\bm{\upmu}=(\omega,\psi,\bm{\updelta})$, where $\bm{\updelta}$ includes the deep parameters that govern the utility functions.

Allowing the $\bm{\upmu}_i$ parameters to be type-specific is not a panacea, but it helps reduce the degree to which variation in the experimental data that is truly driven by heterogeneous preferences or heterogeneous demand for scientific output contaminates our productivity estimates. We further detail the estimation process below.

\subsection{Social Value}\label{subsec:theory_socval}
To incorporate the externalities of knowledge production, which we assume to be net positive, we define the social value produced by each researcher as:
\begin{equation}\label{eq:utmaxproblemsocval}
 \mathcal{W}(\text{\textbf{S}}_i\,,\, \bm{\uptheta}_i\,,\, \bm{\upmu}_i\,,\,\kappa) = \; u_{1i}(M_i) + \kappa u_{2i}(Y^*_i) - u_{3i}(R^*_i\,,\,F^*_i\,,\,D_i) \;,
\end{equation}
where researchers' privately optimal choices and output are given by $R^*_i$, $F^*_i$ and $Y^*_i$. $\kappa$ reflects the size of the externalities associated with researchers' output. Thus, 1/$\kappa$ gives the share of the social value generated by each researcher's output that they themselves capture and $1-1/\kappa$ is the size of the positive externality.

This formulation of externalities is an ad hoc way of implicitly modeling consumer surplus as some constant multiple of producer surplus. However, it has a clear economic interpretation as a measure of appropriation, and we can draw on prior studies that have sought to identify precisely this measure. Studies that focus on commercial innovators have found producers' value capture to be as large as 15\% ($\kappa = 6.7$) and as low as 2\% ($\kappa = 50$) (\citealt{nordhaus2004schumpeterian,jena2008cost,lakdawalla2010economic}). In our empirical analyses, our baseline assumption is $\kappa = 10$, which implies researchers capture 10\% of the social value they create.\footnote{Note also that this approach implies that researchers' private utility reflects the case where $\kappa=1$.} Interestingly, as we will show below, the median researcher is willing to pay approximately \$0.10 dollars to purchase \$1 of research funding, which is a magnitude that is consistent with our assumption that $\kappa = 10$.

\subsection{WTP and Productivity}\label{subsec:theory_wtp}
As described in Section \ref{sec:methodology}, if we can solicit researchers' WTP for some fixed quantity of their inputs, we can estimate their productivity beliefs. Here we outline our specific application of the methodology.

Consider the period prior to production but after which researchers have formed their expectations about their time allocations and, therefore, their expectations about their utility over the 5-year horizon. Now, researchers are offered some alternative contracts by their primary institution that vary either funding guarantees ($G \rightarrow \widetilde{G}$) or administrative duties ($D \rightarrow \widetilde{D}$), but leave the salary unspecified.

For each offer indexed by $j$ and characterized by ($\widetilde{G}_{ij}\,,\, \widetilde{D}_{ij}$), there is a salary ($\widetilde{M}_{ij}=M_{ij}-WTP_{ij}$) that makes the researcher indifferent between their actual contract ($M_i\,,\,G_i\,,\,D_i$) and the offer such that:
\begin{equation}\label{eq:cvthoughtexp_equal}
\mathcal{V}(M_i\,,\, G_i\,,\, D_i \,,\, \bm{\uptheta}_i\,,\, \bm{\upmu} \,,\, ...) = \mathcal{V}(\widetilde{M}_{ij}\,,\, \widetilde{G}_{ij}\,,\, \widetilde{D}_{ij}\,,\, \bm{\uptheta}_i\,,\, \bm{\upmu} \,,\, ...) \;.
\end{equation}

For example, if a researcher is offered more guaranteed funding ($G_i < \widetilde{G}_{ij}$), then they will forecast how this additional funding will lead to changes in their optimal time allocations, their expected input levels, their expected output levels (per their productivity beliefs), and their expected indirect utility. The total increase in their expected indirect utility can be priced by the researcher and stated as a new, lower salary ($M_i > \widetilde{M}_{ij}$).

As shown in Section \ref{sec:methodology}, we can write the alternative salary ($\widetilde{M}_{ij}$) that makes researchers indifferent between their actual position and the offer as a function of known variables and the unknown parameters to be estimated: 
\begin{equation}\label{eq:cvthoughtexp_M}
 \widetilde{M}_{ij} = \Pi(\widetilde{G}_{ij}\,,\, \widetilde{D}_{ij}\,,\,M_i\,,\, G_i\,,\, D_i \,,\,\bm{\uptheta}_i\,,\, \bm{\upmu} \,,\, ...) \;,
\end{equation}
where $\Pi$ is determined by functional forms of, and optimality conditions implied by, Equation \eqref{eq:utmaxproblem}. Thus, if we know the left-hand side of Equation \eqref{eq:cvthoughtexp_M} from some experimental offers, and we also know all components of $\Pi$ except for the parameters $\bm{\upmu}$, then we can estimate those parameters. Appendix \ref{app:estoutline} contains the full details on our estimation routine.

\section{Survey Experiment}\label{sec:experiment}
Here, we describe the thought experiments in the survey (Subsection \ref{subsec:solicitWTP}) and how they connect to the model (Subsection \ref{subsec:modelconnect}). We then report the distribution of responses and conduct tests for face validity, sample selection, and other potential concerns (Subsection \ref{subsec:WTPresults}).

\subsection{Soliciting Willingness to Pay}\label{subsec:solicitWTP}
\paragraph{WTP for Funding and Time} Respondents are presented with four hypothetical scenarios, each offering different trade-offs between salary and inputs. They are asked to imagine that their primary institution has offered them: ($i$) an increase of \$250,000 in guaranteed funding in exchange for a lower salary, ($ii$) an increase of \$1,000,000 in guaranteed funding in exchange for a lower salary, ($iii$) an elimination of all administrative duties in exchange for a lower salary, and ($iv$) an increase in duties by 20 hours per month over a five-year period in exchange for a higher salary.\footnote{Researchers who report no administrative duties are not shown scenario ($iii$).} All of these hypotheticals are posed over a five-year span to fix the time horizon for all respondents. In order to solicit the salary at which they are indifferent between the current position and each offer, respondents are asked to report the lowest offered salary at which they would be willing to accept the offer.\footnote{Pilot tests indicated that researchers could more easily report the lowest salary that could make them take the offer as opposed to the literal salary at indifference, and since indifference is a vanishingly small amount less than this reported value, we treat their answer as the amount at indifference.} Importantly, the survey only solicits researchers' WTP for more funding. This matters because counterfactual reallocations will involve reducing some researchers' funding. In the case of time, the survey solicits both WTP (for more free time) and willingness to accept (WTA; for less free time) and we treat these as symmetric after being filtered through the model. Appendix Figure \ref{fig_thoughtexp} displays examples of how these thought experiments appeared to the survey respondents.

\paragraph{WTP for a Benchmark Good} When soliciting willingness to pay, especially in an un-incentivized manner, it is always possible that respondents under- or overstate their price sensitivity. In order to test for this, we follow \cite{dizon2022improving} and explore respondents' willingness to pay for a ``benchmark good.'' \cite{dizon2022improving} note that, if one solicits a subject's willingness to pay for a good whose value (to the subject) is plausibly uncorrelated with the value of the focal good, then any correlation between the two willingness-to-pay values can be attributed to systematic noise. The survey asks respondents 
to report the maximum amount they would be willing to pay per month for high-speed internet access at their primary residence. We use this as our benchmark good since researchers scientific productivity should plausibly be uncorrelated with their demand for high-speed internet access at home. In Appendix Table \ref{tab_wtp_tests}, we show that respondents' stated WTP for scientific inputs are rarely correlated with their stated WTP for the benchmark good, which suggests a small amount of variation in researchers' answers is attributable to systematic noise.

\subsection{Connection to Model: WTP and Compensating Variation}\label{subsec:modelconnect}
Four survey thought experiments elicit the compensating variation of individual researchers in relation to ($i$) an increase of \$250,000 in guaranteed funding $G_i$, ($ii$) an increase of \$1,000,000 in guaranteed funding $G_i$, ($iii$) a reduction of duties $D_i$ to 0, and ($iv$) an increase in duties by 20 hours per month over a 5-year period. In other words, we ask for income levels $\widehat{M}_{ij}$, where $j=\{1,2,3,4\}$ indexes the four experiments, that would make the researcher's utility in the counterfactual scenario equal to their indirect utility $\mathcal{V}^*_i$ at current allocations. Formally, counterfactual guaranteed funding and duties in the four experiments are:
\begin{equation*}
\begin{aligned}
    & (\widetilde{G}_{i1} \,, \widetilde{D}_{i1}) = \bigl((G_i + \text{\$250,000}) \,, D_i ) \\
     & (\widetilde{G}_{i2} \,, \widetilde{D}_{i2}) = \bigl((G_i + \text{\$1,000,000}) \,, D_i ) \\
     & (\widetilde{G}_{i3} \,, \widetilde{D}_{i3}) = \bigl(G_i \,, 0 \bigr) \\
     & (\widetilde{G}_{i4} \,, \widetilde{D}_{i4}) = \bigl(G_i \,, (D_i + 20 \text{ hours/month} )\bigr) \;,
\end{aligned}
\end{equation*}
where $G_i$ and $D_i$ represent actual states. 

As noted above, the values of $\widehat{M}_{ij}$ reported in these four experiments make the researcher indifferent between all possibilities (per Eq. \ref{eq:cvthoughtexp_equal}), and therefore we can write these reported values to be a known function of observable data and the unknown parameters to be estimated (per Eq. \ref{eq:cvthoughtexp_M}).

\subsection{Survey Experiment Results}\label{subsec:WTPresults}
\begin{figure} \centering
\caption{Willingness to Pay Experiment Responses}\label{fig_hist_wtp}
\subfloat[WTP for +\$1 Research Funding]
{
\label{fig_hist_wtp_dols}
\includegraphics[width=0.495\textwidth, trim=0mm 70mm 0mm 5mm, clip]{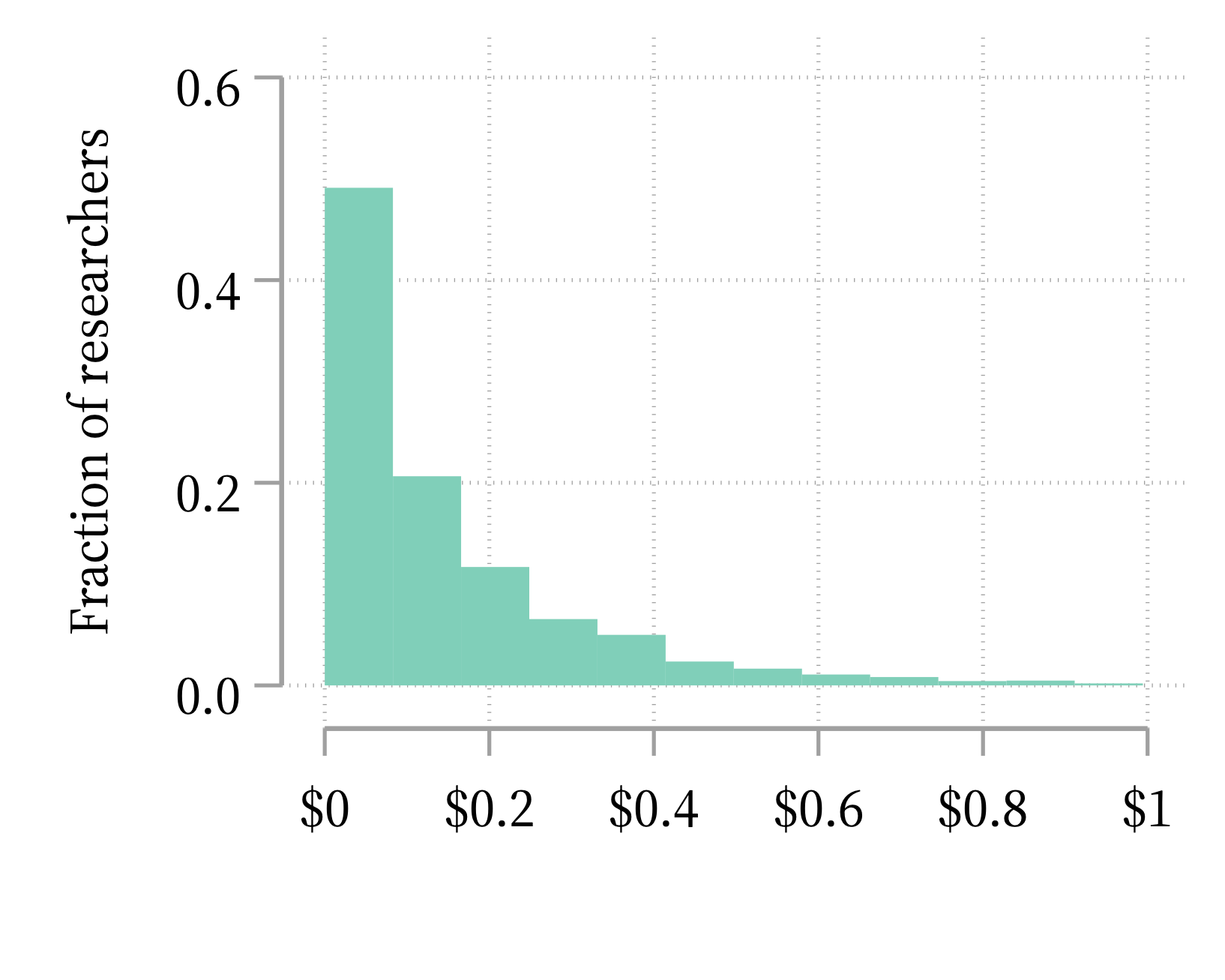}
} 
\subfloat[WTP for --1 hr. Administrative Duties]
{
\label{fig_hist_wtp_hr}
\includegraphics[width=0.495\textwidth, trim=0mm 70mm 0mm 5mm, clip]{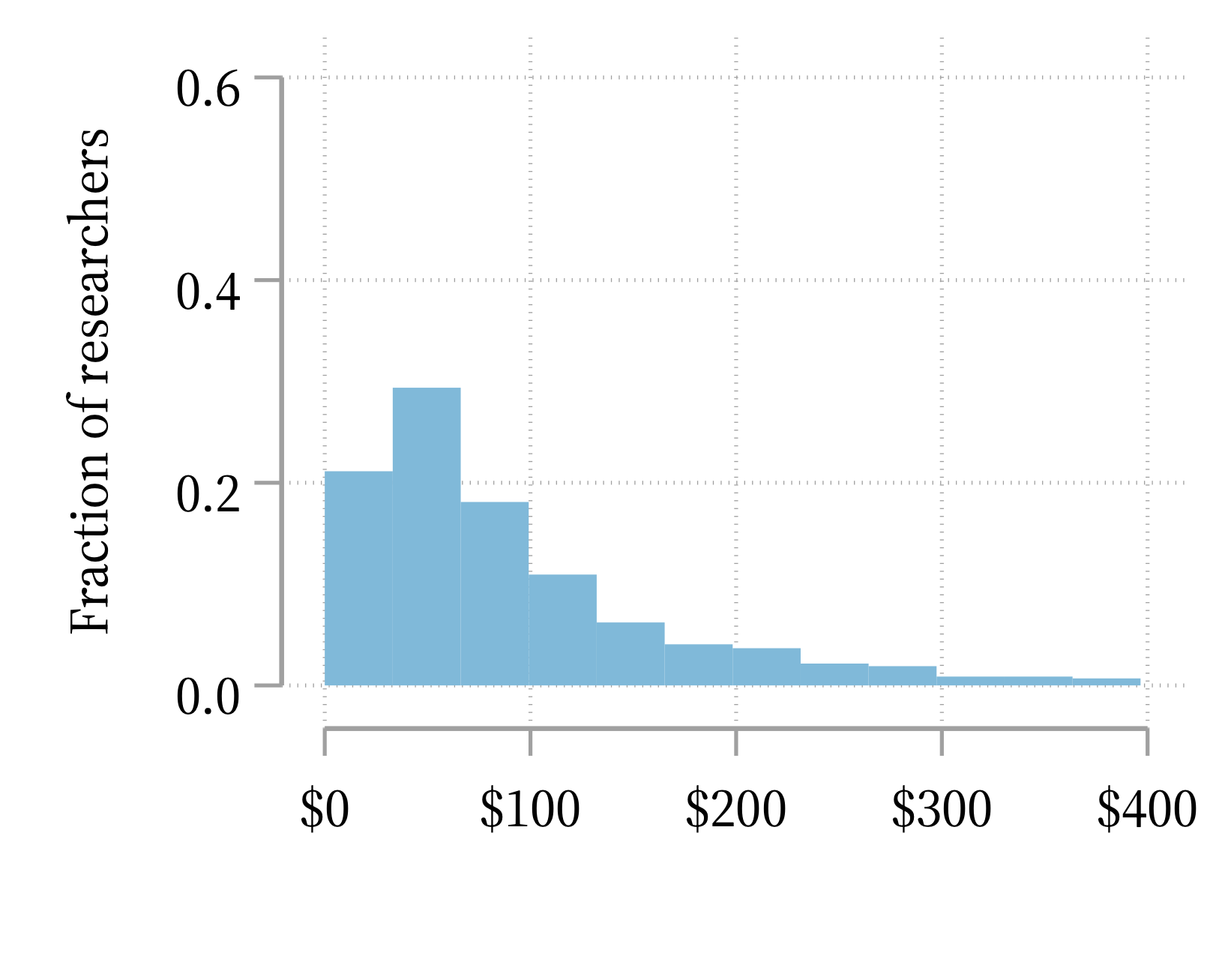}
} \\
\note{\emph{Note}: \input{figtab/note_fig_hist_wtp.tex}}
\end{figure}

Figure \ref{fig_hist_wtp} plots the distribution of WTP for guaranteed research funding and free time (in the form of WTP for less duties or WTA for more duties). The values shown are averaged over the two thought experiments for either factor and converted to a per-dollar basis. As evidenced by Figure \ref{fig_hist_wtp}, there is considerable variation in WTP responses across researchers. Some of this variation reflects different preferences and constraints, but another portion reflects heterogeneous productivity across researchers. The model described above allows us to separate these two forces.

Validating these WTP responses is a difficult but important task. They are the key ingredient of our entire exercise, but they may be driven in part by sample selection effects, behavioral biases, or noise. Although these are practical trade-offs that professors face throughout their careers, systematic data have not been collected in this way before to our knowledge. Thus, it is not clear what plausible variation would look like here. Still, we can conduct some tests motivated by economic and statistical theory to explore how reasonable these distributions of WTP are. 

Motivated by the notion of opportunity costs, we test how WTP varies with researchers' implied hourly wage (per their annual salary divided by their total hours of work). Assuming this implied wage rate is a proxy for researchers' opportunity costs, it should be the case that researchers with higher hourly wages are willing to pay more for their free time. In Appendix Figure \ref{fig_binscat_wtpcorr_hr}, we see that this is indeed the case. Furthermore, it should be the case that, conditional on opportunity costs of time, researchers who expect to spend more time fundraising should have a higher WTP for guaranteed funding. In Appendix Figure \ref{fig_binscat_wtpcorr_dol}, we find this to be true. 

As another test, Appendix Tables \ref{tab_wtp_tests} and \ref{tab_wtp_tests_IMRWTP} describe analyses where we regress researchers' WTP responses on the variables in the model as well as ($i$) the large vector of observable features ($\mathbf{X}_i$), ($ii$) an inverse Mills ratio constructed using the randomized survey participation incentives following \cite{heckman1979sample}, and/or ($iii$) researchers' WTP for the benchmark good (high-speed internet) following \cite{dizon2022improving}. We find the variables in the model can explain 82--96\% of the variation in WTP responses (see Appendix Table \ref{tab_wtp_tests}). This supports the model's ingredients as being key determinants of researchers' answers. When we include the large vector of observable features in addition as additional explanatory features, the $R^2$ statistics increase by only one to four percentage points. Residual variation in WTP responses (conditional on the variables of the model) are not well explained by these heterogeneous observables. This gives us confidence that we are not dramatically mis-specifying heterogeneity in the model.

We also find that the inverse Mills ratio is not a statistically significant predictor of responses (see Appendix Table \ref{tab_wtp_tests_IMRWTP}). Under the assumptions outlined in \cite{heckman1979sample}, this supports the notion that respondents did not differentially select into our sample as a function of their WTP for inputs and suggests some generalizability of our responses. Finally, we find that researchers' WTP for the benchmark good (high-speed internet at their house) is correlated with their WTP for inputs (see Appendix Table \ref{tab_wtp_tests_IMRWTP}). Under the assumption that researchers' demand for home internet is truly uncorrelated with their productivity, this provides some evidence of systematic noise in how respondents are reporting their WTP in these experiments. However, the economic magnitude of this relationship is quite small. Furthermore, the benchmark good WTP is included in our vector of features we use to make the researcher type index (that determines heterogeneity in the consumption parameters), which allows us to control for this to some degree. Overall, despite not being incentivized experiments, researchers' responses behave as expected and are of plausible magnitudes.

\section{Estimates for the Model of Science}\label{sec:estimates}

\subsection{A View of Researchers' Utility Functions}
To illustrate our estimates of the utility function, Appendix Figure \ref{fig_util} shows how an average researcher's utility depends on the levels of the three state variables (salary, administrative duties, guaranteed research funding) and research output. The figure shows the percent change in a researcher's utility as the variable is increased from the 10$^{th}$ percentile level to the 90$^{th}$ percentile level while all other variables and parameters are held fixed at the sample averages.

In terms of magnitude, salary and administrative duties are the most important for researchers' utility. Shifting the average researcher's salary from the 10$^{th}$ percentile to the 90$^{th}$ percentile increases their utility by roughly 60\%. An equivalent relative increase in administrative duties reduces utility by about 20\%. In contrast, similarly scaled increases in guaranteed funding or research output raise utility by only a few percentage points.\footnote{As evidence of fit, the median absolute difference between a researcher's stated WTP and the model's prediction ranges from 6--12 percentage points across the four questions. The numerical estimates of the common parameters and more detailed descriptions of the model's fit are available from the authors upon request.}

\subsection{Productivity Distributions}
\begin{figure} \centering
\caption{Production Function Parameters}\label{fig_hist_pfunc}
\subfloat[Funding intensity, $\gamma$]
{
\label{fig_hist_pfunc_gamma}
\includegraphics[width=0.495\textwidth, trim=0mm 70mm 0mm 5mm, clip]{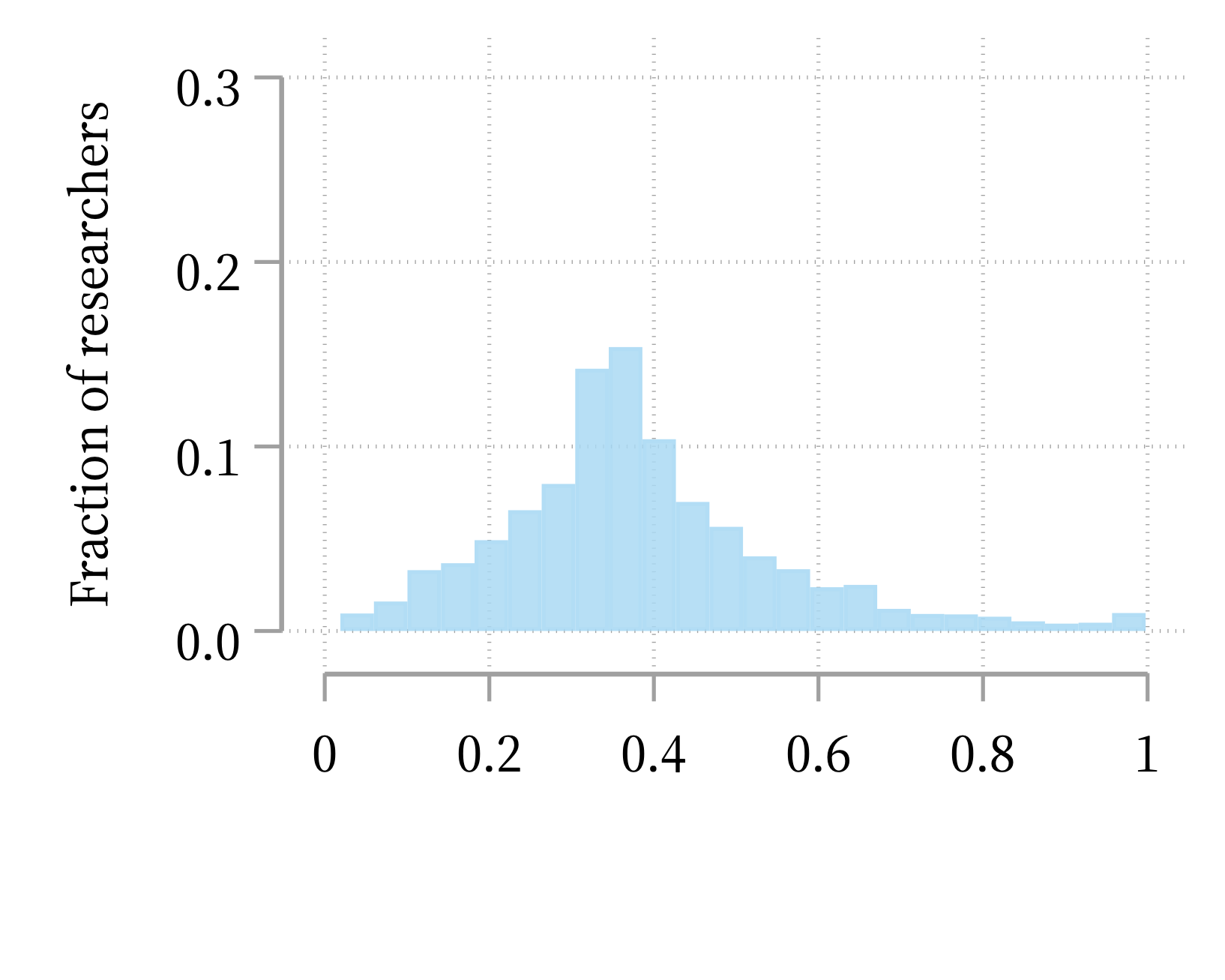}
} 
\subfloat[Scaled TFP, $\alpha$]
{
\label{fig_hist_pfunc_alpha}
\includegraphics[width=0.495\textwidth, trim=0mm 70mm 0mm 5mm, clip]{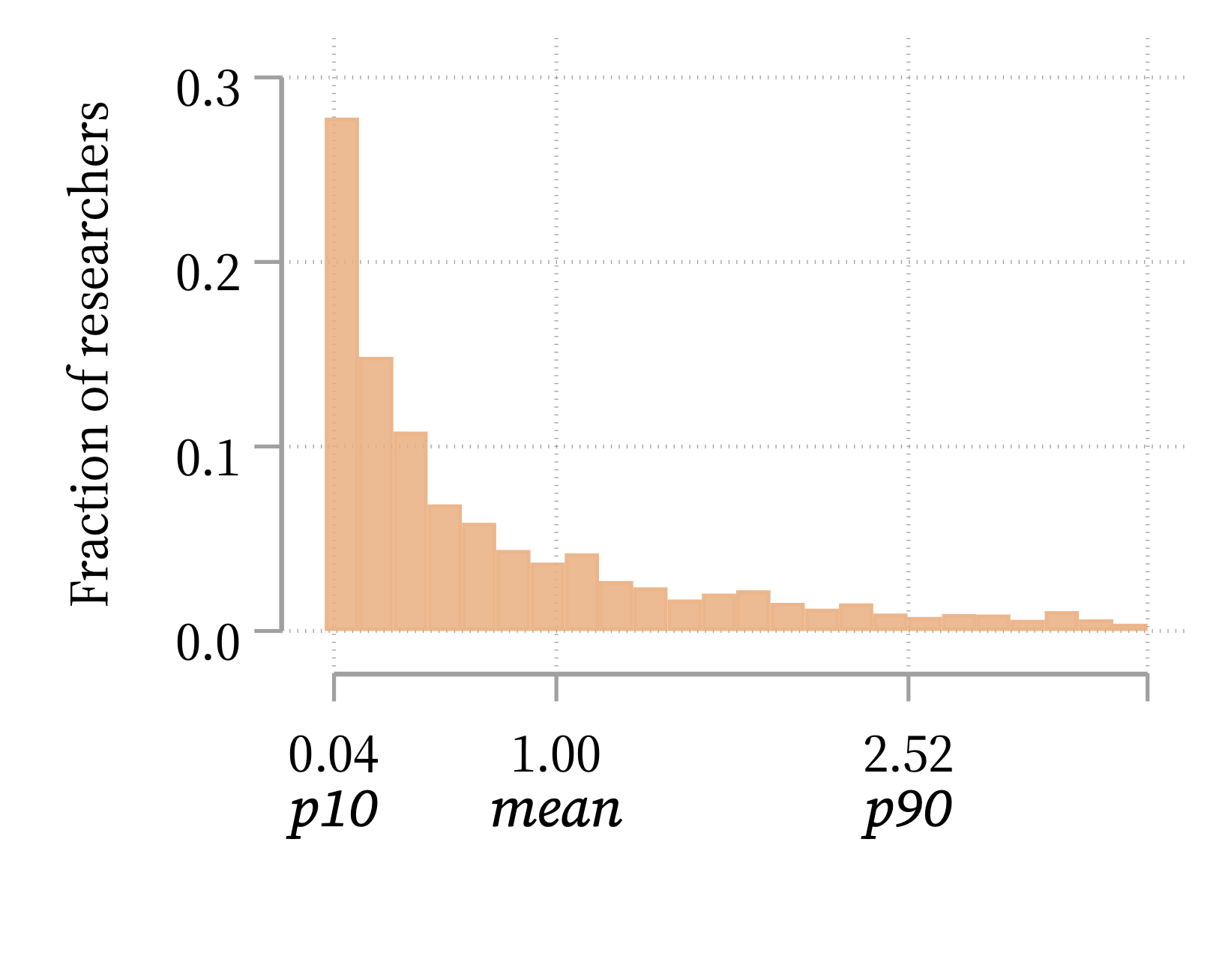}
} \\
\note{\emph{Note}: \input{figtab/note_fig_hist_pfunc.tex}}
\end{figure}

Figure \ref{fig_hist_pfunc} shows the unconditional distributions of the production function parameters $\gamma_i$ (funding intensity) and $\alpha_i$ (TFP). Figure \ref{fig_hist_pfunc_gamma} displays the distribution of the $\gamma_i$ parameter, which describes the relative weight of funding versus research time in researchers' production functions. The distribution highlights a significant degree of heterogeneity across researchers, with roughly 20\% of our sample having a funding intensity either larger than 0.6 or smaller than 0.2.  

As both an interesting exercise and test of face validity, Appendix Figure \ref{fig_gamma_by_field} plots the average funding intensity parameters ($\gamma_i$) for the twenty minor fields represented in the sample. We find that $\gamma_i$ is highest in chemistry and engineering, where research is often capital intensive and involves the use of expensive lab equipment. In contrast, and in line with intuition, the social sciences (e.g., economics, political science) display the lowest funding intensity on average. Notably, Appendix Figure \ref{fig_gamma_by_field} highlights the heterogeneity in funding intensity across researchers even within these narrower fields of study.

Figure \ref{fig_hist_pfunc_alpha} displays the distribution of the $\alpha_i$ parameter, our measure of researchers' TFP. Our estimates reveal a large skew in researchers' beliefs about their productivity.

As one test of face validity, Appendix Table \ref{tab_reg_alphapubcorr} reports regression results that test for associations between researchers' productivity ($\alpha_i$) and their performance per traditional metrics of scientific productivity, which are often simply output levels (e.g., publications, citations). Under some reasonable assumptions, we expect a positive association between these metrics, and we find it.\footnote{Specifically, those assumptions are: ($i$) there is a positive correlation between researchers' input levels and productivity; ($ii$) these traditional output metrics are positively correlated with researchers' true output; ($iii$) researchers' output levels are semi-persistent over time (since we can only work with pre-survey bibliographic data).} Researchers with higher productivity beliefs are more likely to have more actual publication output whether that output is measured using recent publication counts or citation-weighted counts. This provides a signal that our estimates do in fact reflect real productivity differences across researchers.

To understand how variation in TFP affects output, we decompose the variance in log output into components attributable to TFP, input levels, and factor intensity. Note that the variance in log output due to the variance in log TFP is given by: $\text{Var}(\log(\alpha_i)) + 2 \text{Cov}(\log(\alpha_i)\,,\,\gamma_i \log(B_i)) + 2 \text{Cov}(\log(\alpha_i)\,,\,(1-\gamma_i) \log(R_i))$. Using this relationship, we estimate that 46
\unskip\% of the variance in output across researchers is due to the variance in TFP. Without adjusting for the covariances between TFP and the input levels and funding-intensity parameter, we obtain an estimate of 80
\unskip\%. This finding indicates that more productive researchers obtain more inputs that are well-suited to their production functions, and it motivates questions related to allocative efficiency that we return to in the next section. 

\begin{figure} \centering
\caption{TFP Distributions}\label{fig_alpha_distcompare}
\includegraphics[height=0.5\textwidth, trim=0mm 20mm 0mm 5mm, clip]{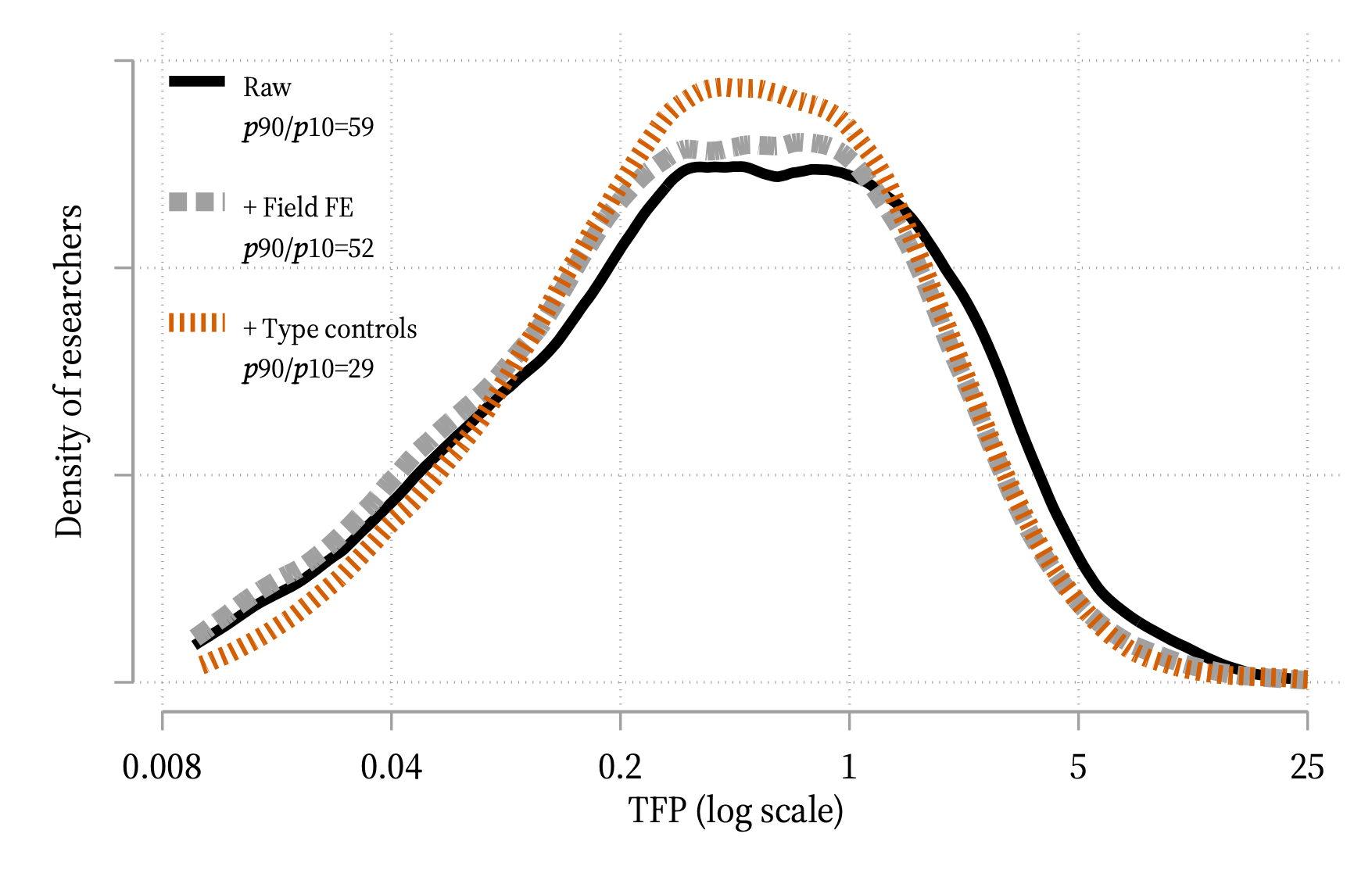}
\-\\
\note{\emph{Note}: \input{figtab/note_fig_alpha_distcompare.tex}}
\end{figure}

While Figure \ref{fig_hist_pfunc_alpha} shows a substantial degree of heterogeneity, some fraction of this variation may be due to differences in the demand for their output or the nature of science within their fields of study; for a similar reason, most studies on the industrial organization of firms report productivity distributions based only on within-industry variation. Thus, Figure \ref{fig_alpha_distcompare} shows the distribution of researcher productivity ($\alpha_i$) after various controls for field-specific (or ``industry-specific'') variation are introduced.

First, Figure \ref{fig_alpha_distcompare} shows the raw distribution of productivity levels, which mirrors the distribution shown in Figure \ref{fig_hist_pfunc_alpha} but now on a logarithmic scale. Next, we regress researchers' TFP estimates on a set of major-field fixed effects, and we report the distribution of residual productivity levels. Lastly, we also condition productivity on the full vector of covariates used to generate the researcher type index.

Figure \ref{fig_alpha_distcompare} also reports the ratio of the 90$^{th}$ and 10$^{th}$ percentiles, the ``90--10 TFP ratio'' measure of variance commonly reported in traditional, firm-level studies. The 90--10 TFP ratio of the raw productivity estimates is substantial, but, even after conditioning on the additional controls, we still find a large dispersion. The 90$^{th}$ percentile researcher believes they are roughly 30 times as productive as the 10$^{th}$ percentile researcher.

It is difficult to benchmark this dispersion. Firm-level estimates are primarily from manufacturing sectors, which typically involve 90--10 TFP ratios on the scale of 1.5$\times$ to 5$\times$ (\citealt{syverson2011determines}). To date, most worker-level productivity estimates tend to be similarly dispersed (\citealt{hoffman2024people}); however, those estimates do not come from occupations as complex and creatively oriented as science. One notable example is \cite{sackman1968exploratory} who used direct observation to estimate software engineers' productivity and found that some individuals were as much as an order of magnitude faster on coding tasks than others. Interestingly, this is the same approximate scale that we identify with our sample of researchers.\footnote{For more evidence of face validity, we note that the 90-10 ratio for field-normalized citations in our sample is 56, which is very similar to the degree of productivity dispersion we estimate.} 

Motivated by the prevalence of power laws in the tails of distributions (\citealt{gabaix2009power}), we examine the distribution of TFP among the most productive researchers. Appendix Figure \ref{fig_topa_powerlaw} shows the traditional log rank versus log value plots along with the regression results following the approach described by \cite{gabaix2009power}. We find clear evidence of power laws when focusing on the top 20\% or top 1\% of researchers per their TFP. The regressions yield power law exponents of approximately 2--3, which are significantly larger than Zipf's law (exponent of 1) but are on the same scale as the power law exponents observed in income distributions and stock prices (\citealt{gabaix2009power}). Future work that estimates high-skilled workers' productivity (and not simply their output) in other settings would be worthwhile.

\section{Counterfactuals and Allocative Efficiency}\label{sec:efficiency}

\subsection{Overview}
In this section, we use our productivity estimates to analyze allocative efficiency through the lens of our model. Specifically, we treat guaranteed funding ($G$) and duties ($D$) as policy levers a planner can adjust to maximize a given objective. Importantly, we search for the allocations of guaranteed funding and duties that maximize an objective after accounting for researchers' endogenous behavioral responses to the planner's decisions. The margins for behavioral responses are total hours worked, time spent fundraising (to obtain additional funds), and time spent on research (to directly produce output). In reality, researchers may endogenously adjust along many other margins (e.g., the types of projects they pursue), but these complexities are beyond the scope of our model. Still, adjustments to time allocations are likely a first-order response, and ignoring other margins allows us to keep the problem tractable.

Within the literature on misallocation, tests of efficiency are performed using either: ($i$) \emph{relative} benchmarks, where allocative efficiency across multiple markets is equated, which reveals how much of the gap in aggregate output between markets is due to allocative frictions (e.g., \citealt{hsieh2009misallocation}); and ($ii$) \emph{absolute} benchmarks, which compare the actual output level to that which a social planner could achieve (e.g., \citealt{petrin2013estimating,asker2019mis}). Our approach allows us to perform both of these types of tests.\footnote{However, we do not explicitly model the sources of inefficiencies that give rise to wedges between actual an optimal input allocations. But, in Sections \ref{sec_friction_general} and \ref{sec_friction_casestudy} we identify and investigate features of producers that are correlated with the size of their input wedge.}

We consider two alternative objectives. First, we estimate the allocation that maximizes total output. Second, we optimize for a utilitarian objective of maximizing researchers' aggregate utility. To aid interpretation, we conduct an exercise where we estimate how much total research funding must be increased \emph{using actual allocations} to achieve the same growth in scientific output that we are able to achieve \emph{using actual funding levels} in alternative allocations. Beyond this, we explore a range of alternative constraints on how inputs are reallocated to draw broader conclusions about the gains from reallocation. Appendix \ref{app:moremodel} contains further details on how we specify the optimization problems and constraints.

The externalities of science loom large in these counterfactuals. As noted, our approach assumes all researchers' production involves the same relative amount of externalities (per the common parameter $\kappa$ that multiplies their private value from output into social value). In practice, externalities likely vary greatly across fields, which is why we primarily focus on reallocations \emph{within} the five major fields of researchers in our sample (i.e., Engineering, Math and related; Humanities and related; Medical and Health Sciences; Natural Sciences; Social Sciences). We also report results where we condition researchers' actual and counterfactual outcomes (i.e., output, utility) on the large vector of covariates used in the researcher type index. This approach allows us to isolate gains from reallocation between researchers with similar observable features.

\subsection{Summary of Counterfactuals}\label{subsec_countarfactuals_summary}
Table \ref{tab_maincounterfac_summaries} details the changes we estimate after reallocating inputs to maximize total output (Cols. 2--3) or researchers' total utility (Cols. 4--5); Column (1) describes the actual allocation of inputs for reference. In all cases of Table \ref{tab_maincounterfac_summaries}, we hold the total amount of funding in each major field fixed, we only reallocate within fields, and social value is evaluated at $\kappa=10$ (implying that scientists capture 10\% of the value of their output). Given our limited ability to incorporate heterogeneous preferences into the model, Columns (3) and (5) report changes in output and welfare after removing variation in those metrics correlated with the covariates used to construct the researcher type index.\footnote{In those cases, we regress researcher-level output or welfare metrics on the model variables and the full vector of covariates used in the type index, and then we subtract out variation in the metric predicted by the covariates conditional on the model variables. We include the model variables as controls because the productivity parameters are (partially) determined by them and so we do not want to remove variation in outcomes due to productivity differences.}

\begin{table}[h!] \centering \small
\caption{Outcomes given Actual and Optimized Allocations}\label{tab_maincounterfac_summaries}
\-\\
\begin{tabular}{l r c r r r r } 
 \hline\hline 
 & \multicolumn{1}{c}{Current} & & \multicolumn{4}{c}{Optimized} \\ 
 & \multicolumn{1}{c}{allocation} & & \multicolumn{4}{c}{allocations} \\ \cline{2-2} \cline{4-7} 
 & \multicolumn{1}{c}{(1)} & & \multicolumn{1}{c}{(2)} & \multicolumn{1}{c}{(3)} & \multicolumn{1}{c}{(4)} & \multicolumn{1}{c}{(5)} \\ 
 \hline\underline{\emph{Research inputs}}&&&&&& \\
Research hrs./week, avg.&18.5&&+27\%&+27\%&+14\%&+14\% \\
Research hrs./week, s.d.&9.6&&+43\%&+43\%&+43\%&+43\% \\
Budget \$-K/year, avg.&147.1&&0\%&0\%&0\%&0\% \\
Budget \$-K/year, s.d.&206.9&&--18\%&--18\%&--20\%&--20\% \\
&&&&&& \\
\underline{\emph{Research output}}&&&&&& \\
Output, avg.&\emph{\footnotesize{n.r.}}&&+160\%&+140\%&+160\%&+130\% \\
Output, s.d.&\emph{\footnotesize{n.r.}}&&--5.2\%&--7.9\%&--5.4\%&--5.7\% \\
Output per hr.&\emph{\footnotesize{n.r.}}&&+100\%&+91\%&+130\%&+110\% \\
&&&&&& \\
\underline{\emph{Welfare}}&&&&&& \\
Researcher utility, avg.&\emph{\footnotesize{n.r.}}&&+4.5\%&+3.3\%&+4.7\%&+3.2\% \\
Researcher utility, s.d.&\emph{\footnotesize{n.r.}}&&--5.2\%&--7.9\%&--5.4\%&--5.7\% \\
Researcher utility per hr.&\emph{\footnotesize{n.r.}}&&+25\%&+24\%&+16\%&+15\% \\
&&&&&& \\
Social value, avg.&\emph{\footnotesize{n.r.}}&&+16\%&+5.7\%&+16\%&+5.2\% \\
Social value, s.d.&\emph{\footnotesize{n.r.}}&&+21\%&--6.5\%&+20\%&--3.4\% \\
Social value per hr.&\emph{\footnotesize{n.r.}}&&+34\%&+27\%&+26\%&+19\% \\
\hline
Objective, max&&&$ Y$&$ Y$&$ \mathcal{V}$&$ \mathcal{V}$ \\
Type controls&&&&$ \checkmark$&&$ \checkmark$ \\
\hline
 \hline 
 \end{tabular}
\-\\
\note{\emph{Note}: Reports summary statistics for inputs under actualallocations (Col. 1). The first three sets of rows in Columns 2--5 report the percentage change in research inputs (\emph{Research inputs}), outputs (\emph{Research outputs}), and utility (\emph{Welfare}) under alternative allocations; estimates are rounded to aid in comparison. The bottom sets of rows outline the objective of the counterfactuals explored in Columns 2--5. The two different objectives explored are maximizing output ($ Y$) or researchers' private utility ($ \mathcal{V}$). Models with \emph{Type controls} report output and welfare changes after removing residual variation due to covariates used to construct the researcher type index. All optimized allocations allow for researchers' behavioral responses.
}
\end{table}

\paragraph{Input Reallocation} The first four rows of Table \ref{tab_maincounterfac_summaries} report how reallocation changes the equilibrium distribution of inputs. For these rows, Columns (2--3) are identical since they are based on the same counterfactual and we do not include any controls for these statistics; likewise for Columns (4--5). Compared to the status quo, the counterfactual allocations lead to more research time on average (approximately 20\%), they increase the variance in research time (approximately 40\%), and they decrease the variance in research budgets (approximately --20\%).

In the Appendix, we illustrate the actual and optimal input distributions as well as Lorenz curves (see Figure \ref{fig_lorenz_actvopt}) to show how more or less unequal the distributions are under actual and optimal allocations. In general, the model opts to make the distribution of duties more unequal, while the inequality of the distribution of guaranteed funding is relatively unchanged. After researchers' behavioral responses to these reallocations, there is little change in the inequality of the distribution of research time and total funding (see Figure \ref{fig_lorenz_actvopt}).

\paragraph{Gains from Reallocations} Overall, the model suggests that there are large gains in output to be had from alternative allocations. Whether the objective is to maximize output or researchers' private utility, the new allocations yield roughly 130--160\% more output. Given the changes in research time, this translates into significant welfare gains on the scale of 3--4\% for researchers and 5--15\% for society. The finding that both objectives yield qualitatively similar gains is primarily driven by the fact that researchers have control over their time. Whatever the planner hopes to maximize, the approach is to ensure that the most productive researchers are incentivized to spend more time working on their science.

To contextualize these gains, we estimate how much more funding under actual allocations would be necessary to achieve the same growth in scientific output that our counterfactual allocations achieve using current funding levels. For simplicity, we assume these additional funds are injected into the market as proportional increases in guaranteed funds ($G$) for all researchers. Specifically, we solve for the percentage increase in researchers' $G$ that ultimately yields the same total growth in output reported in Table \ref{tab_maincounterfac_summaries}. Importantly, a 1\% increase in $G$ for all researchers can yield a $\lesseqgtr$1\% increase in aggregate funding $B$ and, in turn, output $Y$ even in the absence of a behavioral response despite the fact that we have specified production as constant returns to scale. We formally describe this mechanical composition effect in Appendix \ref{subsec_compeffect}. In short, since guaranteed funding $G$ is (weakly) less than total funding $B$, the percentage change in a scientists' total funding and output will vary across researchers depending on correlations between initial output shares, funding intensity parameters ($\gamma_i$), and the share of funding from guarantees ($G_i/B_i$); see Appendix \ref{subsec_compeffect} for more discussion. 

Table \ref{tab_alternatives} reports these estimates under both a ``\emph{Mechanical}'' scenario that does not incorporate researchers' behavioral responses and a ``\emph{Behavioral}'' scenario that does. Allowing for the behavioral response, this exercise indicates that funding guarantees would need to grow by roughly 200\% for the actual allocations to achieve the same growth in output that we observe in the counterfactuals. In the case allowing a behavioral response, this corresponds to total research budgets increasing by roughly 40\%. On an annual basis this amounts to roughly \$60,000 per researcher. Scaling these results from our sample to the entire population of researchers targeted by the survey, this implies the gains from a more efficient allocation are equivalent to a funding increase of roughly \$14 billion per year.

\begin{table}\centering \small
\caption{Growth in Funding with Actual Allocations Needed to Produce the Same Output \\ as Optimized Allocation of Actual Funding}\label{tab_alternatives}
\-\\
\begin{tabular}{l r r } 
 \hline\hline 
 & \multicolumn{1}{c}{Mechanical} & \multicolumn{1}{c}{Behavioral}  \\ 
 \hline\emph{\underline{Relative growth in funding}}&& \\
Guaranteed, $ G$&210\%&206\% \\
Total, $ B$&85\%&41\% \\
&& \\
\emph{\underline{Absolute growth in total funding, $ B$}}&& \\
Sample average&+\$126K&+\$60K \\
Sample total&+\$503M&+\$240M \\
Population total&+\$30,186M&+\$14,444M \\

 \hline\hline 
 \end{tabular}
\-\\
\note{\emph{Note}: Reports the increase in guaranteed ($ G$) and total ($ B$) research funding under actual allocations necessary to achieve the same growth in output achieved by the reallocation of actual input levels that maximizes researchers' utility. Sample averages on a per-researcher basis are reported in addition to sample totals (summing over all in-sample researchers) and implied population totals (scaling the sample up to the population size per the survey response rate). The \emph{Mechanical} scenario does not allow researchers to re-optimize their time allocations and the \emph{Behavioral} scenario does.
}
\end{table}

Notably, allowing for the behavioral response is economically important. More guaranteed research funding (which substitutes for fundraising and complements researcher time) leads researchers to spend less time fundraising and more time on their research (approximately +25\% in the aggregate). Without accounting for researchers' endogenous response to budget growth, we underestimate the impact of growing the budget by roughly two-fold.

\paragraph{Additional Results} In the Appendix, we report results from alternative counterfactual exercises (Appendix Table \ref{tab_counterfac_summaries}). There are a few findings of note. First, the allocation of duties is only meaningfully relevant for influencing researchers' utility, whereas the allocation of funding is what has the main influence on output.\footnote{When only duties are reallocated, researchers' utility increases by roughly 3\%, but output only increases by roughly 1\%. When only funding is reallocated, researchers' utility increases by only 1\%, but output increases by 150\%. Maximizing output or researchers' utility yield relatively similar outcomes} Second, and rather interestingly, when we allow the total research budget to be unconstrained, and therefore only researchers' fundraising decisions constrain the size of the budget, the total research budget grows only about 15\%. While we caution against interpreting this result too seriously, but it suggests that the total research budget is not far from the optimum given the (fixed) size of the research workforce. Of course, this ignores dynamic concerns which may certainly be relevant. In other unreported analyses, we find qualitatively similar outcomes when we alter our assumptions about input caps, preference heterogeneity, and reallocation constraints, which suggest our results do not hinge on any specific modeling assumption.\footnote{These additional specification tests are available from the authors on request.}

\subsection{Input Wedges: Actual Versus Optimal}\label{sec_friction_general}
The results summarized thus far suggest that reallocating researchers' time and funding constraints can yield significant gains in output and welfare. In order to better understand how these gains are achieved, we now focus on a single counterfactual specification and compare the actual and optimized input levels. For simplicity, we focus for the remainder of this section on the scenario where duties and funding are jointly optimized to maximize output, which corresponds to the counterfactual described in Columns (2--3) of Table \ref{tab_maincounterfac_summaries}.

To better understand how actual and optimal input levels correlate at the individual level, we estimate a series of regressions of the following form:
\begin{equation}\label{eq:wedge_predict}
\text{Actual Input Level}_i = a + \beta \text{Optimal Input Level}_i + \delta Z_i + \epsilon_i \;,
\end{equation}
which relates researchers' actual and optimal input levels possibly conditioning on one (or more) covariate $Z$. If allocations were perfectly efficient, such a regression would yield an estimate $\widehat{\beta}=1$ (with no standard error) since actual levels would equal optimal levels. If allocations are not efficient, then $\widehat{\beta}<1$.

\begin{figure} \centering
\caption{Actual and Optimal Input Level Correlations}\label{fig_binscat_actvopt}
\subfloat[Research Time]
{
\label{fig_binscat_actvopt_C_Ri}
\includegraphics[width=0.475\textwidth, trim=0mm 10mm 0mm 5mm, clip]{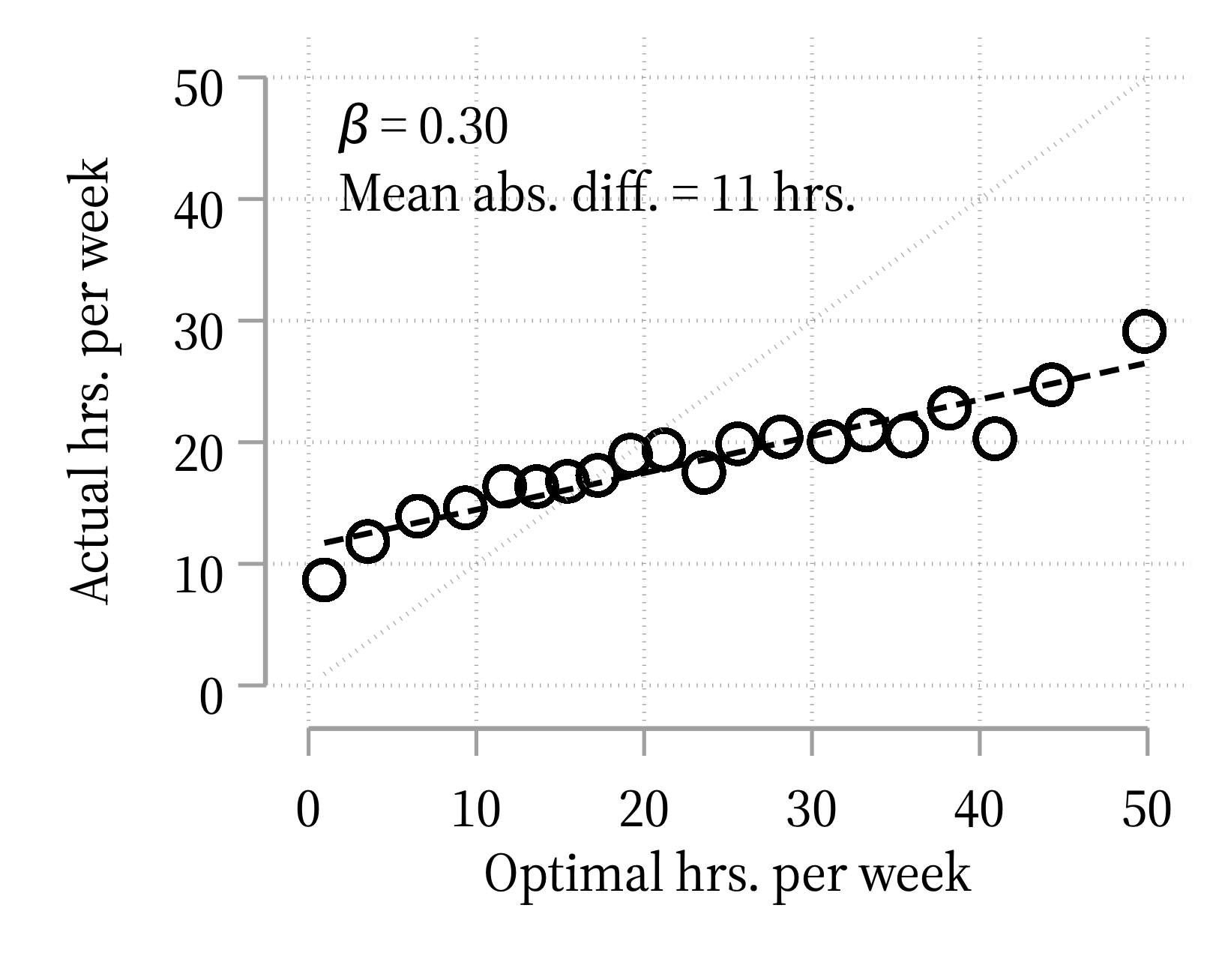}
} 
\subfloat[Total Research Funding]
{
\label{fig_binscat_actvopt_C_Bi}
\includegraphics[width=0.475\textwidth, trim=0mm 10mm 0mm 5mm, clip]{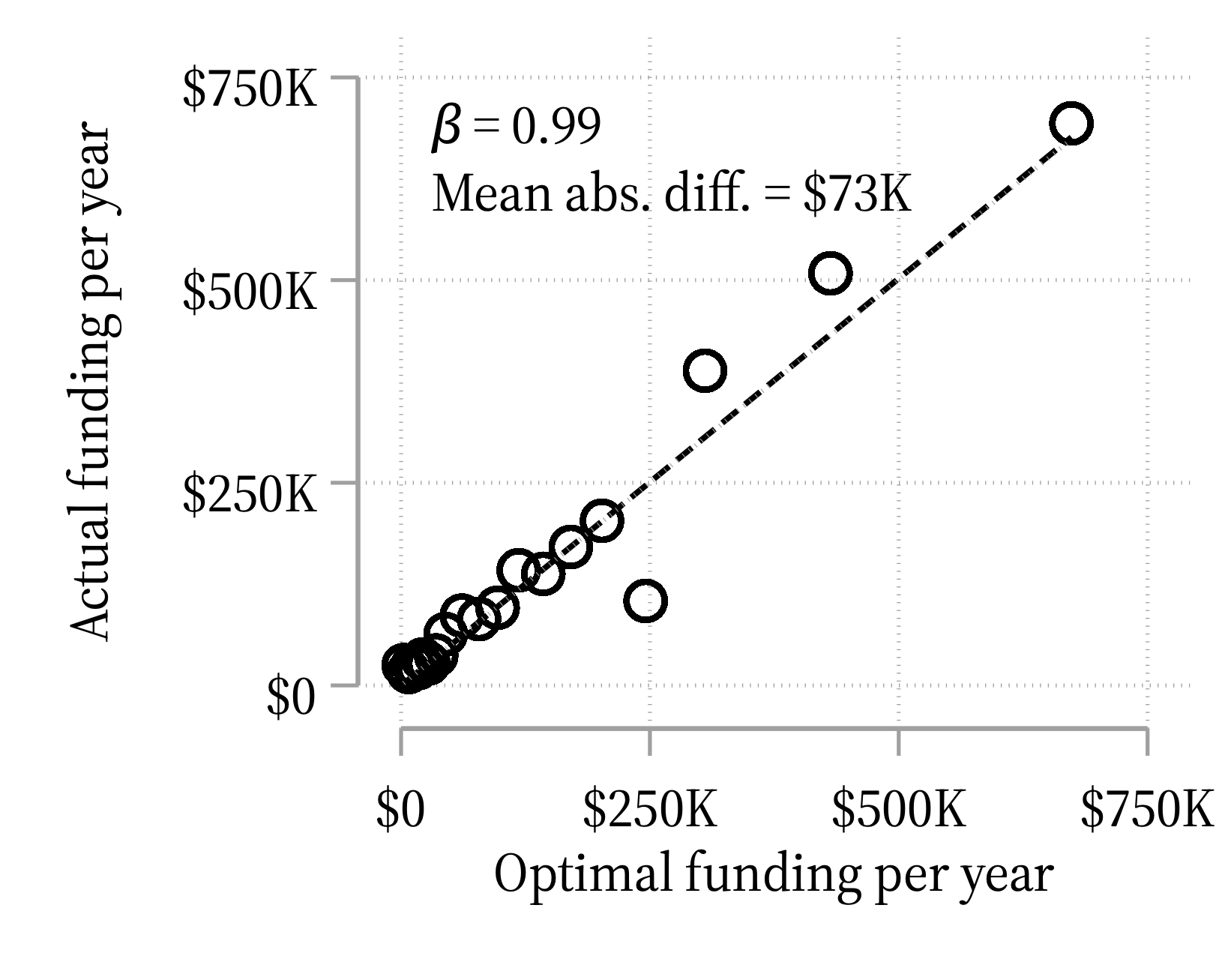}
}
\-\\
\note{\emph{Note}: \input{figtab/note_fig_binscat_actvopt.tex}}
\end{figure}

Focusing first on estimating Equation \ref{eq:wedge_predict} with no covariates, Figure \ref{fig_binscat_actvopt} reports estimates of $\beta$ on a binned scatterplot of the data. For both inputs, there is a clear positive relationship between researchers' actual and optimal levels. For every hour a researcher \emph{should} commit to research, they actually commit 0.3 hours on average. Interestingly, for every one dollar in total funding the researcher \emph{should} have, they actually have nearly one dollar (0.987) on average. The figure also reports the mean absolute differences between actual and optimal levels, which are approximately 10 hours of research time and \$70,000 in funding. For reference, these values are both roughly half the sample means.

Appendix Table \ref{tab_wedgereg_full} Column (1) replicates these univariate regressions, and then Columns (2--5) include additional sets of covariates, which are all subsets of the vector of covariates used in the researcher type index: researchers' professional positions (e.g., rank, tenure status); their subjective measures of their research output (i.e., as reported in Table \ref{tab_sumstat_qualoutput}); and their socio-demographic features (e.g., age, gender, household structure). 

Examining the $R^2$ statistics reported in Table \ref{tab_wedgereg_full} Columns (2--5) as we include different sets of covariates generally reveals small increases in $R^2$ compared to Column (1). At most, the inclusion of the full vector of covariates increases the $R^2$ by 4--14 percentage points. We interpret this pattern as indicating that misallocation using these sorts of observable features is a challenging exercise, and that the degree of misspecification in the model (along these specific dimensions) is relatively small. The latter gives us confidence in our results, and the former has the interesting implication that predicting which researchers are over- or under-resourced based only on their observables may be a challenging exercise.

\subsection{Case Studies of Potential Frictions}\label{sec_friction_casestudy}
When estimating regressions of the form in Equation \ref{eq:wedge_predict}, it is important to be careful when interpreting the $\delta$ coefficient. A covariate $Z$ may be a significant predictor of actual levels conditional on optimal levels (i.e., $\widehat{\delta}\neq0$) for two reasons: ($i$) the feature truly is a predictor of misallocation as implied by the model, in which case a positive (negative) association indicates that the feature is predictive of a researcher being over-resourced (under-resourced) due to a friction; ($ii$) there is misspecification in the model and the feature describes some heterogeneous preferences or demand variation that we have failed to capture, which has led to bias in our productivity estimates. We cannot separate these two possibilities. 

Here, we embrace the interpretation that our estimates of $\delta$ are indicative of a friction (and not misspecification) only for two features for which misspecification is unlikely to be a major concern, but these results should still be interpreted cautiously.

First, we focus on a highly scrutinized feature: gender. A large body of work has documented a wide range of biases and frictions facing female researchers when it comes to the acquisition of inputs (or credit) for their science.\footnote{See, for example, \cite{witteman2019gender,kim2021women}.} But while the vast majority of this work rejects null hypotheses and finds female researchers are under-resourced, they typically cannot formally quantify \emph{how much} female researchers are under-resourced. Our model and approach allow us to do just that.

In Appendix Table \ref{tab_wedgereg_full}, Columns (4--5) report the results from estimating regressions of the form shown in Equation \ref{eq:wedge_predict}, where $Z_i$ is an indicator for researchers who self-report as female. Whether we only include other socio-demographic covariates (Col. 4) or the full set of covariates (Col. 5), we estimate a statistically significant negative association indicating female researchers are under-resourced. They spend approximately 1 hour fewer on their research per week and have roughly \$10,000 less in total research funding annually. Both of these wedges are approximately 10\% of the sample mean. 

Next, we focus on another question that has received much attention by meta-science scholars: the Matthew effect. In general, the Matthew effect posits that researchers amass resources beyond what their productivity warrants due to their social status (\citealt{merton1968matthew}). The specific version of this effect that we can test for is the degree to which the use of grant dollars and publication outcomes as productivity proxies distorts the allocation of inputs (e.g., \citealt{lee2013bias,gralka2019measure}). Again, we run regressions of the form shown in Equation \ref{eq:wedge_predict}, now including a vector of variables describing researchers' recent grant funding and publication or citation output. As reported in Appendix Table \ref{tab_reg_productivityproxies}, we find some evidence of statistically significant distortions whereby a one standard deviation increase in these traditional proxies leads to over-resourcing on the scale of roughly 0.1--0.2 standard deviations (approximately 5--10\% relative to the sample means). 

These two cases reveal statistically significant predictors of the input wedges that are plausibly due to frictions in the allocation process. Researchers are more likely to be under-resourced if they are female, and they are more likely to be over-resourced if their observable input and output measures are higher. Still, the $R^2$ statistics shown in Table \ref{tab_wedgereg_full} convey a seemingly novel point---these observable features, which have received so much attention thus far, explain a very small amount of the misallocation implied by the model. It appears difficult to use standard observable features to predict which researchers are over or under-resourced.

\subsection{Output Differences across Major Fields}
Our final exercise seeks to understand the determinants of scientific output differences across the major fields of researchers. Doing so takes a strong stance on the comparability of output across fields, which is debatable. Thus, we treat this exercise as more speculative.

In general, the aggregate output of a market depends on input levels (i.e., the number of researchers, their funding, and their time spent on research), productivity levels (i.e., researchers' TFP), and the market's allocative efficiency. Here, we explore the relative importance of these three dimensions.

\begin{figure}[h!] \centering
\caption{Across Field Output Comparisons}\label{fig_output_decomp}
\includegraphics[width=\textwidth, trim=0mm 10mm 0mm 5mm, clip]{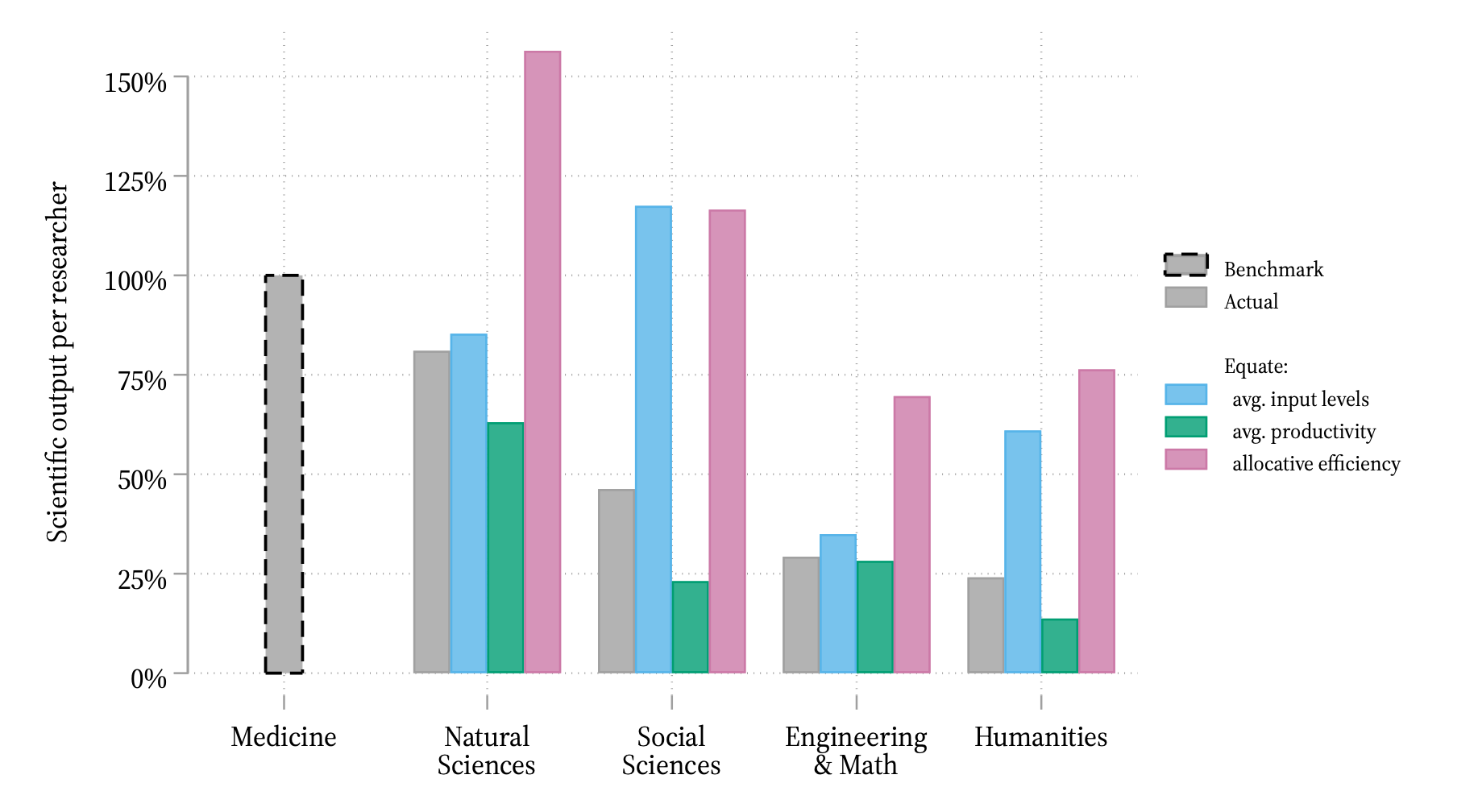}
\-\\
\note{\emph{Note}: \input{figtab/note_fig_output_decomp.tex}}
\end{figure}

First, we divide all fields' total output by the number of researchers in that field in order to compare the per capita output only. Next, we use the field with the most output, Medicine, as a benchmark and consider the other fields' output in percentage terms relative to Medicine's actual total (per capita) output. The gray bars in Figure \ref{fig_output_decomp} plot the total scientific output implied by the data and our productivity estimates across the five major fields of study. The four comparison fields have aggregate output levels that are roughly 25--75\% that of Medicine.

Medicine is the most resourced field, so our first test is to equate input levels across fields. In this case, aggregate output in the Social Sciences and the Humanities more than doubles. But, except for the Social Sciences most of the gap in aggregate output between each field and Medicine remains.

Average TFP levels across the fields are relatively similar such that equating productivity has little impact on differences in aggregate output. In fact, the average TFP level in Medicine is slightly lower than that of all other fields, which may be related to the size of the field.

Lastly, we use our estimates from the counterfactual where we maximize output within each of the five major fields to estimate aggregate output gaps in the scenario where all fields are equal in their allocative efficiency (i.e., output is maximized).\footnote{Since this leads to output growth in all fields, we re-scale aggregate levels so that the level observed in the field of Medicine remains the benchmark at 100\%.} Here, we see a reordering of fields in terms of their per capita output. The Natural and Social sciences would produce more output than Medicine, and the other two fields would rise to producing nearly 75\% the output of Medicine. In this sense, differences in allocative efficiency appear to be the most important determinant of differences in aggregate output across fields of science.

We can only speculate as to why, based on our sample of medical researchers, the field of medicine appears to be the best at allocating resources efficiently. One might think that medical researchers spend more time engaged in fundraising efforts and that process facilitates positive selection on productivity; however, medical researchers exhibit quite average levels of fundraising effort.\footnote{The field-specific averages for fundraising hours per week are as follows: Engineering \& Math = 6.1; Humanities = 2.7; Medicine = 4.9; Natural sciences = 6.7; Social Sciences = 2.5.}  The mean duty level for medical researchers is slightly lower relative to other fields, and the variance is much larger. This sort of allocation is consistent with the logic of our model and results --- in a world with highly dispersed productivity, the planner should move non-research duties onto as few, low productivity (in the scientific sense) researchers. Of course there are both practical constraints and unmodeled objectives. For example, high-productivity researchers could plausibly have the largest externalities from their other duties (e.g., teaching) and the allocations we observe are balancing social value from research with social value from those other duties. Further work on understanding across-field differences in allocation mechanisms in science seems warranted.

\section{Discussion}\label{sec:discuss}
Studies of industrial markets continue to find that a substantial share of the variation in output across sectors and regions can be attributed to differences in allocative efficiency. In this paper, we use a new approach to productivity estimation to study the market for science. Our approach allows us to estimate researchers' productivity beliefs without observing output quantities or input prices. We find actual input allocations to be positively correlated with those that maximize plausible objectives, but we also find large gains to be had from more efficient allocations. In the counterfactuals we explore, total scientific output per research-hour or per research-dollar could be more than doubled.

Our model abstracts away from many important considerations that likely drive actual allocations (e.g., dynamic returns), and our data relies on researchers' stated preferences. These and other limitations of our approach motivate us to interpret these magnitudes as plausible upper bounds on the gains from more efficient use of scientific inputs. Still, these large potential gains provide a glimmer of hope amid declining R\&D productivity across most sectors of the economy (\citealt{bloom2020ideas}) and the persistently growing burden of knowledge that raises the cost of conducting frontier science (\citealt{jones2009burden}). Furthermore, our approach to identifying productive scientists may prove useful in talent selection processes more generally (e.g., \citealt{agarwal2020invisible}).

Our approach has deep roots in the economics of labor (i.e., surveys of time use and work-leisure trade-offs), industrial organization (i.e., production function estimation), marketing (i.e., using surveys to identify demand functions), and macroeconomics (i.e., models of factor misallocation). By drawing on insights from these fields, our methodology allows us to overcome many of the challenges that have long plagued our understanding of productivity and efficiency in science.

Our analyses reveal researchers' \emph{beliefs} about their productivity and therefore their \emph{beliefs} about how well inputs are allocated. This is a crucial limitation, since there are clearly many potential biases affecting these beliefs. Still, science is inherently about forecasting uncertain outcomes, and, therefore, the optimal mechanisms for identifying productive researchers and allocating them more inputs will need to tackle this challenge of engaging with researchers' forecasts of their productivity. Specifically, one important next step in this line of work will be to ensure that the producers being studied report their willingness-to-pay for inputs truthfully. Of course, the theoretical underpinnings of how to elicit true willingness-to-pay estimates have long been established (e.g., \citealt{becker1964measuring}). However, the magnitudes of costs in our are on the scale of tens to hundreds of thousands of dollars. Thus, developing the practical details of incentivizing truthful responses from producers at this scale would be a fruitful endeavor.

More broadly, our approach may prove useful in other settings. There are many markets populated by a large number of producers acquiring inputs in a highly decentralized way to produce outputs that are not easy to observe. For example, in developing economies, accurate producer-level data on outputs can be difficult to obtain (e.g., \citealt{tybout2000manufacturing}); in entrepreneurship, many organizations never produce observable output before exiting (e.g., \citealt{decker2014role}); and in nonprofit sectors, the output may be so high-dimensional that reaching consensus on a suitable proxy is challenging (e.g., \citealt{philipson2001medical}). In each of these examples, our methodology could provide a way forward to better understand the distribution and determinants of productivity and efficiency.


\begin{spacing}{1.25}
\bibliography{\bib}
\end{spacing}

\newpage
\section*{Supplemental Appendix}
\appendix

\section{Example Applications of Methodology}\label{app:method}
\setcounter{table}{0}
\setcounter{figure}{0}
\setcounter{equation}{0} 
\renewcommand{\theequation}{A\arabic{equation}}
\renewcommand{\thetable}{A\arabic{table}}
\renewcommand{\thefigure}{A\arabic{figure}}

\subsection{Application 1: Constant Returns to Scale and Convex Costs}
Consider first the case of a firm that operates a linear production function in a single input, e.g., labor, and faces convex adjustment costs because, e.g., there are hiring and firing frictions. The optimization problem is:
\begin{equation}
\begin{aligned}
   \max_{l_i} \ \ \alpha_i l_i -  w l_i - c l_i^{\psi}
\end{aligned}
\end{equation}
with $b(\cdot) = \alpha_i l_i + m_i$ and $m_i=0$, $c(\cdot) =  w l_i + c l_i^{\psi}$ ($c>0$ and $\psi>1$), and $\bm{\upmu}_i = \bm{\upmu} = (w, c, \psi)$.
The optimality condition of the problem is:
\begin{equation}
    \alpha_i - w - c \psi l_i^{\psi-1} = 0
\end{equation}
In the first step of the estimation procedure, we evaluate the optimality condition at observed allocations to characterize $\alpha_i$ as a function of parameters and observed allocations:
\begin{equation}
    \alpha_i(\hat{l}_i,w,c,\psi) = w + c \psi \hat{l}_i^{\psi-1}
\end{equation}
where $\hat{l}_i$ denotes the observed input allocation. Moreover, conditional on attributes, parameters, and prices, the solution is determined by:
\begin{equation}
    l^*_i = \left( \frac{\alpha_i - w}{\psi c} \right)^{\frac{1}{\psi-1}}
\end{equation}
In the second step, we implement the thought experiment and we offer to the firm $\Delta>0$ extra units of labor, which the firm can hire by just paying $WTP_i$ dollars and without incurring the convex adjustment cost. Of course, the firm can re-optimize the quantity of labor that hires at market conditions. Therefore, the problem is:
\begin{equation}
\begin{aligned}
   \max_{\tilde{l}_i} \ \ \alpha_i (\tilde{l}_i+\Delta) -  w \tilde{l}_i - c \tilde{l}_i^{\psi} - WTP_i
\end{aligned}
\end{equation}
and the solution:
\begin{equation}
    \tilde{l}^*_i = \left( \frac{\alpha_i - w}{\psi c} \right)^{\frac{1}{\psi-1}} = l^*_i
\end{equation}
Replacing the solution in the profit function and equating profits at current allocations and in the thought experiment, we obtain:
\begin{equation}
    WTP_i = \alpha_i(\hat{l}_i,w,c,\psi)  ( \tilde{l}^*_i + \Delta) - w \tilde{l}^*_i - c \left(\tilde{l}^*_i\right)^\psi - \alpha_i(\hat{l}_i,w,c,\psi)   l^* + w l^*  + c \left(l^{*}\right)^\psi
\end{equation}
Using the fact that $\tilde{l}^*_i =l^*_i$, we obtain:
\begin{equation}
    WTP_i = \alpha_i(\hat{l}_i,w,c,\psi) \Delta = \left( w + c \psi \hat{l}_i^{\psi-1} \right) \Delta
\end{equation}
which allows the identification of $w,c,\psi$ and thus $\alpha_i(\hat{l}_i,w,c,\psi)$, given $\hat{l}_i$ and $\Delta$. The same logic applies to the case of decreasing returns to scale, namely:
\begin{equation*}
   \max_{l_i} \ \ \alpha_i l_i^{\beta} -  w l_i - c l_i^{\psi}
\end{equation*}
with $v(\cdot)= \alpha_i l_i^{\beta} + m_i$ for $m_i=0$ and  $\beta\in(0,1)$, $c(\cdot) =  w l_i + c l_i^{\psi}$ for $c>0$ and $\psi>1$, and $\bm{\upmu}_i = \bm{\upmu} = (\beta, w, c, \psi) $. In the general case, the optimal input choice does not admit a closed-form expression, but we can show that $\frac{\partial \tilde{l}^*_i}{\partial \Delta} \neq -1$, i.e., the additional input does not perfectly crowd out the quantity of input sourced in the market. This is a sufficient condition such that $WTP_i$ does depend on $\alpha_i$ and thus on all estimands $\bm{\upmu}$.

\subsection{Application 2: Decreasing Returns to Scale and Linear Costs}
Here, we show an example of how a linear cost schedule renders our approach unable to recover productivity. The setting is:
\begin{equation*}
   \max_{l_i} \ \ \alpha_i l_i^{\beta} -  w l_i
\end{equation*}
i.e., the problem coincides with the previous application for $\psi=0$. The first order condition is $\beta \alpha_i l^{\beta-1}_i = w$, which identifies:
\begin{equation}
    \alpha_i(\hat{l}_i,\beta,w) = \left( \frac{w}{\beta} \right) \hat{l}^{1-\beta}_i 
\end{equation}
and determines the optimal allocation as:
\begin{equation}
    l^*_i = \left( \frac{\beta \alpha_i}{w} \right)^{\frac{1}{1-\beta}}
\end{equation}
Under the thought experiment, the optimization problem becomes:
\begin{equation*}
   \max_{\tilde{l}_i} \ \ \alpha_i (\tilde{l}_i+
   \Delta) ^{\beta} -  w \tilde{l}_i
\end{equation*}
which implies: 
\begin{equation}
    \tilde{l}^*_i = \left( \frac{\beta \alpha_i}{w} \right)^{\frac{1}{1-\beta}} - \Delta = l^*_i - \Delta
\end{equation}
Replacing in the profit function and deriving $WTP_i$, we obtain:
\begin{equation*}
\begin{aligned}
    WTP_i = & \alpha_i (\tilde{l}^*_i +\Delta)^{\beta} - w \tilde{l}^*_i  - \alpha_i l^*_i + w  l^*_i \\
    = & \alpha_i (l^*_i - \Delta +\Delta)^{\beta} - w (l^*_i-\delta)  - \alpha_i l^*_i + w  l^*_i \\
    = & w \Delta
\end{aligned}
\end{equation*}
where from the first to the second line we replace $\tilde{l}^*_i = l^*_i - \Delta$. Because the willingness to pay just depends on $w$, $\beta$ cannot be identified, which implies that also $\alpha_i$ is not pinned down.

\subsection{Application 3: Individuals as Producers with Utility Function}
Consider the problem of an agent that gets utility from income $m_i$ and output $\alpha_i d_i^{\beta} l_i^{1-\beta}$---produced using a fixed input $d_i$ and labor $l_i$---and gets disutility from working. The problem is:
\begin{equation}
\begin{aligned}
    \max_{l_i} \ \ \ln m_i + \left( \alpha_i d_i^{\beta} l_i^{1-\beta} \right)^{\eta} - \phi l_i^{\psi}
\end{aligned} 
\end{equation}
with $b(\cdot) = \ln mi_i + \left( \alpha_i l_i^{\beta} d_i^{1-\beta} \right)^\eta$ for $\beta\in(0,1)$ and $ \eta \in (0,1)$, $c(\cdot) =  \phi l_i^{\psi}$ for $\phi>0,\psi>1$, and $\bm{\upmu}_i = \bm{\upmu} = (\beta, \eta, \phi, \psi)$.
The optimality condition is:
\begin{equation}
    \eta (1-\beta) \alpha_i^\eta d_i^{\eta \beta} l_i^{(1-\beta)\eta-1} = \phi \psi l_i^{\psi-1}
\end{equation}
Therefore:
\begin{equation}
    \alpha_i (\hat{l}_i,d_i,\beta,\eta,\phi,\psi) = \left( \frac{\phi \psi \hat{l}_i^{\psi-1}}{\eta (1-\beta) d_i^\eta \hat{l}_i^{\eta(1-\beta)-1}} \right)^{\frac{1}{\eta}}
\end{equation}
and 
\begin{equation}
    l^*_i = \left( \frac{\eta (1-\beta) \alpha_i^{\eta} d_i^{\eta \beta}}{\phi \psi} \right)^{\frac{1}{\psi - \eta (1-\beta)}}
\end{equation}
In the setting of a thought experiment where $\Delta$ additional units of the fixed input are offered to the agent, the problem becomes:
\begin{equation}
\begin{aligned}
    \max_{\tilde{l}_i} \ \ \ln \left( m_i-WTP_i\right) + \left( \alpha_i (d_i+\Delta)^{\beta} \tilde{l}_i^{1-\beta} \right)^{\eta} - \phi \tilde{l}_i^{\psi}
\end{aligned} 
\end{equation}
with 
\begin{equation*}
    \tilde{l}^*_i = \left( \frac{\eta (1-\beta) \alpha_i^{\eta} (d_i+\Delta)^{\eta \beta}}{\phi \psi} \right)^{\frac{1}{\psi - \eta (1-\beta)}}
\end{equation*}
Therefore, $WTP_i$ solves:
\begin{equation}
\begin{aligned}
    & \ln \left(m_i - WTP_i \right) + \alpha_i^\eta \left( \frac{\eta (1-\beta)}{\phi \psi} \right)^{\frac{\eta (1-\beta)}{\psi - \eta(1-\beta)}} (d_i + \Delta)^{\frac{\beta \psi \eta}{\psi - \eta(1-\beta)}} - \phi \left( \frac{\eta (1-\beta) \alpha_i^{\eta} (d_i+\Delta)^{\eta \beta}}{\phi \psi} \right)^{\frac{\psi}{\psi - \eta (1-\beta)}} + \\
    & - \ln m_i - \alpha_i^\eta \left( \frac{\eta (1-\beta)}{\phi \psi} \right)^{\frac{\eta (1-\beta)}{\psi - \eta(1-\beta)}} d_i^{\frac{\beta \psi \eta}{\psi - \eta(1-\beta)}} - \phi \left( \frac{\eta (1-\beta) \alpha_i^{\eta} d_i^{\eta \beta}}{\phi \psi} \right)^{\frac{\psi}{\psi - \eta (1-\beta)}} = 0
\end{aligned}
\end{equation}
and depends on all parameters.

\clearpage
\section{Additional Survey Statistics and Comparisons}
\setcounter{table}{0}
\setcounter{figure}{0}
\setcounter{equation}{0} 
\renewcommand{\theequation}{B\arabic{equation}}
\renewcommand{\thetable}{B\arabic{table}}
\renewcommand{\thefigure}{B\arabic{figure}}
Table \ref{tab_responserate_byincent} shows the effects of the randomized participation incentives and reminders on survey completion. Figures \ref{fig_nonresponse_dim}$-$\ref{fig_nonresponse_herd} illustrate sample representativeness and Figure \ref{fig_salcompare_pub_slf} documents the alignment between self- and publicly-reported salaries (see \cite{myers2023new} for more on the sample). Table \ref{tab_sumstat_othervar1} describes the summary statistics for the sample on numerous dimensions. Table \ref{tab_pwcorr_qualoutput} shows the pairwise correlations of the subjective output measures.

\begin{table}[htb!]\centering \footnotesize
\caption{Survey Completion per Randomized Treatments}\label{tab_responserate_byincent}
\-\\
{
\def\sym#1{\ifmmode^{#1}\else\(^{#1}\)\fi}
\begin{tabular}{l*{1}{c}}
\hline\hline
                    &\multicolumn{1}{c}{(1)}         \\
\hline
Either incentive    &     0.00297\sym{**} \\
                    &   (0.00117)         \\
[1em]
Both incentives     &     0.00797\sym{***}\\
                    &   (0.00141)         \\
[1em]
1 reminder          &      0.0112\sym{***}\\
                    &   (0.00114)         \\
[1em]
2 reminders         &      0.0188\sym{***}\\
                    &   (0.00119)         \\
[1em]
Constant            &      0.0197\sym{***}\\
                    &   (0.00107)         \\
\hline
$ N$ obs.           &     131,672         \\
\hline\hline
\end{tabular}
}
\-\\
\note{\emph{Note}: Reports estimates from a regression of a binary indicator of survey completion on binary indicators for the randomized incentives and reminders including observations for all researchers emailed. Robust standard errors reported; $^{*} p<0.10, ^{**} p<0.05, ^{***} p<0.01$.
}
\end{table}

\begin{figure}[htb!] \centering
\caption{Sample Representativeness per Publication and Grant Funding Measures}\label{fig_nonresponse_dim}
\label{fig_nonresponse_dima}
\includegraphics[width=0.8\textwidth, trim=0mm 10mm 0mm 5mm, clip]{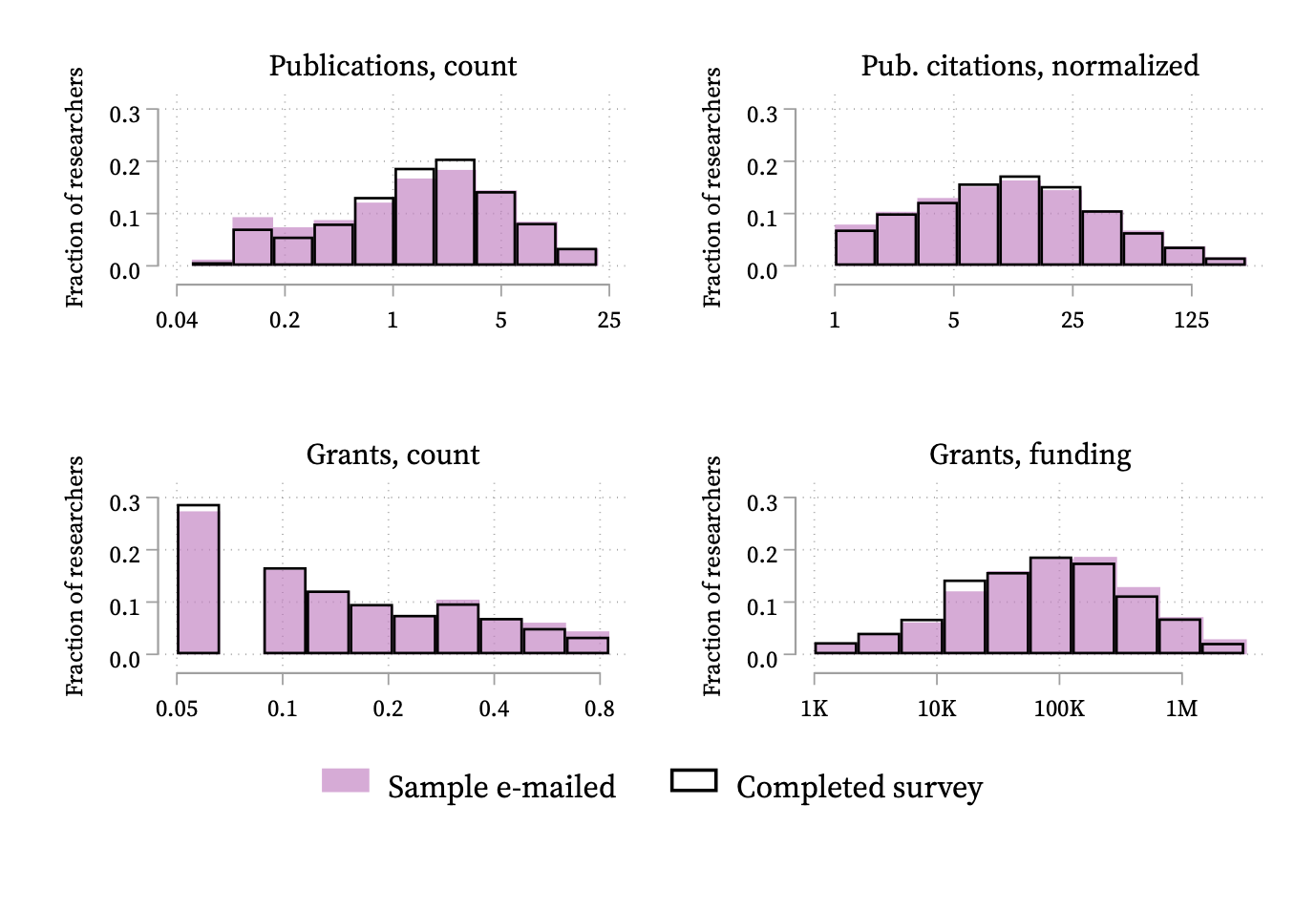}
\\
\note(\emph{Note}: Shows the distribution of publication and grant outcomes split by whether the researcher was emailed (i.e., the full sample) versus those who completed the survey; note the log $ x$ axes. See \cite{myers2023new} for regression-based estimates of the differences.
\end{figure}

\begin{figure}[htb!] \centering
\caption{Sample Representativeness per Institutional Funding Measures}\label{fig_nonresponse_herd}
\label{fig_nonresponse_herda}
\includegraphics[width=\textwidth, trim=0mm 10mm 0mm 5mm, clip]{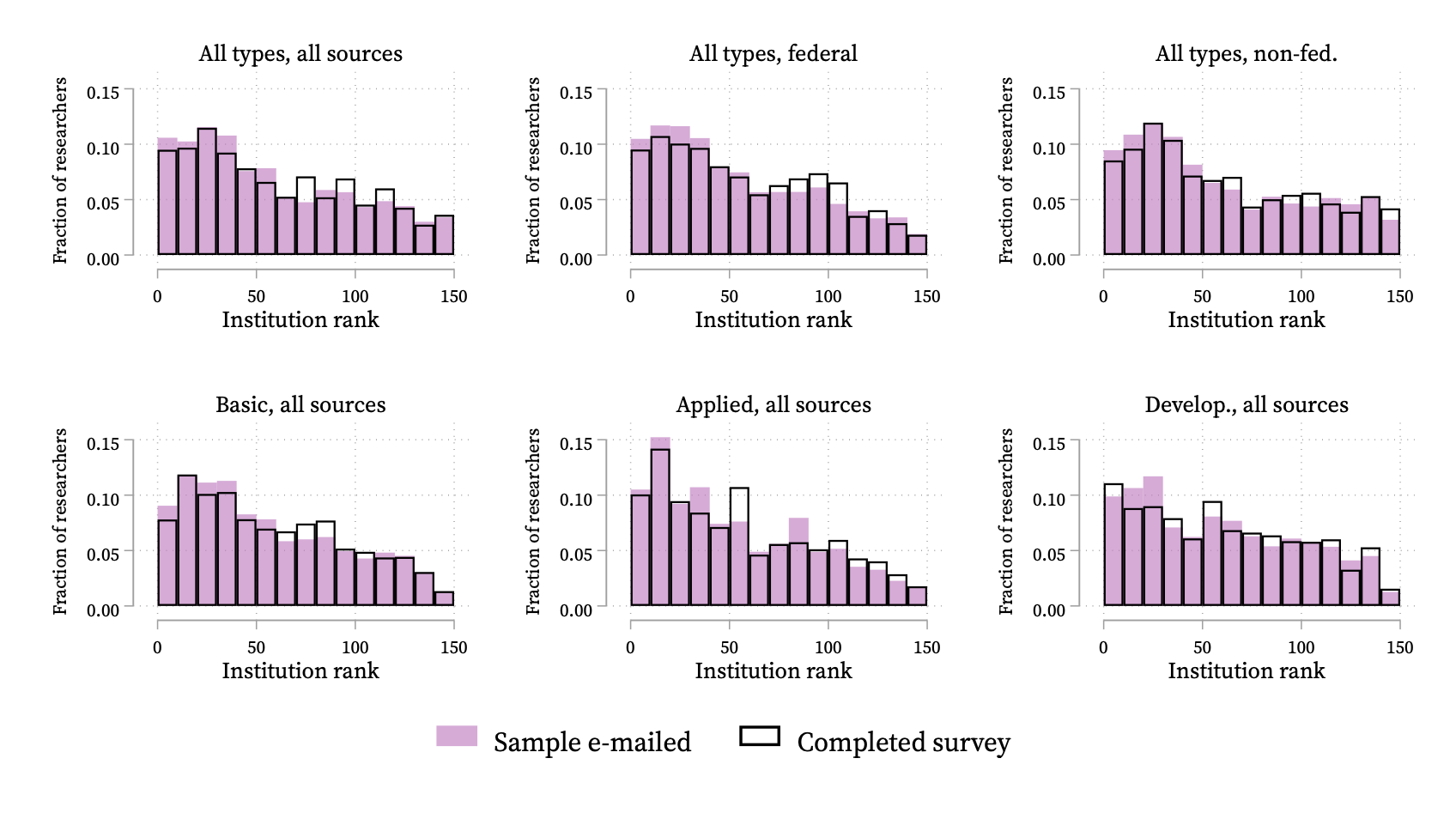}
\\
\note{\emph{Note}: Shows the distribution of institution-level funding split by whether the researcher was emailed (i.e., the full sample) versus those who completed the survey. See \cite{myers2023new} for regression-based estimates of the differences.}
\end{figure}

\begin{figure}[htb!] \centering
\caption{Correlation of Self- and Publicly-reported Annual Salaries}\label{fig_salcompare_pub_slf}
\includegraphics[height=0.5\textwidth, trim=0mm 10mm 0mm 5mm, clip]{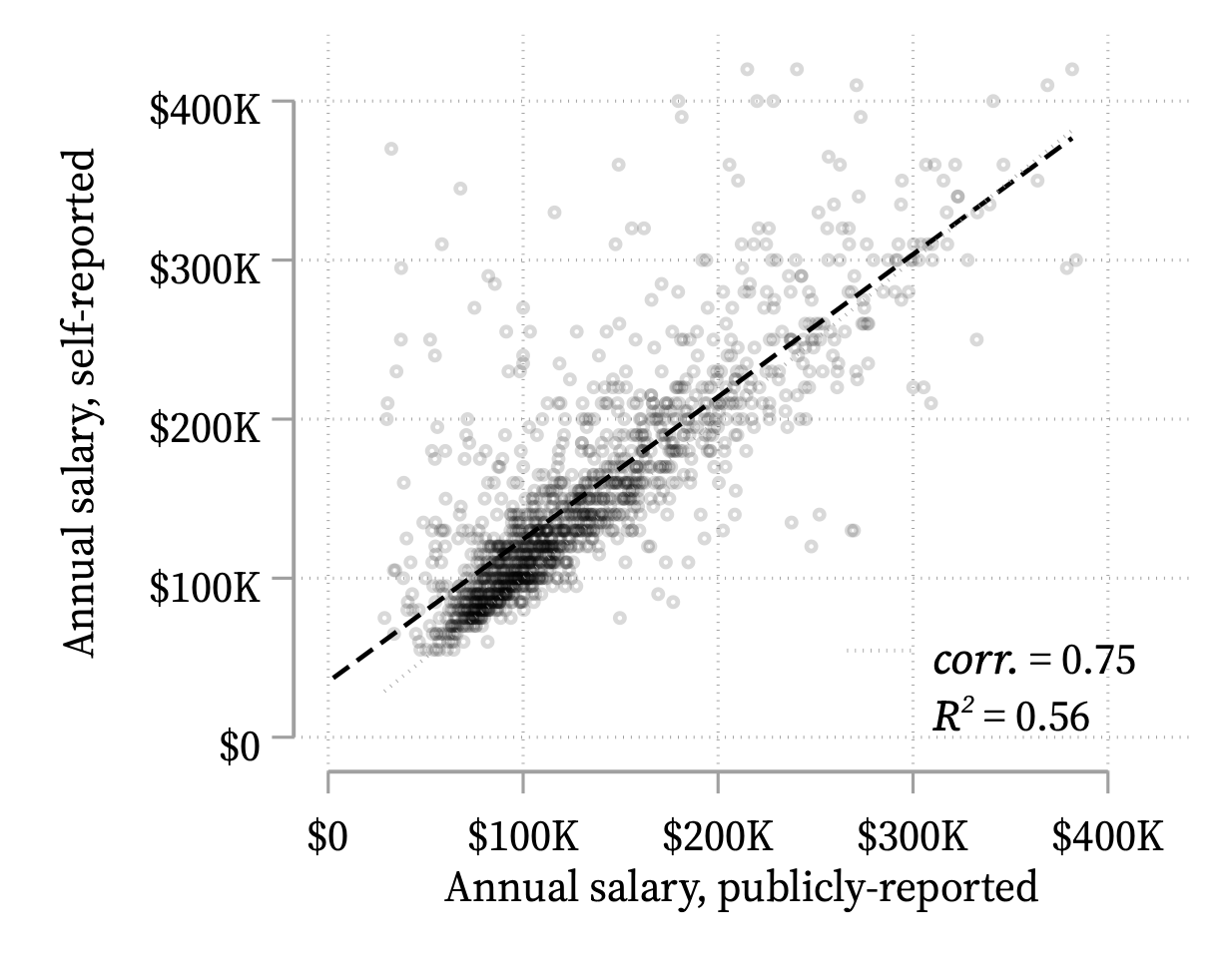}
\-\\
\note{\emph{Note}: \input{figtab/note_fig_salcompare_pub_slf.tex}}
\end{figure}

\begin{table}[htbp!]\centering \footnotesize
\caption{Summary Statistics of Other Variables}\label{tab_sumstat_othervar1}
{
\def\sym#1{\ifmmode^{#1}\else\(^{#1}\)\fi}
\begin{tabular}{l*{1}{rrrrrr}}
\hline\hline
                    & \emph{mean}& \emph{s.d.}\\
\hline
\underline{\emph{From HERD: Institution-level R\&D}}&            &            \\
Total R\&D, \$M    &      611.32&      451.18\\
R\&D per researcher, \$M&        0.61&        0.86\\
Share federal gov.'t R\&D, [0,1]&        0.52&        0.12\\
Share basic R\&D, [0,1]&        0.63&        0.20\\
\underline{\emph{From Dimensions: Research output}}&            &            \\
Publications per year&        5.28&        7.09\\
Citations per year  &       23.07&       49.10\\
Co-authors per publication per year&        9.09&       68.40\\
\underline{\emph{Position details}}&            &            \\
Assistant professor, \{0,1\}&        0.26&        0.44\\
Associate professor, \{0,1\}&        0.26&        0.44\\
Full professor, \{0,1\}&        0.41&        0.49\\
Other rank, \{0,1\} &        0.07&        0.25\\
Not on tenure track, \{0,1\}&        0.19&        0.39\\
Pre-tenure, \{0,1\} &        0.22&        0.42\\
Tenured, \{0,1\}    &        0.58&        0.49\\
Years until next contract eval.&        3.68&        1.82\\
Duration of contract&        2.45&        1.93\\
\underline{\emph{Gender identity}, \{0,1\}}&            &            \\
Female              &        0.40&        0.49\\
Male                &        0.55&        0.50\\
Other or N.R       &        0.05&        0.22\\
\underline{\emph{Racial/ethnic identity}, \{0,1\}}&            &            \\
Asian               &        0.13&        0.33\\
Black               &        0.03&        0.18\\
Hispanic            &        0.06&        0.23\\
White               &        0.77&        0.42\\
Other or N.R       &        0.05&        0.21\\
\underline{\emph{Citizenship}, \{0,1\}}&            &            \\
Citizen, domestic-born&        0.71&        0.45\\
Citizen or perm resident, foreign-born&        0.24&        0.43\\
Other or N.R citizenship&        0.05&        0.21\\
1st--3rd generation in U.S.&        0.30&        0.46\\
Other or N.R generation in U.S.&        0.70&        0.46\\
\underline{\emph{Other covariates}}&            &            \\
Age                 &       48.78&       12.05\\
Household total income&  260,336.00&  215,388.31\\
Married or domestic partnership, \{0,1\}&        0.82&        0.39\\
Single, \{0,1\}     &        0.13&        0.34\\
Other or N.R relationship, \{0,1\}&        0.05&        0.22\\
Dependents in household&        0.98&        1.13\\
Risk-taking in personal life, [0,10]&        5.26&        2.13\\
\hline\hline
\end{tabular}
}
\note{\emph{Note}: Reports summary statistics for 4,003 researcher-level observations. \emph{From HERD} indicates variables from \cite{nsf2023herd}. \emph{From Dimensions} indicates variables from the \cite{dimension2018data} dataset. \emph{N.R.} stands for Not Reported.
}
\end{table}

\begin{table}[htbp!]\centering \footnotesize
\caption{Pairwise Correlations of Subjective Output Measures}\label{tab_pwcorr_qualoutput}
\-\\
{
\def\sym#1{\ifmmode^{#1}\else\(^{#1}\)\fi}
\begin{tabular}{l*{8}{l}}
\hline\hline
                &\multicolumn{8}{c}{}                                                                                   \\
                & Articles   &    Books   &  Methods   & Products   & Academic   &   Policy   & Business   &   Public   \\
\hline
Articles        &        &            &            &            &            &            &            &            \\
Books           &    --0.13$^{s}$&        &            &            &            &            &            &            \\
Methods         &     {\hspace{1.7mm}}0.08$^{s}$&    --0.20$^{s}$&        &            &            &            &            &            \\
Products        &    --0.13$^{s}$&    --0.12$^{s}$&     {\hspace{1.7mm}}0.23$^{s}$&        &            &            &            &            \\
Academic        &     {\hspace{1.7mm}}0.48$^{s}$&    --0.01   &     {\hspace{1.7mm}}0.02   &    --0.24$^{s}$&        &            &            &            \\
Policy          &     {\hspace{1.7mm}}0.01   &    --0.00   &     {\hspace{1.7mm}}0.10$^{s}$&     {\hspace{1.7mm}}0.25$^{s}$&    --0.07$^{s}$&        &            &            \\
Business        &    --0.03   &    --0.07$^{s}$&     {\hspace{1.7mm}}0.20$^{s}$&     {\hspace{1.7mm}}0.38$^{s}$&    --0.12$^{s}$&     {\hspace{1.7mm}}0.24$^{s}$&        &            \\
Public          &    --0.12$^{s}$&     {\hspace{1.7mm}}0.19$^{s}$&     {\hspace{1.7mm}}0.04   &     {\hspace{1.7mm}}0.21$^{s}$&    --0.20$^{s}$&     {\hspace{1.7mm}}0.28$^{s}$&     {\hspace{1.7mm}}0.13$^{s}$&        \\
\hline\hline
\end{tabular}
}
\-\\
\note{\emph{Note}: Reports pairwise correlations for the four subjective measures of researchers' intended output types (Journal articles; Books; Materials or Methods; Products) and the four subjective measures of researchers' intended audiences (Academic peers; Policymakers; Businesses and organizations; General public); $^{s} p<0.01$.
}
\end{table}

\clearpage
\section{Additional Model and Experiment Details}\label{app:moremodel}
\setcounter{table}{0}
\setcounter{figure}{0}
\setcounter{equation}{0} 
\renewcommand{\theequation}{C\arabic{equation}}
\renewcommand{\thetable}{C\arabic{table}}
\renewcommand{\thefigure}{C\arabic{figure}}

\subsection{Derivation of Policy Functions}\label{app:policyfct}
The policy functions $\mathcal{R}(\text{\textbf{S}}_i,\bm{\uptheta}_i,\bm{\upmu}_i)$, $\mathcal{F}(\text{\textbf{S}}_i,\bm{\uptheta}_i,\bm{\upmu}_i)$ and $\mathcal{H}(\text{\textbf{S}}_i,\bm{\uptheta}_i,\bm{\upmu}_i)$ characterize the solution $(R^*_i, F^*_i, H^*_i)$ to problem \eqref{eq:utmaxproblem} for each individual scientist $i$ as a function of states $\text{\textbf{S}}_i$, attributes $\bm{\uptheta}_i$, and parameters $\bm{\upmu}_i$. They determine indirect utility $\mathcal{V}^*_i = \mathcal{V}(\text{\textbf{S}}_i, \bm{\uptheta}_i , \bm{\upmu}_i)$ at current allocations. Substituting the constraints, the utility maximization problem is:
\begin{equation}\label{eq:utmaxproblem_app}
\begin{aligned}
    \mathcal{V}(\cdot) = \max_{F_i,H_i} u_{1i}(M_i) + u_{2i}(\alpha_i \left( B_{\text{min}} + G_i + \phi_i F_i \right)^{\gamma_i} \left(H_i-F_i-D_i\right)^{1-\gamma_i} ) - u_{3i}(H_i,D_i) \;,
\end{aligned}
\end{equation}
with the policy functions $\mathcal{F}(\cdot,\cdot,\cdot)$ and $\mathcal{H}(\cdot,\cdot,\cdot)$ solving the optimality conditions:
\begin{eqnarray}
    \frac{\partial u_{2,i} (Y_i)}{\partial Y_i} \frac{\partial Y_i}{\partial F_i} + \lambda_{F,i} & = 0 \;\; \\ \label{eq:focF}
    \frac{\partial u_{2,i} (Y_i)}{\partial Y_i} \frac{\partial Y_i}{\partial H_i} - \frac{\partial u_{3,i}(H_i,D_i)}{\partial H_i} - \lambda_{H,i} & = 0 \;, \label{eq:focH}
\end{eqnarray}
and $\mathcal{R}(\cdot,\cdot,\cdot)$ being residually determined based on the time-constraint \eqref{eq:utmaxproblem_H}. To derive the policy functions, we start from model's optimality conditions \eqref{eq:focF} and \eqref{eq:focH} and evaluate them using the functional forms described in the main text. We obtain the conditions:
\begin{eqnarray}
    Y_i^{1-\eta_i} \Big[ \gamma_i \phi_i (B_{\text{min}}+G_i+\phi_i F_i )^{-1} - (1-\gamma_i) (H_i-D_i-F_i)^{-1} \Big] + \lambda_{F,i} & = 0 \;\; \\ \label{eq:focF_fct}
    Y_i^{1-\eta_i} (H_i-D_i-F_i)^{-1} (1-\gamma_i) - \psi (H_i-D_i+D_i^{\xi_i})^{\zeta_i} - \lambda_{H,i} & = 0 \;. \label{eq:focH_fct}
\end{eqnarray}
We characterize the policy functions in two intervals of $H_i$. The first is $H_i \in ( D_i, H_{i,F>0} ]$ and the second is $H_i\in (H_{i,F>0}, H_{\text{max}}]$, depending on whether the optimal fundraising time allocation is a corner solution ($F_i=0$) or not ($F_i>0$). We identify the threshold $H_{i,F>0}$ below which optimal fundraising is zero by solving \eqref{eq:focF_fct} for $F_i$ as a function of $H_i$ assuming that $F_i$ is strictly positive, i.e., $\lambda_{F,i}=0$:
\begin{equation}\label{eq:FofH_interior}
    F_i = \gamma_i (H_i-D_i) - \frac{1-\gamma_i}{\phi_i} (B_{\text{min}}+G_i) \;.
\end{equation}
Therefore, \eqref{eq:FofH_interior} implies that fundraising time is strictly positive as long as:
\begin{equation}
    H_i > D_i + \frac{1-\gamma_i}{\gamma_i \phi_i} (B_{\text{min}}+G_i) \;,
\end{equation}
which identifies the threshold $H_{i,F>0}$ that is strictly larger than $D_i$ as long as $\gamma_i<1$. As a consequence, we can solve for optimal hours $H_i$ using equation \eqref{eq:focH_fct}, which takes the piece-wise functional form:
\begin{equation}\label{eq:solH}
    \begin{cases}
        (1-\gamma_i) [ \alpha_i (B_{\text{min}}+G_i)^{\gamma_i} ]^{1-\eta_i} (H_i-D_i)^{(1-\gamma_i)(1-\eta_i)-1} - \psi (H_i-D_i+D_i^{\xi_i})^{\zeta_i} - \lambda_{H,i} = 0 & \text{if} \ H_i\leq H_{i,F>0} \\
        \Big[ \alpha_i (\phi_i \gamma_i)^{\gamma_i} (1-\gamma_i)^{1-\gamma_i} \Big]^{1-\eta_i} \Big( H_i - D_i \frac{B_{\text{min}}+G_i}{\phi_i} \Big)^{-\eta_i} - \psi (H_i-D_i+D_i^{\xi_i})^{\zeta_i} - \lambda_{H,i} = 0  & \text{if} \ H_i > H_{i,F>0}
    \end{cases}
\end{equation}
and defines an implicit solution $\mathcal{H}_i$ for $H_i$ as a function of all parameters and state variables. \eqref{eq:solH} is continuous but not differentiable at $H_{i,F>0}$. Moreover, the Lagrange multiplier $\lambda_{H,i}$ equals zero for $H_i<H_{\text{max}}$ and $H_i$ is always strictly larger than $D_i$ because $\lim_{H_i\rightarrow D_i} \eqref{eq:solH}=+\infty$. Hence, the non-negativity constraint on research time is always met. Given the policy function $\mathcal{H}_i$ for hours, the functions defining optimal allocation to fundraising time and research time are:
\begin{equation}
\mathcal{F}_i = 
    \begin{cases}
        0 & \text{if} \ \mathcal{H}_i\leq H_{i,F>0} \\
        \gamma_i (\mathcal{H}_i-D_i) - \frac{1-\gamma_i}{\phi_i} (B_{\text{min}}+G_i) & \text{if} \ \mathcal{H}_i > H_{i,F>0}
    \end{cases}
\end{equation}
and
\begin{equation}
    \mathcal{R}_i = \mathcal{H}_i - D_i - \mathcal{F}_i \;.
\end{equation}

\subsection{Reducing Dimensions of Heterogeneity: \texorpdfstring{$k$}{\emph{k}}-means Clustering Results}\label{app:kmeans}

\begin{figure}[htb!] \centering
\caption{Researcher Heterogeneity}\label{fig_hist_hetindex}
\includegraphics[height=0.5\textwidth, trim=0mm 10mm 0mm 5mm, clip]{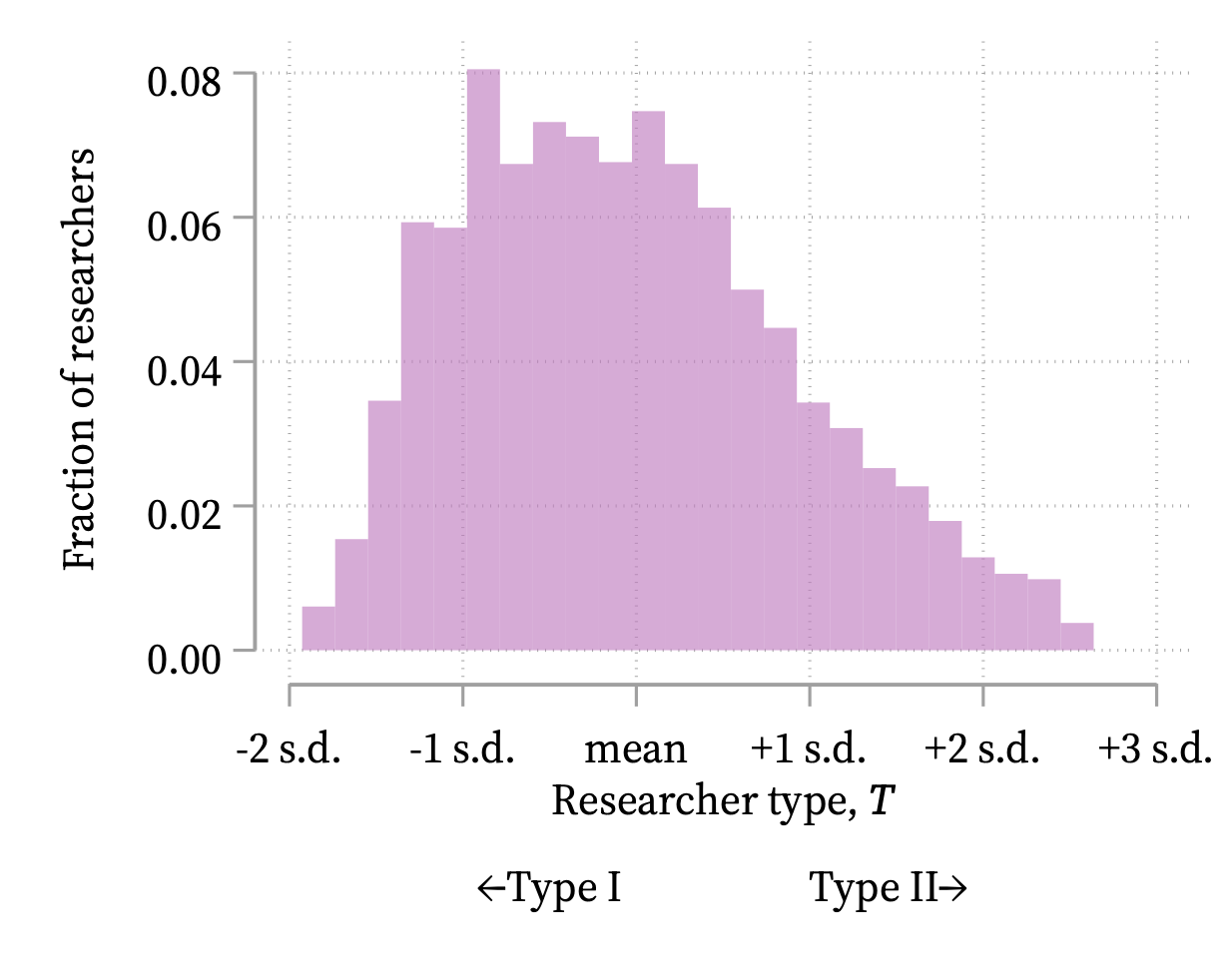}
\-\\
\note{\emph{Note}: \input{figtab/note_fig_hist_hetindex.tex}}
\end{figure}

Ideally, our model would allow for rich heterogeneity in researchers' preferences (i.e., the $u_{1,i}(\cdot)$, $u_{2,i}(\cdot)$, and $u_{3,i}(\cdot)$ functions). This would help ensure that (true) variation in preferences would not mistakenly be attributed as (estimated) variation in productivity. We lack enough data to estimate researcher-specific preference functions, but we do have the large vector of observables from the survey $\mathbf{X}_i$.\footnote{We could allow the parameters that govern the preference functions to depend on deep parameters and these observables. For example, for the preference function parameter $\sigma_i$, we could assume that $\sigma_i = \exp(\delta_{\sigma,0} + \sum_x X_i \delta_{\sigma,x})$. However, this presents some practical estimation challenges (i.e., estimating approximately (35$\times$3=) 105 additional parameters; the $\delta_{\sigma,x}$ parameters in the previous example).} This motivates the approach outlined in Section \ref{sec:model}. We use $k$-means clustering to collapse the multi-dimensional heterogeneity from the observables $\mathbf{X}_i$ into a single-dimensional index, $T_i$. Specifically, we assume there are two clusters of researcher types ($k$=2), and we use $k$-means clustering to estimate each researcher's euclidean distance from one of those groups; we use that distance as $T_i$, which provides a smooth, continuous measure of heterogeneity. Thus, for the preference function parameter $\sigma_i$, we assume that $\sigma_i = \exp(\delta_{\sigma,0} + T_i \delta_{\sigma})$. Figure \ref{fig_hist_hetindex} shows the distribution of the one-dimensional euclidean similarity scores. Researchers with larger values of $T_i$ tend to be, among other things, older, white, males that are full professors in the humanities and natural sciences. A table reporting the mean differences between the two types of researchers is available upon request. 

\subsection{Estimation Outline}\label{app:estoutline}

We infer individual attributes and parameters using a mix of calibration and estimation. First, we set minimum funding $B_{min}=\$5,000$ and fix total hours endowment $H_{max}$ to 62 work hours per week (approximately the 90th percentile of the observed work hours). Second, we estimate individual attributes and common parameters of the utility function, including deep parameters that determine individual-specific parameters in $\bm{\upmu}_i$ as functions of researcher's type $T_i$. We use survey information on hours worked, research time, fundraising time, guaranteed funding, expected additional funds raised, duties, and the salaries reported in the alternative offer experiments, which we map to model counterparts $(H_i,R_i\,,\,F_i\,,\,G_i\,,\,\phi_i F_i\,,\, D_i\,,\, \{ M_{ij}\}_{j=1}^4)$, respectively. 

Conditional on some common parameters, we can infer individual-specific attributes $\bm{\uptheta}_i$ from researchers' optimality conditions and model's structure. One important note is that we have separate estimation routines for inferring individual attributes $\bm{\uptheta}_i$ among researchers with non-zero fundraising time ($F_i>0$) and those at a corner solution ($F_i=0$). For the former group, we first infer fundraising ability $\phi_i$ by exploiting the identity equating the observed, expected additional funding ($EG_i$) to $\phi_i F_i$. Therefore, $\forall i =1,...,N_F$:
\begin{equation}\label{eq:phicalibr}
 \widehat{\phi}_i = \frac{EG_i}{F_i} \;.
\end{equation}
Expression (\ref{eq:phicalibr}) is well-defined if and only if observed $F_i>0$, which is why the inference strategy must differ for researchers with $F_i=0$ at observed allocations.\footnote{For numerical stability, we constrain $\widehat{\phi}_i \geq 1$.} As a second step, conditional on $\widehat{\phi}_i$, we infer factor shares $\widehat{\gamma_i}$ from researchers' first order condition in fundraising time, which we derive in Appendix \ref{app:policyfct}. For the chosen functional forms, this takes the expression:
\begin{equation}\label{eq:gammacalibr}
 \widehat{\gamma}_i = \frac{B_i}{B_i + \phi_i R_i} = \frac{B_{\text{min}}+G_i + \widehat{\phi}_i F_i}{B_{\text{min}}+G_i + \widehat{\phi}_i F_i + \widehat{\phi}_i R_i } \;.
\end{equation}
Equation (\ref{eq:gammacalibr}) states that $\gamma_i$ constitutes the weight of total funding over the total dollar-value of inputs used in scientific activity, with the last term of the denominator being the dollar-valued opportunity cost of research time relative to fundraising. Finally, the optimality condition for total hours worked determines productivity $\widehat{\alpha}_i$ as a function of parameters, observed allocations, and $\widehat{\phi}_i$ and $\widehat{\gamma}_i$:
\begin{equation}\label{eq:alphahat}
 \eta_i (1-\widehat{\gamma}_i) \widehat{\alpha}_i^{1-\eta_i} (B_{\text{min}} + G_i + \widehat{\phi}_i F_i)^{(1-\eta_i) \widehat{ \gamma}_i } R_i^{(1-\eta_i) (1-\widehat{\gamma}_i) -1} = \psi (H_i-D_i+D_i^{\xi_i})^{\zeta_i} \;,
\end{equation}
This equation holds exactly if $H_i<H_{\text{max}}$. Therefore, Equations \eqref{eq:phicalibr}, \eqref{eq:gammacalibr}, and \eqref{eq:alphahat} determine individual attributes as functions of parameters and observed allocations for researchers with positive fundraising time. Unfortunately, for the group of researchers reporting zero fundraising time, we can neither infer $\widehat{\phi}_i$ from Equation \eqref{eq:phicalibr}, nor can we compute $\widehat{\gamma}_i$.\footnote{The Lagrange multiplier $\lambda_{F,i}$ is strictly positive and unknown.} Therefore, we assume that $\phi_i$ and $\gamma_i$ are parametric polynomial functions of state variables, and we estimate these functions using the other sub-sample of researchers with $F_i>0$.\footnote{We use the following specifications: $\phi_i = \exp \left\{ \sum_{p=1}^3 \beta_{G,p} G^p_i + \sum_{p=1}^3 \beta_{D,p} D^p_i + \sum_{p=1}^3 \beta_{M,p} M^p_i \right\}$, and $\gamma_i = \left( 1+ \exp\left\{ \sum_{p=1}^3 \iota_{G,p} G^p_i + \sum_{p=1}^3 \iota_{D,p} D^p_i + \sum_{p=1}^3 \iota_{M,p} M^p_i \right\} \right)^{-1}$, which we estimate using poisson and logistic regressions. Whenever the fitted $(\widehat{\phi}_i,\widehat{\gamma}_i)$ combination would imply, given the observed state variables and for a specific vector of parameters $\bm{\upmu}_i$, that the optimal (observed) time allocation to fundraising would be strictly positive, we re-scale $\widehat{\phi}_i$ below the individual-specific lower bound below which the optimal fundraising time at observed states is indeed null.}

Finally, given the estimates attributes $\widehat{\phi}_i$ and $\widehat{\gamma}_i$, the state variables, and the vector of common parameters $\widehat{\bm{\upmu}}$, we infer productivity beliefs $\widehat{\alpha}_i$ through Equation \eqref{eq:alphahat}. Given estimates of individual attributes $\widehat{\bm{\uptheta}}_i = (\widehat{\alpha}_i,\widehat{\gamma}_i,\widehat{\phi}_i)$ as functions of parameters $\bm{\upmu}=(\omega, \psi,\bm{\updelta})$, calibrated values $(B_{\text{min}}, H_{\text{max}})$, and observed allocations, we estimate the vector $\bm{\upmu}$ by generalized method of moments as:
\begin{equation}\label{eq:gmmlossfct}
 \widehat{\bm{\upmu}} = \arg \min_{\bm{\upmu}} \mathcal{L}(\bm{\upmu}) = \arg \min_{\bm{\upmu}} \sum_{i=1}^{N} \sum_{j=1}^4 \left( M^{\text{obs}}_{ij} - \widehat{M}_{ij} ( \text{\textbf{S}}_{ij}, \widehat{\bm{\uptheta}}_i (\bm{\upmu}_i(\bm{\upmu},T_i)) , \bm{\upmu}_i(\bm{\upmu},T_i) ) \right)^2 \;,
\end{equation}
where $M_{ij}$ is researcher $i$'s answer to thought experiment $j$ in the data and $\widehat{M}_{ij}$ is its model-based counterpart, which is itself an implicit function of the vector of counterfactual states $\text{\textbf{S}}_i$, researcher's type $T_i$, and of the estimand $\bm{\upmu}$ which determines individual attributes $\widehat{\bm{\uptheta}}_i$ and individual-specific parameters $\bm{\upmu}_i$.\footnote{To save notation, we omit that $\widehat{M}_{ij}$ are also a function of calibrated $(B_{\text{min}},H_{\text{max}})$.} Parameter estimates $\widehat{\bm{\upmu}}$ solve \eqref{eq:gmmlossfct} conditional on parameter restrictions specified in Section \ref{subsec:theory}. Appendix \ref{app:estalgo} provides additional details on the search algorithm.

\subsection{Estimation Algorithm}\label{app:estalgo}
Here, we describe the estimation of the common parameters \\ $\bm{\upmu} = (\omega,\psi,\delta_{\sigma,0},\delta_{\sigma,1},\delta_{\eta,0},\delta_{\eta,1},\delta_{\xi,0},\delta_{\xi,1},\delta_{\zeta,0},\delta_{\zeta,1})$. First, we define a grid of parameter values at which we perform a preliminary evaluation of the loss function in \eqref{eq:gmmlossfct}. The grid is defined by the Cartesian product of the following:
\begin{equation*}
\begin{aligned}
    & \bm{\omega}^{(0)} = [0.1, 1, 10]\;,  \quad \bm{\psi}^{(0)} = [0.00001, 1, 10]\;,  \\
    & \bm{\delta}_{\sigma,0}^{(0)} = [-100, 0]\;,  \quad \bm{\delta}_{\sigma,1}^{(0)} = [0]\;,  \\
    & \bm{\delta}_{\eta,0}^{(0)} = [\ln(0.8)-\ln(0.2), 0, \ln(0.2)-\ln(0.8)]\;, \quad \bm{\delta}_{\eta,1}^{(0)} = [0]  \\
    & \bm{\delta}_{\xi,0}^{(0)} = [-11.5,  \ln(2)]\;, \quad \bm{\delta}_{\xi,1}^{(0)} = [0]\;,  \\
    & \bm{\delta}_{\zeta,0}^{(0)} = [-11.5, 0]\;, \quad \bm{\delta}_{\zeta,1}^{(0)} = [0]\;.  \\
\end{aligned}
\end{equation*}
Second, we select the grid-points where the value of the loss function is within 0.5\% of the minimum, and we use them as initial points for the numerical solution. We perform a preliminary ``Nelder-Mead simplex direct search'' with a high tolerance on the loss function and standard tolerance on parameter values. We then select the parameters vectors where the value of the loss function is within 1\% of the minimum, and we use them as initial points with the same search algorithm and with lower loss function tolerance. We repeat this step twice until we are left with a single candidate optimal parameter vector.

\subsection{Objectives and Constraints for Counterfactuals}
We first analyze a setting where social planner's objective is to maximize field-specific aggregate scientist's utility by reallocating guaranteed funding $G_i$ and administrative duties $D_i$ within the major field. We define the output function
\begin{equation*}
    \mathcal{Y}_i(\widetilde{G}_i,\widetilde{D}_i,\pi_f)=\alpha_i \left( B_{\text{min}} + G_i + \phi_i \mathcal{F}(\widetilde{G}_i,\widetilde{D}_i,\pi_f) \right)^{\gamma_i} \mathcal{R}(\widetilde{G}_i,\widetilde{D}_i,\pi_f)^{1-\gamma_i} \, 
\end{equation*}
where the adjustment factor $\pi_f$ to fundraising ability in field $f$ enforces the constraint that additional funding in the counterfactual allocation must equal observed funding. For convenience, in this section we omit from the notation the dependence of the policy functions on attributes $\widehat{\bm{\uptheta}}_i$ and parameters $\widehat{\bm{\upmu}}_i$. Therefore, the problem in field $f$ is:
\begin{equation}\label{eq:planner_maxU}
    \begin{aligned}
        \max_{(\widetilde{G}_i, \widetilde{D}_i )_{i =1}^{N_f}} & \sum_{i=1}^{N_f} \biggl( \kappa \frac{\mathcal{Y}_i(\widetilde{G}_i,\widetilde{D}_i,\pi_f)^{1-\eta_i}}{1-\eta_i} - \psi \frac{(\mathcal{H}_i(\widetilde{G}_i,\widetilde{D}_i,\pi_f)-\widetilde{D_i}+\widetilde{D}_i^{\xi_i})^{1+\zeta_i}}{1+\zeta_i} \biggr) \\
        & \text{subject to} \\
        & \sum_{i=1}^{N_f} \widetilde{G}_i = \widehat{G}^{tot} \;\;\; [\text{Multiplier}  \ \lambda_{G,f}] \\
        &\sum_{i=1}^{N_f} \widetilde{D}_i = \widehat{D}^{tot} \;\;\; [\text{Multiplier}  \ \lambda_{D,f}] \\
        & \sum_{i=1}^{N_f} EG_i = \pi_f \sum_{i=1}^{N_f}  \widehat{\phi}_i \mathcal{F}_i(\widetilde{G}_i,\widetilde{D}_i,\pi_f) \;, 
    \end{aligned}
\end{equation}
where $N_f$ is the number of scientists in field $f$. We also impose non-negativity constraints on individual $\widetilde{G}_i$ and $\widetilde{D}_i$ allocations. The individual research output function $\mathcal{Y}_i= \widehat{\alpha}_i \mathcal{B}_i^{\widehat{\gamma}_i} \mathcal{R}_i^{1-\widehat{\gamma}_i} $ and hours $\mathcal{H}_i$ make explicit that both are functions of the considered instruments $(\widetilde{G}_i, \widetilde{D}_i)$. The last constraint requires that total additional funding in the observed allocations. Therefore, all the policy functions not only vary with policy levers $(\widetilde{G}_i,\widetilde{D}_i)$ but also with $\pi_f$, which can be interpreted as an endogenous adjustment to fundraising probability. When the new allocation violates the constraint, fundraising probability uniformly shrinks, thus reducing additional funding both directly and indirectly through the behavioral decline in fundraising hours. 

In the optimal allocations, the planner seeks to equate the marginal utility of guaranteed funding and duties across individuals, conditional on the non-negativity constraint and the additional funding constraint. Therefore, the solution satisfies the following equations for each $i=1,...,N_f$:
\begin{eqnarray}
    & \frac{d \mathcal{V}_i}{d \widetilde{G}_i} = \kappa \mathcal{Y}_i^{-\eta_i} \frac{d \mathcal{Y}_i}{d \widetilde{G}_i} - \psi (\mathcal{H}_i-\widetilde{D}_i+\widetilde{D}_i^{\xi_i})^{\zeta_i} \frac{\partial \mathcal{H}_i}{\partial \widetilde{G}_i} = \lambda_{G,f} - \lambda_{G^*,i} \label{eq:focG_maxU} \\
    & \frac{d \mathcal{V}_i}{d \widetilde{D}_i} = \kappa \mathcal{Y}_i^{-\eta_i} \frac{d \mathcal{Y}_i}{d \widetilde{D}_i} - \psi (\mathcal{H}_i-\widetilde{D}_i+D_i^{\xi_i})^{\zeta_i} \Bigl( \frac{\partial \mathcal{H}_i}{\partial \widetilde{D}_i} - 1 + \xi_i \widetilde{D}_i^{\xi_i-1} \Bigr)  = \lambda_{D,f} - \lambda_{D^*,i} \;, \label{eq:focD_maxU}
\end{eqnarray}
where the marginal product of guaranteed funds $\frac{d \mathcal{Y}_i}{d \widetilde{G}_i}$ and the (negative) marginal product of duties $\frac{d \mathcal{Y}_i}{d \widetilde{D}_i} $ are determined by equations \eqref{eq:dYdG} and \eqref{eq:dYdD}, respectively:
\begin{eqnarray}
    & \frac{d \mathcal{Y}_i}{d G_i} = \widehat{\alpha}_i \mathcal{B}_i^{\widehat{\gamma}_i} \mathcal{R}_i^{1-\widehat{\gamma}_i} \Bigl( \widehat{\gamma}_i \mathcal{B}_i^{-1} \frac{\partial \mathcal{B}_i}{\partial \widetilde{G_i}} + (1-\widehat{\gamma}_i) \mathcal{R}_i^{-1} \frac{\partial \mathcal{R}_i}{\partial \widetilde{G_i}} \Bigr) \label{eq:dYdG} \\
    & \frac{d \mathcal{Y}_i}{d D_i} = \widehat{\alpha}_i \mathcal{B}_i^{\widehat{\gamma}_i} \mathcal{R}_i^{1-\widehat{\gamma}_i} \Bigl( \widehat{\gamma}_i \mathcal{B}_i^{-1} \frac{\partial \mathcal{B}_i}{\partial \widetilde{D_i}} + (1-\widehat{\gamma}_i) \mathcal{R}_i^{-1} \frac{\partial \mathcal{R}_i}{\partial \widetilde{D_i}} \Bigr)  \;.\label{eq:dYdD}
\end{eqnarray}
Total funding as a function of state variables is defined by $\mathcal{B}_i(G_i,D_i,\pi_f) = B_{\text{min}} + G_i + \pi_f \widehat{\phi}_i \mathcal{F}_i (G_i,D_i,\pi_f)$.

\begin{figure}[htb] \centering
\caption{Example Image of Survey Experiment}\label{fig_thoughtexp}
\includegraphics[height=0.5\textwidth, trim=0mm 0mm 0mm 0mm, clip]{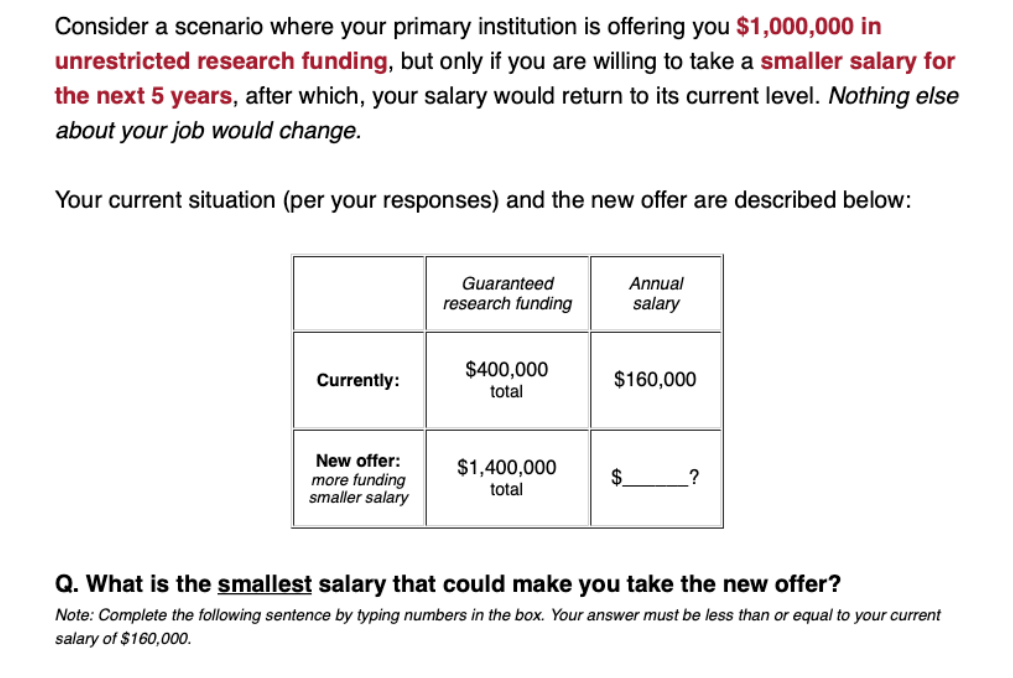}
\-\\
\note{\emph{Note}: Shows a screenshot of the survey experiment designed to solicit researchers' willingness to trade off their salary for additional guaranteed funding.}
\end{figure}

\begin{figure}[htb] \centering
\caption{Correlations of Implied Valuations}\label{fig_binscat_wtpcorr}
\subfloat[Less Administrative Duties]
{
\label{fig_binscat_wtpcorr_hr}
\includegraphics[width=0.475\textwidth, trim=0mm 10mm 0mm 5mm, clip]{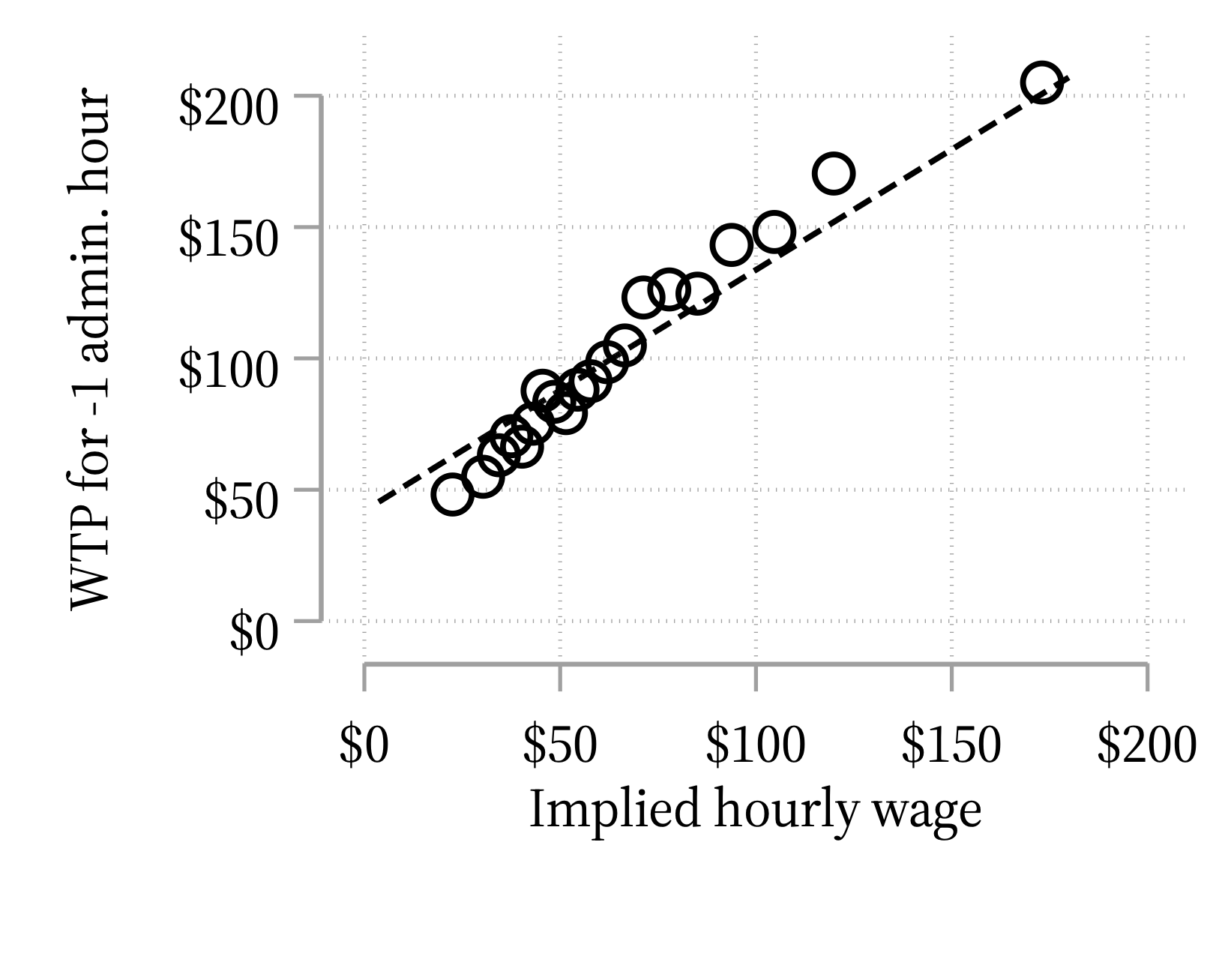}
} 
\subfloat[More Research Funding]
{
\label{fig_binscat_wtpcorr_dol}
\includegraphics[width=0.475\textwidth, trim=0mm 10mm 0mm 5mm, clip]{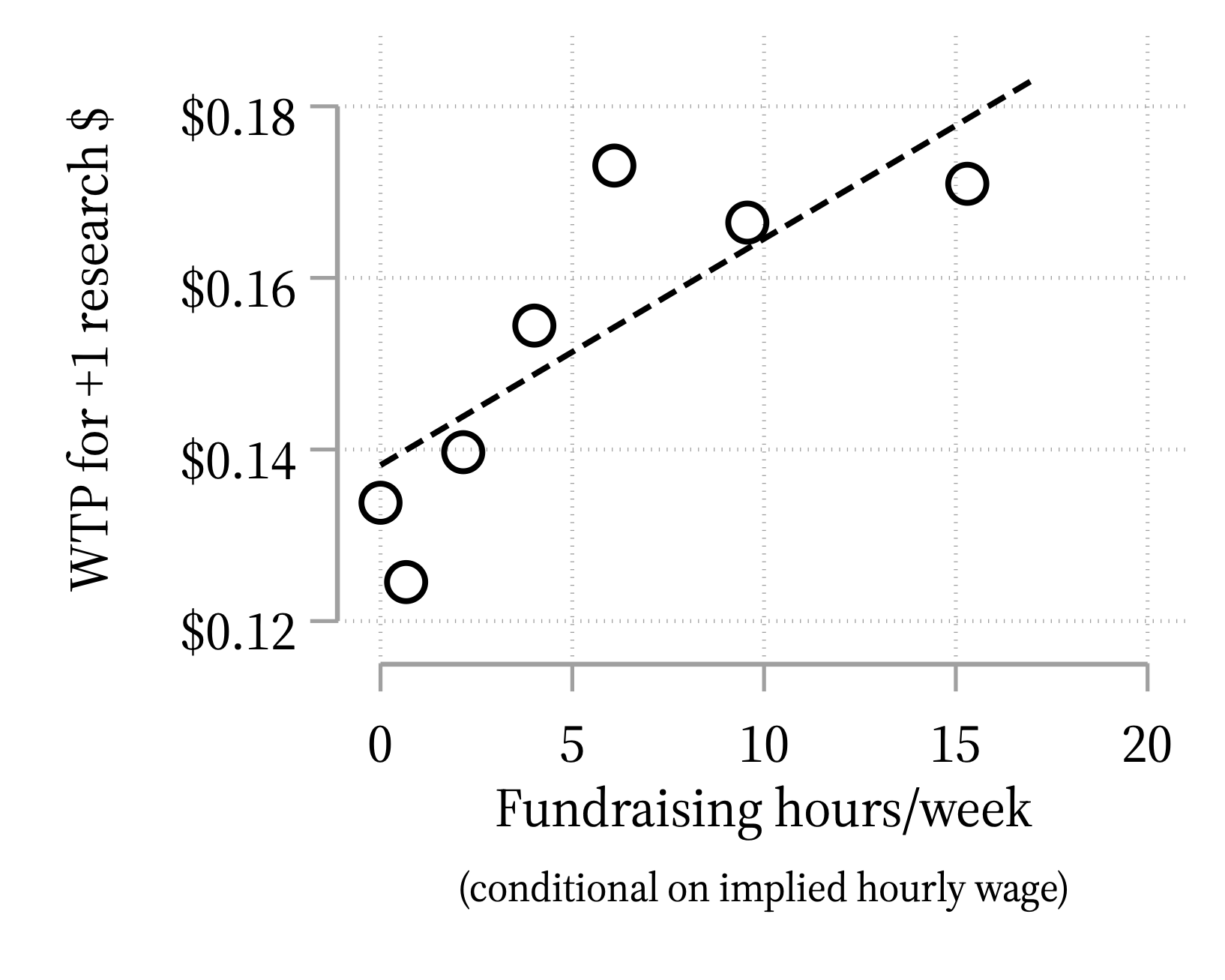}
}
\-\\
\note{\emph{Note}: \input{figtab/note_fig_binscat_wtpcorr.tex}}
\end{figure}

\begin{table}[htb]\centering \footnotesize
\caption{Proportion of Response Variation Correlated with Model and Non-model Variables}\label{tab_wtp_tests}
\-\\
{
\def\sym#1{\ifmmode^{#1}\else\(^{#1}\)\fi}
\begin{tabular}{l*{11}{c}}
\hline\hline
            & \multicolumn{2}{c}{WTP +\$250K} & & \multicolumn{2}{c}{WTP +\$1M} & & \multicolumn{2}{c}{WTP +admin.} & & \multicolumn{2}{c}{WTP --admin.} \\ \cline{2-3} \cline{5-6} \cline{8-9} \cline{11-12} &\multicolumn{1}{c}{(1)}         &\multicolumn{1}{c}{(2)}         &            &\multicolumn{1}{c}{(3)}         &\multicolumn{1}{c}{(4)}         &            &\multicolumn{1}{c}{(5)}         &\multicolumn{1}{c}{(6)}         &            &\multicolumn{1}{c}{(7)}         &\multicolumn{1}{c}{(8)}         \\
\hline

\\ $ R^2$   &        0.95         &        0.96         &            &        0.87         &        0.89         &            &        0.82         &        0.86         &            &        0.96         &        0.97         \\
\\ \hline Model vars.&$ \checkmark$         &$ \checkmark$         &            &$ \checkmark$         &$ \checkmark$         &            &$ \checkmark$         &$ \checkmark$         &            &$ \checkmark$         &$ \checkmark$         \\
$ \mathbf{X}$&                     &$ \checkmark$         &            &                     &$ \checkmark$         &            &                     &$ \checkmark$         &            &                     &$ \checkmark$         \\
$ N$ obs.   &       4,003         &       4,003         &            &       4,003         &       4,003         &            &       4,003         &       4,003         &            &       4,003         &       4,003         \\
\hline\hline
\end{tabular}
}
\-\\
\note{\emph{Note}: Reports the $ R^2$ statistics from regressions of researchers' responses to the four experiments (i.e., their willingness to pay for the alternative scenarios) on different combinations of variables: \emph{Model vars.} includes the state and choice variables of the model; \emph{$ \mathbf{X}$} includes the full vector of variables that comprise the type index.
}
\end{table}

\begin{table}[htb]\centering \footnotesize
\caption{Potential Role of Non-response Bias and Stated Preference WTP Bias}\label{tab_wtp_tests_IMRWTP}
\-\\
{
\def\sym#1{\ifmmode^{#1}\else\(^{#1}\)\fi}
\begin{tabular}{l*{11}{c}}
\hline\hline
                    & \multicolumn{2}{c}{WTP +\$250K} & & \multicolumn{2}{c}{WTP +\$1M} & & \multicolumn{2}{c}{WTP +admin.} & & \multicolumn{2}{c}{WTP --admin.} \\ \cline{2-3} \cline{5-6} \cline{8-9} \cline{11-12} &\multicolumn{1}{c}{(1)}         &\multicolumn{1}{c}{(2)}         &            &\multicolumn{1}{c}{(3)}         &\multicolumn{1}{c}{(4)}         &            &\multicolumn{1}{c}{(5)}         &\multicolumn{1}{c}{(6)}         &            &\multicolumn{1}{c}{(7)}         &\multicolumn{1}{c}{(8)}         \\
\hline
IMR                 &                     &   --0.000858         &            &                     &    --0.00391         &            &                     &    --0.00466         &            &                     &   --0.000835         \\
                    &                     &   (0.00379)         &            &                     &   (0.00593)         &            &                     &   (0.00684)         &            &                     &   (0.00327)         \\
[1em]
Benchmark           &                     &     --0.0126\sym{***}&            &                     &     --0.0258\sym{***}&            &                     &      0.0134\sym{*}  &            &                     &     --0.0102\sym{***}\\
                    &                     &   (0.00411)         &            &                     &   (0.00644)         &            &                     &   (0.00742)         &            &                     &   (0.00355)         \\
\hline
$ R^2$              &        0.95         &        0.95         &            &        0.87         &        0.87         &            &        0.82         &        0.82         &            &        0.96         &        0.96         \\
\hline\hline
\end{tabular}
}
\-\\
\note{\emph{Note}: Reports the estimates from regressions of researchers' responses to the four experiments (i.e., their willingness to pay for the alternative scenarios) on a control for non-response bias in the form of the Inverse Mills Ratio (IMR) and a control for bias in individuals' stated WTP in the form of their WTP for the benchmark good (i.e., high-speed internet at home); all variables are standardized. \emph{Model vars.} includes the state and choice variables of the model. Robust standard errors reported; $^{*} p<0.10, ^{**} p<0.05, ^{***} p<0.01$.
}
\end{table}

\clearpage
\section{Additional Productivity and Efficiency Results}
\setcounter{table}{0}
\setcounter{figure}{0}
\setcounter{equation}{0} 
\renewcommand{\theequation}{D\arabic{equation}}
\renewcommand{\thetable}{D\arabic{table}}
\renewcommand{\thefigure}{D\arabic{figure}}
\subsection{Additional Tables and Figures}

\begin{figure} \centering
\caption{Utility Function Visualization}\label{fig_util}
\subfloat[Salary, $M$]
{
\label{fig_util_M}
\includegraphics[width=0.475\textwidth, trim=0mm 10mm 0mm 5mm, clip]{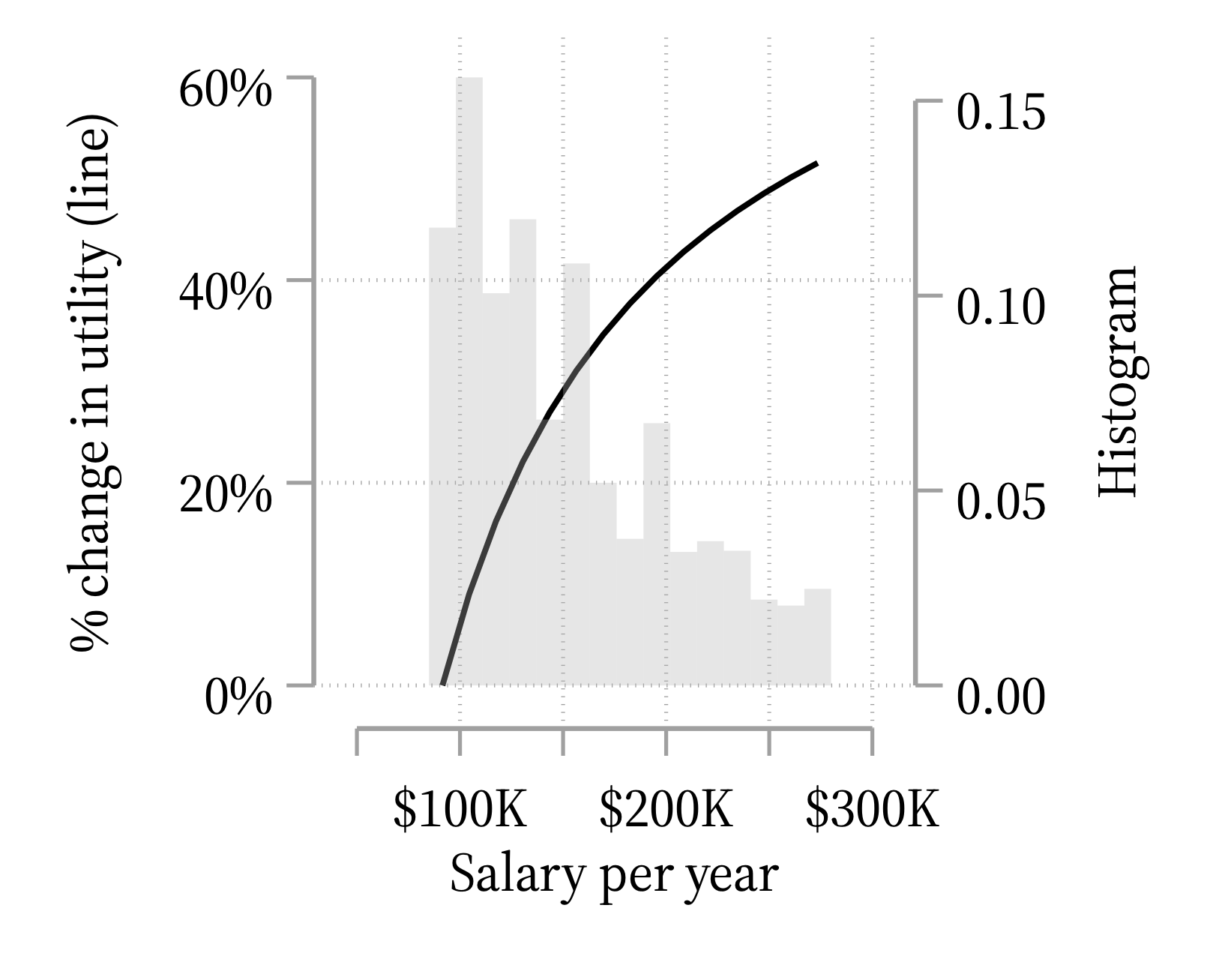}
} 
\subfloat[Administrative Duties, $D$]
{
\label{fig_util_A}
\includegraphics[width=0.475\textwidth, trim=0mm 10mm 0mm 5mm, clip]{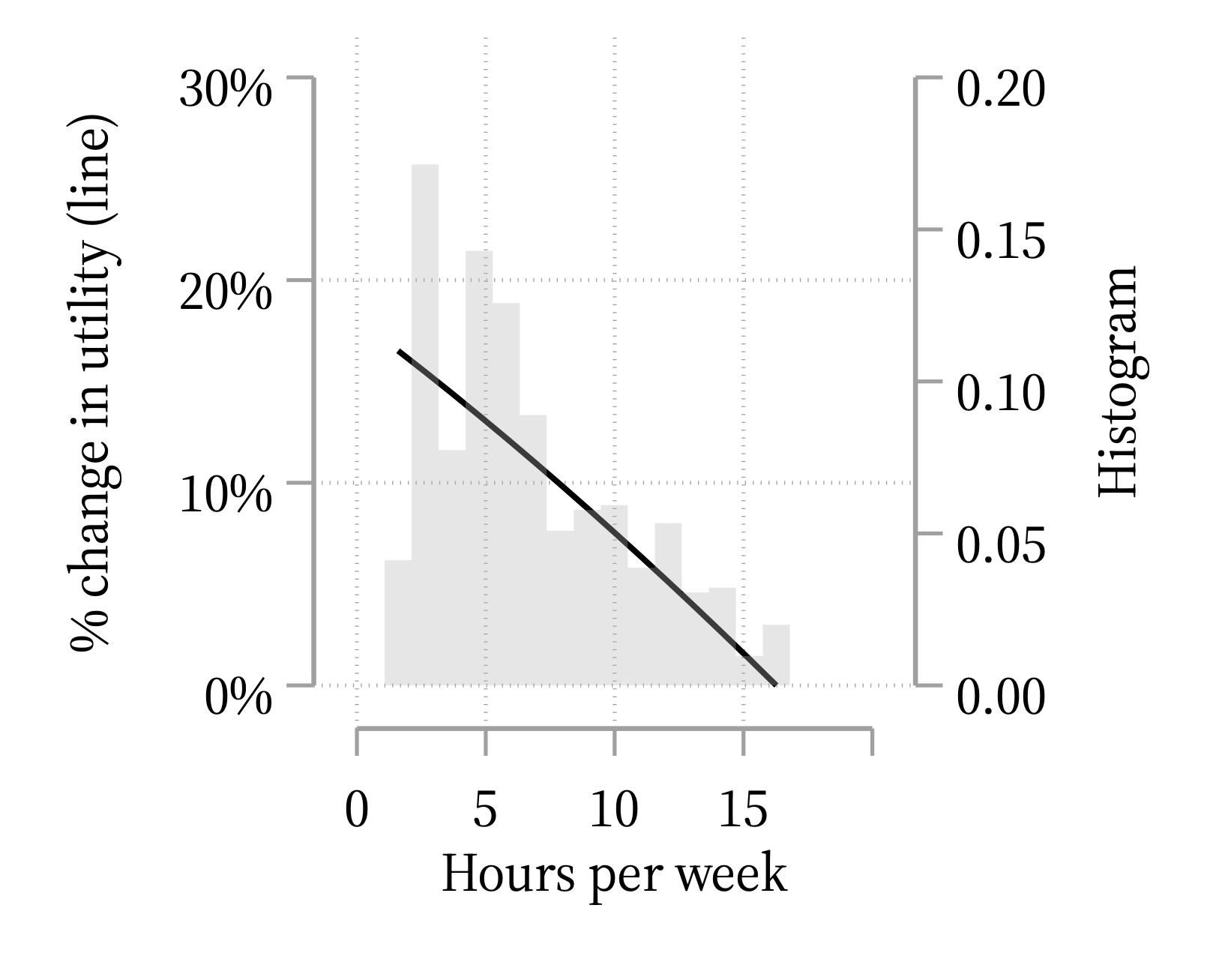}
}
\\ \-\\ \-\\
\subfloat[Guaranteed Research Budget, $G$]
{
\label{fig_util_G}
\includegraphics[width=0.475\textwidth, trim=0mm 10mm 0mm 5mm, clip]{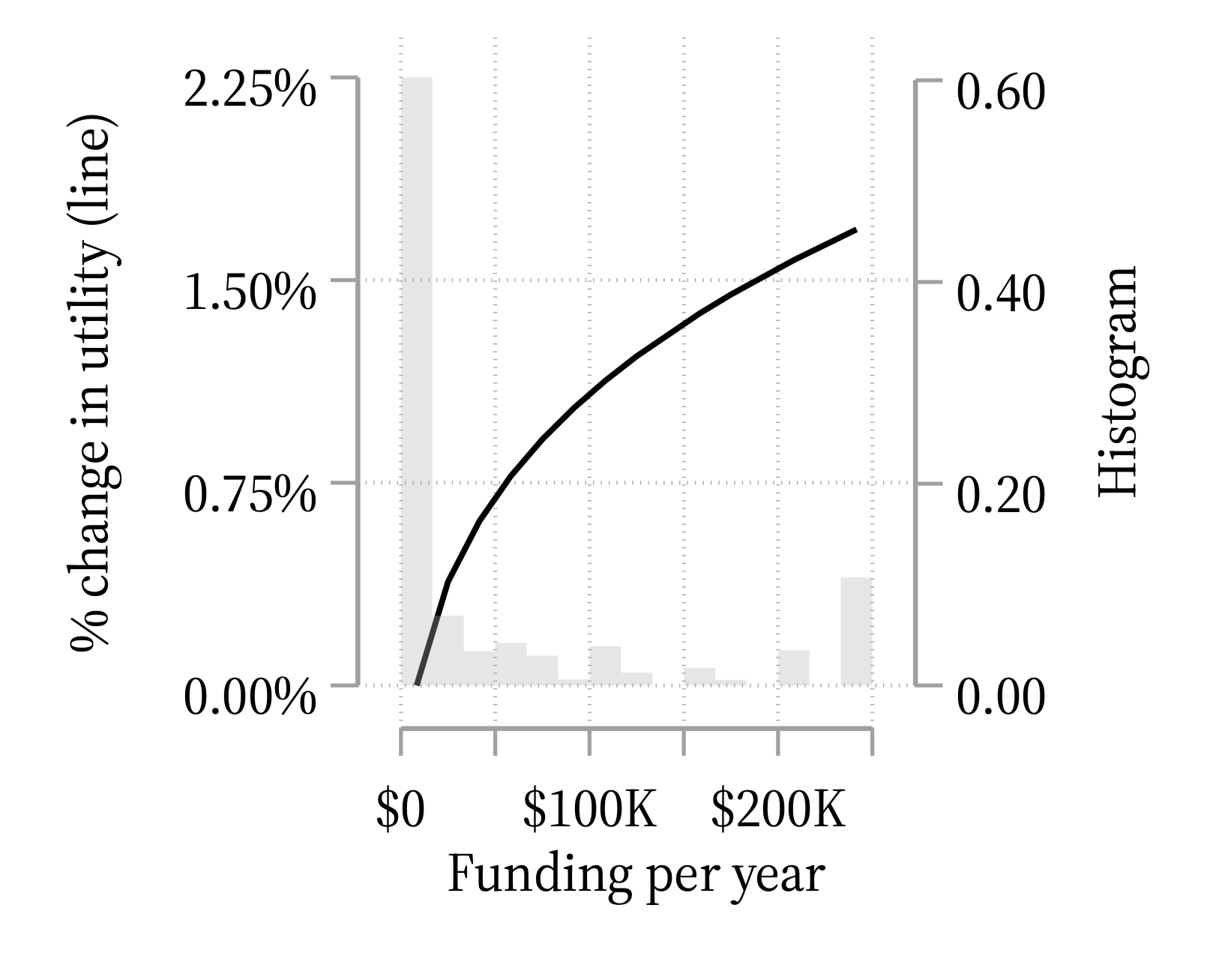}
}
\subfloat[Research Output, $Y$]
{
\label{fig_util_Y}
\includegraphics[width=0.475\textwidth, trim=0mm 10mm 0mm 5mm, clip]{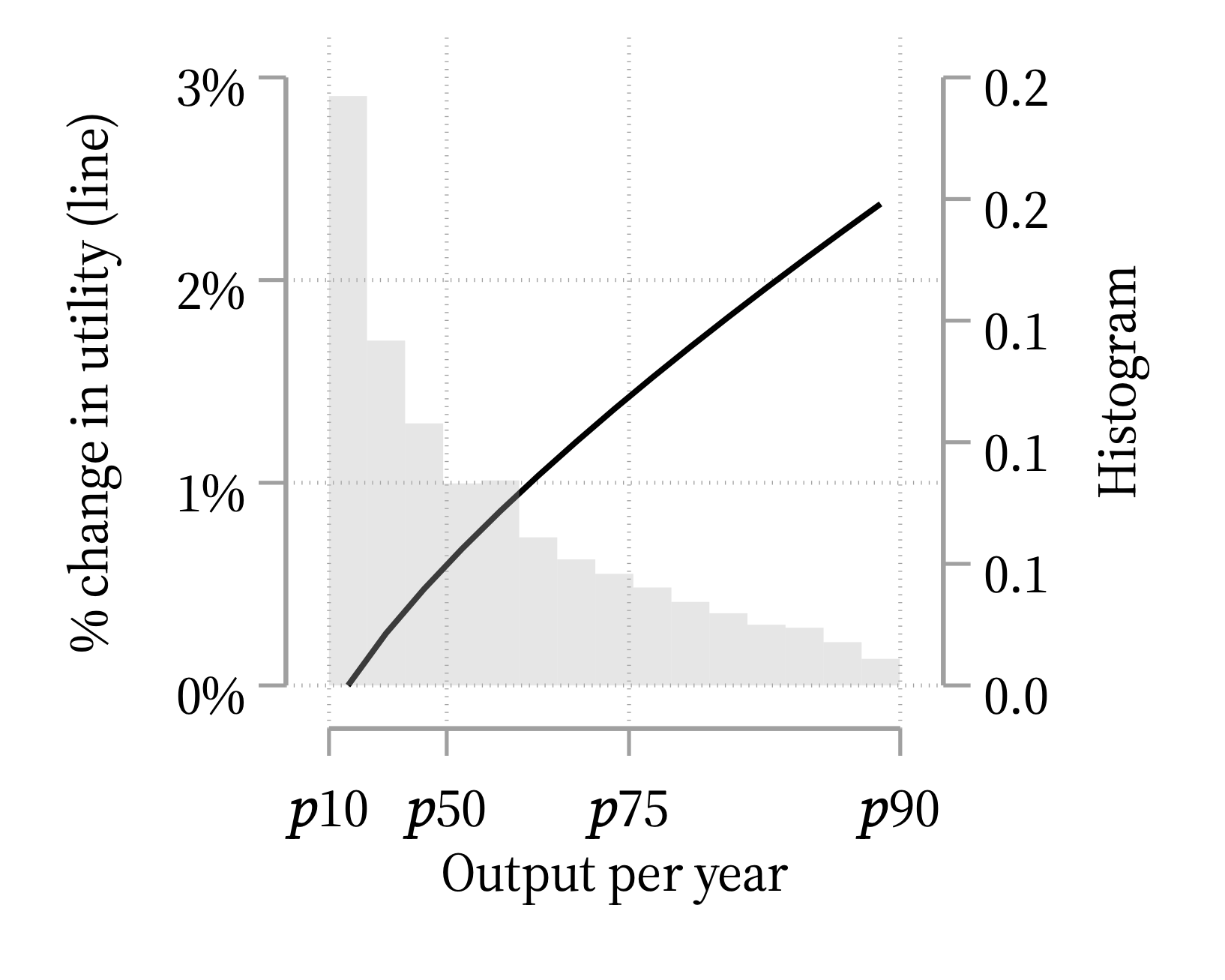}
}
\-\\
\note{\emph{Note}: \input{figtab/note_fig_util.tex} For simplicity, behavioral responses where researchers re-optimize are not incorporated here.}
\end{figure}

\begin{figure}[htbp] \centering
\caption{Research Funding Intensity ($\gamma$) by Minor Field of Study}\label{fig_gamma_by_field}
\includegraphics[height=0.5\textwidth, trim=0mm 10mm 0mm 5mm, clip]{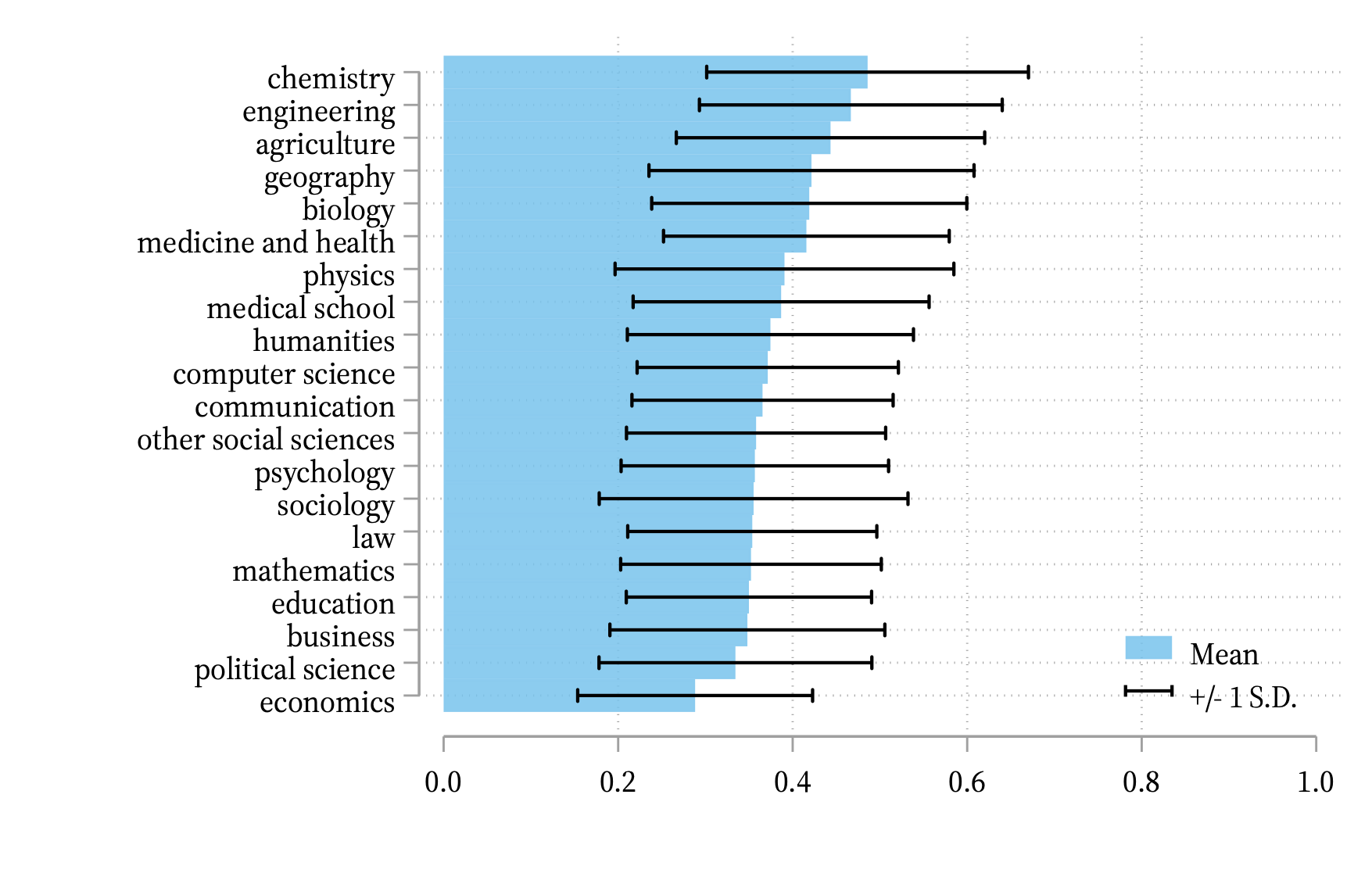}
\-\\
\note{\emph{Note}: \input{figtab/note_fig_gamma_by_field.tex}}
\end{figure}

\begin{figure} \centering
\caption{Power Laws in the TFP Tail}\label{fig_topa_powerlaw}
\subfloat[Top 20\% Researchers]
{
\label{fig_topa_powerlaw20}
\includegraphics[width=0.495\textwidth, trim=5mm 20mm 0mm 5mm, clip]{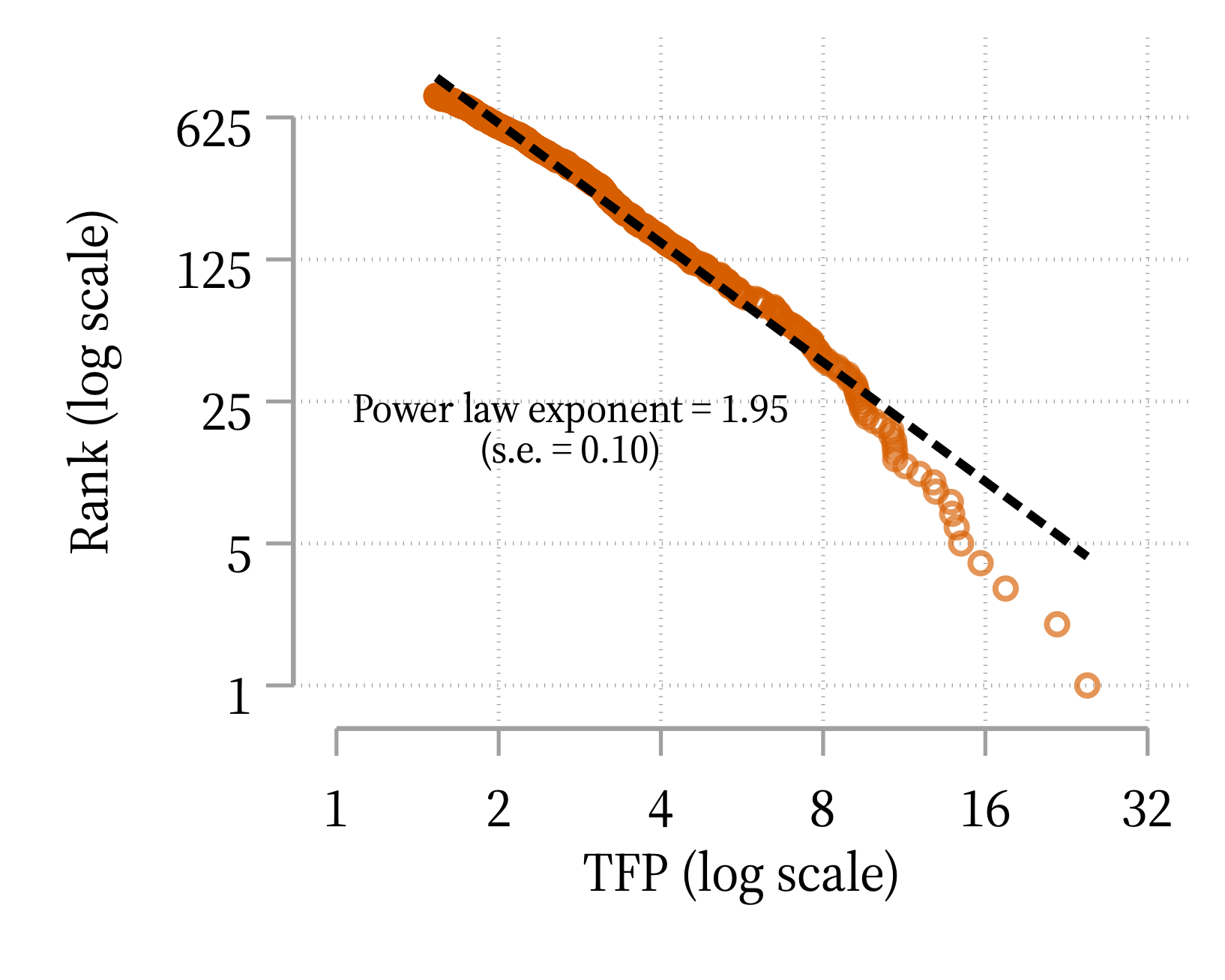}
} 
\subfloat[Top 1\% Researchers]
{
\label{fig_topa_powerlaw01}
\includegraphics[width=0.495\textwidth, trim=5mm 20mm 0mm 5mm, clip]{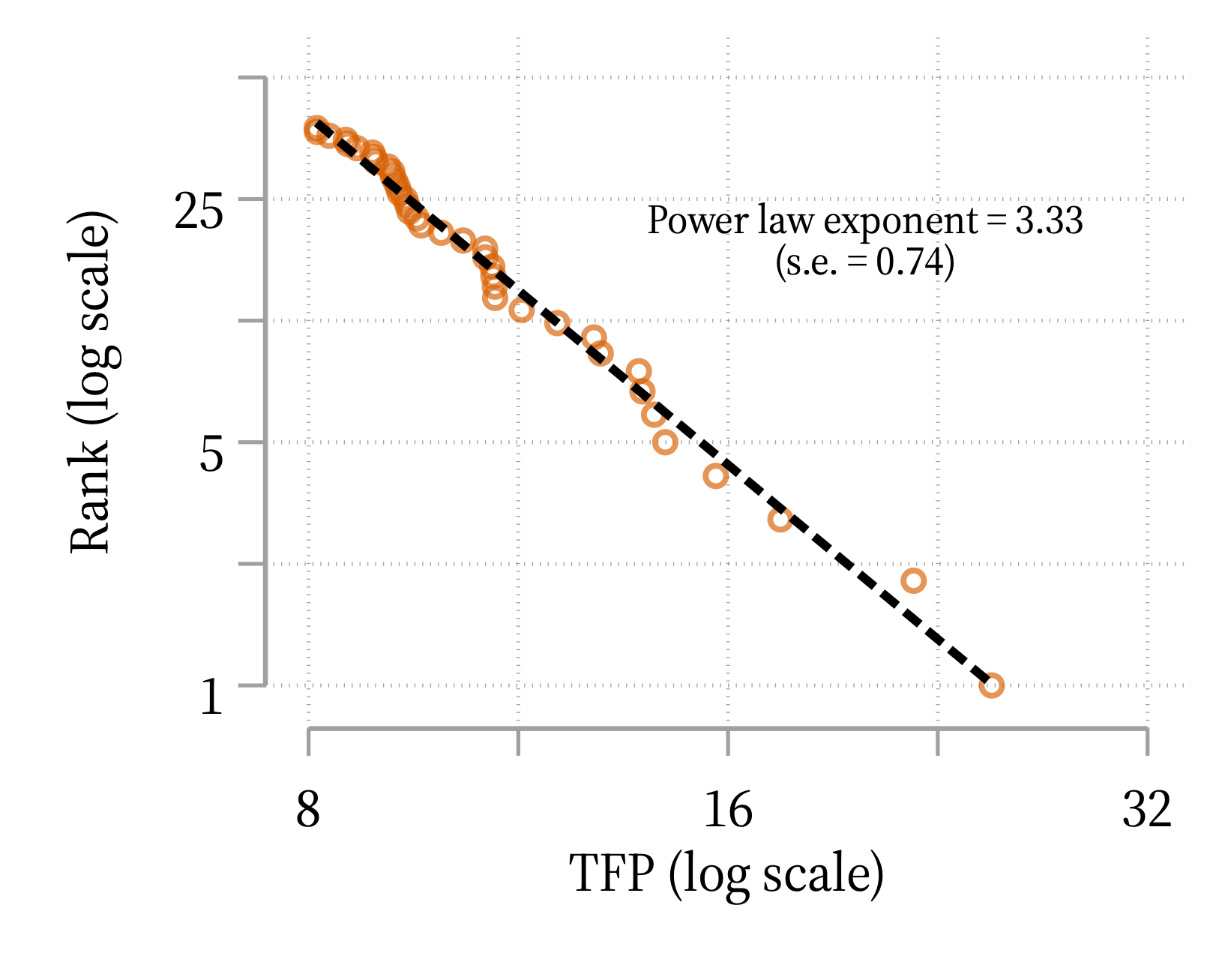}
} \\
\note{\emph{Note}: \input{figtab/note_fig_topa_powerlaw.tex}}
\end{figure}

\begin{table}[htbp]\centering \footnotesize
\caption{Observable Output Correlations}\label{tab_reg_alphapubcorr}
\-\\
{
\def\sym#1{\ifmmode^{#1}\else\(^{#1}\)\fi}
\begin{tabular}{l*{5}{c}}
\hline\hline
            &\multicolumn{2}{c}{Recent publications,} & &\multicolumn{2}{c}{Recent publications,} \\ 
 &\multicolumn{2}{c}{count} & &\multicolumn{2}{c}{cite-weighted} \\ \cline{2-3} \cline{5-6} 
 &\multicolumn{1}{c}{(1)}         &\multicolumn{1}{c}{(2)}         &            &\multicolumn{1}{c}{(3)}         &\multicolumn{1}{c}{(4)}         \\
\hline
            &                     &                     &            &                     &                     \\
$ \log(\alpha) $&       0.201\sym{***}&       0.176\sym{***}&            &       0.276\sym{***}&       0.254\sym{***}\\
            &    (0.0406)         &    (0.0397)         &            &    (0.0648)         &    (0.0648)         \\
[1em]
$ \log(\text{recent grant \$}) $&                     &      0.0220\sym{***}&            &                     &      0.0174\sym{***}\\
            &                     &   (0.00350)         &            &                     &   (0.00525)         \\
\hline
Field--FE, $ \gamma, \phi, \mathbf{X}$--index&$ \checkmark$         &$ \checkmark$         &            &$ \checkmark$         &$ \checkmark$         \\
$ R^2$      &        0.18         &        0.20         &            &        0.15         &        0.16         \\
$ N$ obs.   &       2,703         &       2,703         &            &       2,703         &       2,703         \\
\hline\hline
\end{tabular}
}
\-\\
\note{\emph{Note}: Reports the estimates from regressions of researchers' publication measures (including publications from 2018--2022) on their estimated research productivity ($ \alpha$) as well as vector of controls that includes major field fixed effects (\emph{Field-FE}) and the researchers' funding intensity ($ \gamma$), fundraising efficiency ($ \phi$), their type ($ \mathbf{X}$-\emph{index}), and, in some specifications, a control for their publicly observable research grant funding over the same period. Robust standard errors reported; $^{*} p<0.10, ^{**} p<0.05, ^{***} p<0.01$.
}
\end{table}

\begin{figure}[htbp] \centering
\caption{Lorenz Curves for Actual and Optimal Input Levels}\label{fig_lorenz_actvopt}
\subfloat[Administrative Duties]
{
\label{fig_lorenz_actvopt_S_Ai}
\includegraphics[width=0.475\textwidth, trim=0mm 10mm 0mm 5mm, clip]{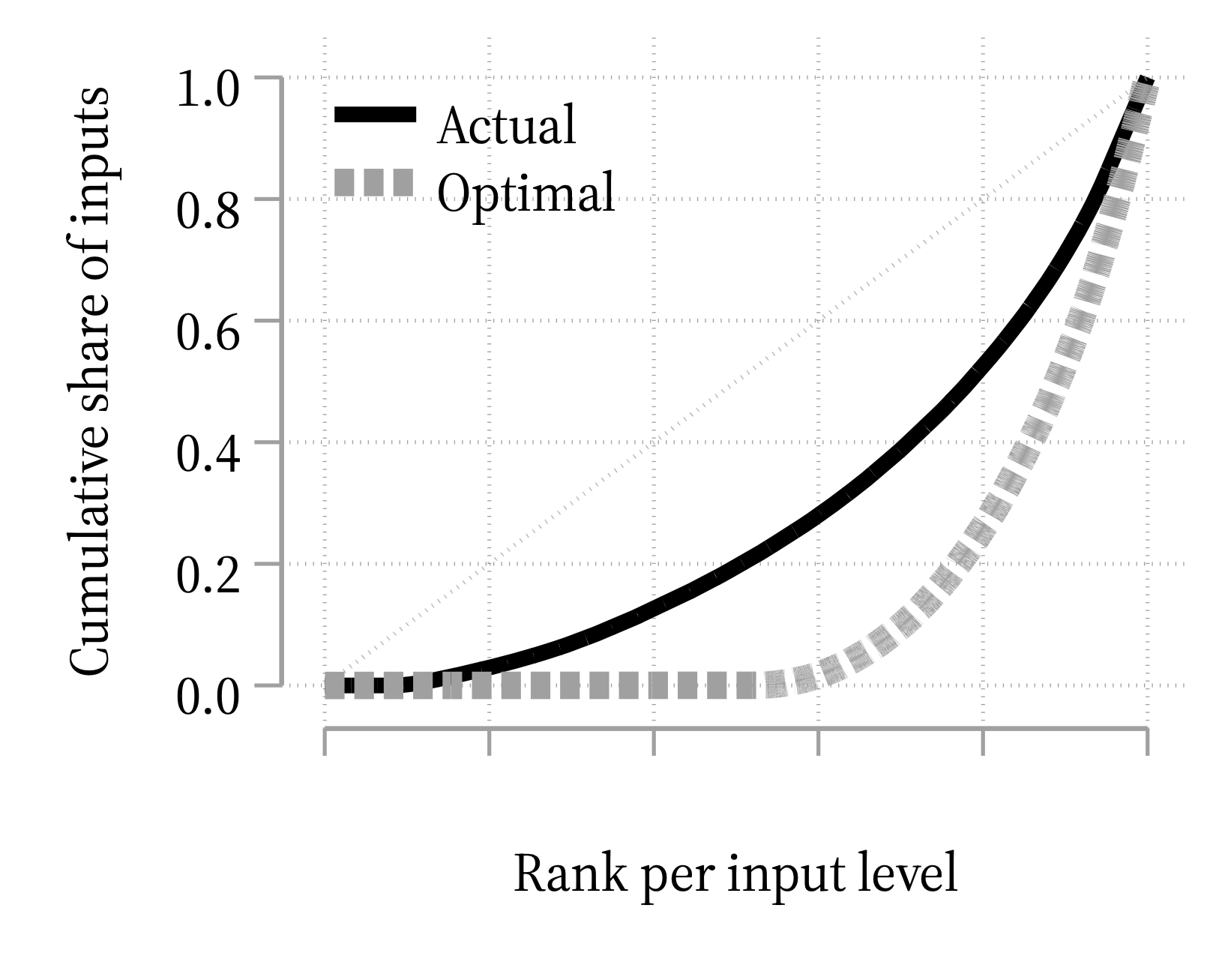}
} 
\subfloat[Guaranteed Funding]
{
\label{fig_lorenz_actvopt_S_Gi}
\includegraphics[width=0.475\textwidth, trim=0mm 10mm 0mm 5mm, clip]{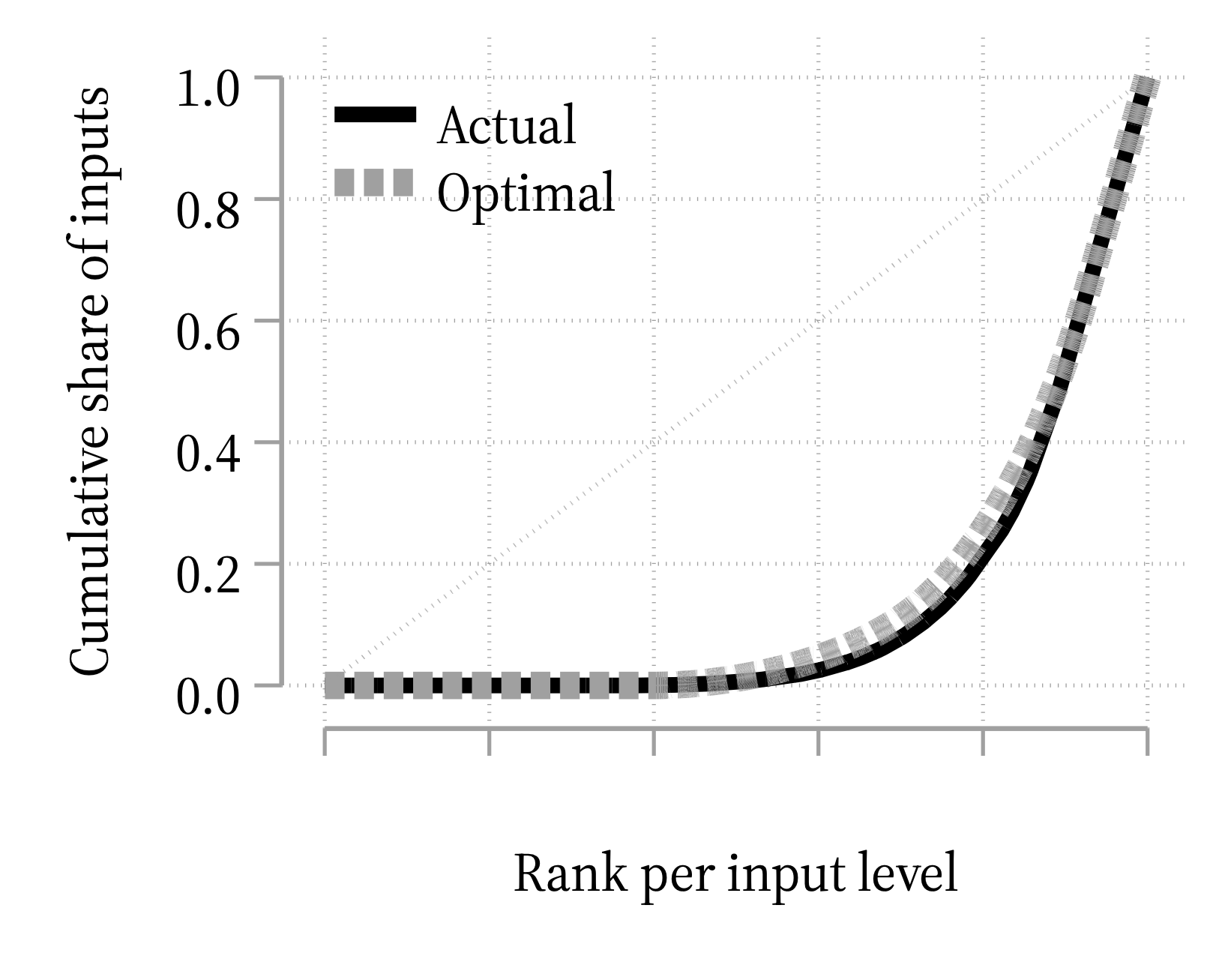}
}
\\ \-\\ \-\\
\subfloat[Research Time]
{
\label{fig_lorenz_actvopt_C_Ri}
\includegraphics[width=0.475\textwidth, trim=0mm 10mm 0mm 5mm, clip]{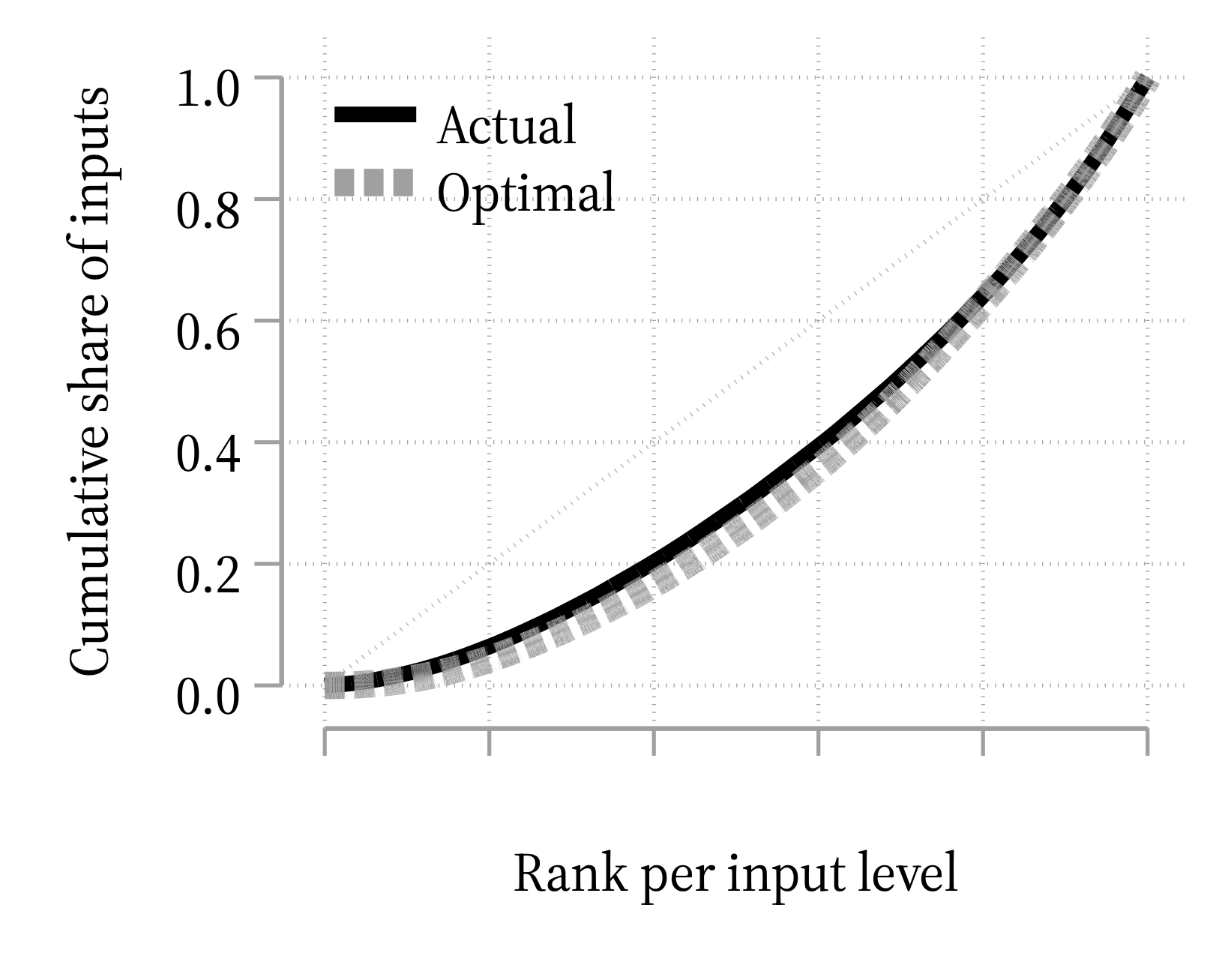}
} 
\subfloat[Total Research Funding]
{
\label{fig_lorenz_actvopt_C_Bi}
\includegraphics[width=0.475\textwidth, trim=0mm 10mm 0mm 5mm, clip]{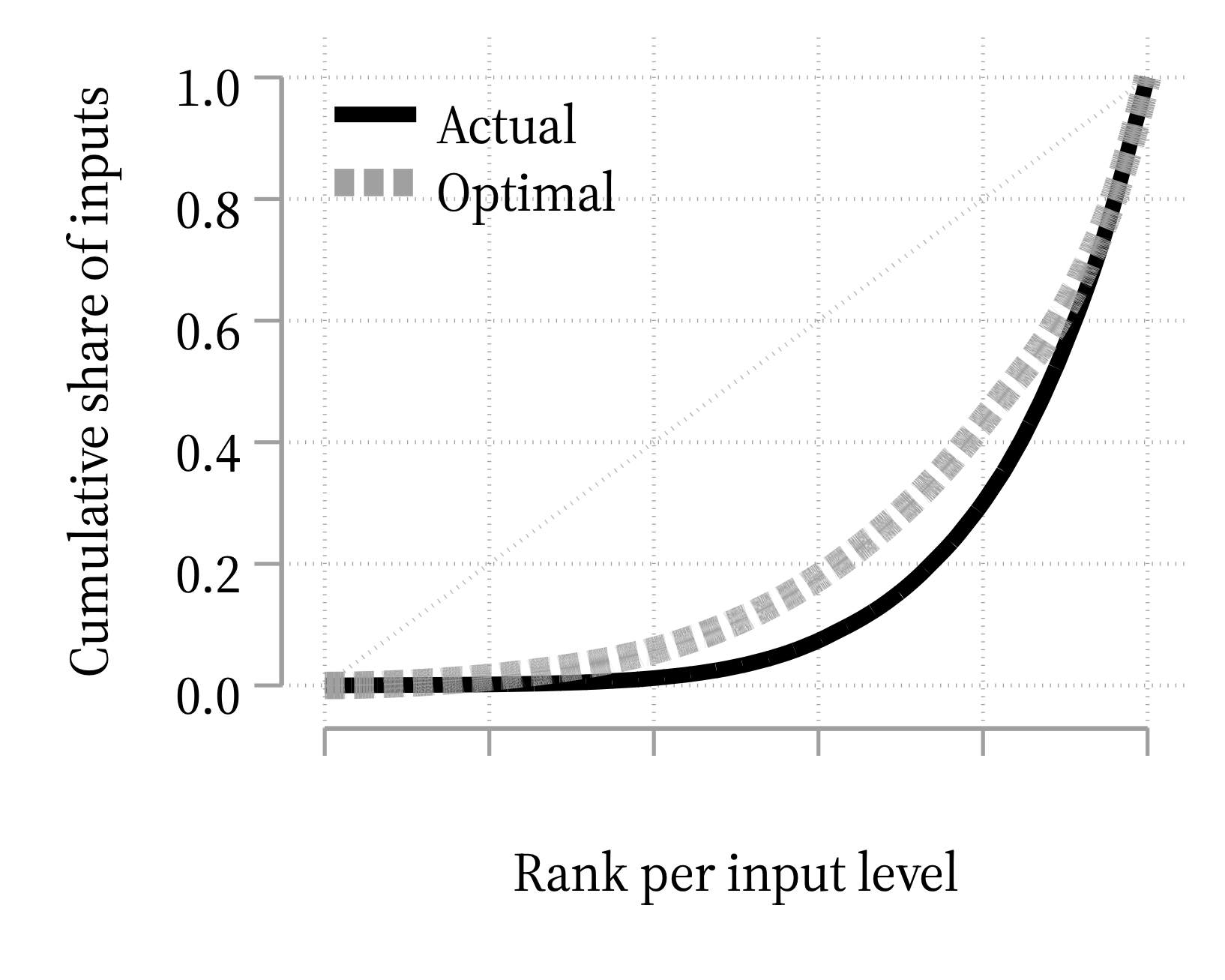}
}
\-\\
\note{\emph{Note}: \input{figtab/note_fig_lorenz_actvopt.tex}}
\end{figure}

\begin{table}[htbp]\centering \footnotesize
\caption{Production with Actual and Optimized Allocations---Alternative Counterfactuals}\label{tab_counterfac_summaries}
\-\\
\begin{tabular}{l r c r r r r r } 
 \hline\hline 
 & \multicolumn{1}{c}{Current} & & \multicolumn{5}{c}{Optimized} \\ 
 & \multicolumn{1}{c}{allocation} & & \multicolumn{5}{c}{allocations} \\ \cline{2-2} \cline{4-8} 
 & \multicolumn{1}{c}{(1)} & & \multicolumn{1}{c}{(2)} & \multicolumn{1}{c}{(3)} & \multicolumn{1}{c}{(4)} & \multicolumn{1}{c}{(5)} & \multicolumn{1}{c}{(6)}  \\ 
 \hline\underline{\emph{Research inputs}}&&&&&&& \\
Research hrs./week, avg.&18.5&&--11\%&+18\%&+14\%&+25\%&+16\% \\
Research hrs./week, s.d.&9.6&&+6.0\%&+18\%&+43\%&+39\%&+41\% \\
Budget \$-K/year, avg.&147.1&&0\%&0\%&0\%&0\%&+14\% \\
Budget \$-K/year, s.d.&206.9&&+14\%&--24\%&--20\%&--20\%&--3.9\% \\
\hline
 \hline 
 \end{tabular}
\-\\
\note{\emph{Note}: Reports summary statistics for inputs under actual allocations (Col. 1). The first three sets of rows in Columns 2--6 report the percentage change in research inputs (\emph{Research inputs}), outputs (\emph{Research outputs}), and utility (\emph{Welfare}) under alternative allocations; estimates are rounded to aid in comparison. The \emph{Input Reallocation} rows report the amount of inputs reallocated expressed as a percentage of the total level of the input (e.g., 50\% implies that half of all dollars are moved from one researcher to another). The bottom sets of rows outline the objective and constraints of the five different counterfactual allocations explored in Columns 2--6. The two different objectives explored are maximizing researchers' private utility ($ \mathcal{V}$) or output ($ Y$). $ D$ refers to administrative duties, and $ G$ refers to guaranteed research funding. \emph{ Unconstrained} $ B$ indicates the scenario when the total research budget is left unconstrained and so the total amount of funding in the market is limited only by researchers' fundraising choices. All optimized allocations allow for researchers' behavioral responses after $ D$ and/or $ G$ have been reallocated.
}
\end{table}

\clearpage\begin{table}[htb]\centering \footnotesize
\caption{Input Wedges and Gender Differences}\label{tab_wedgereg_full}
\-\\
{
\def\sym#1{\ifmmode^{#1}\else\(^{#1}\)\fi}
\begin{tabular}{l*{5}{c}}
\hline\hline
                    &\multicolumn{1}{c}{(1)}         &\multicolumn{1}{c}{(2)}         &\multicolumn{1}{c}{(3)}         &\multicolumn{1}{c}{(4)}         &\multicolumn{1}{c}{(5)}         \\
\hline
& \\ 
 \multicolumn{5}{l}{\emph{\underline{Panel (a): Actual research time}}} \\ 
 & \\ 
 Optimal level             &       0.301\sym{***}&       0.294\sym{***}&       0.300\sym{***}&       0.352\sym{***}&       0.349\sym{***}\\
                    &   (0.00970)         &   (0.00989)         &   (0.00991)         &   (0.00985)         &    (0.0102)         \\
[1em]
Female              &                     &                     &                     &      --1.333\sym{***}&      --0.885\sym{***}\\
                    &                     &                     &                     &     (0.286)         &     (0.279)         \\
\hline\hline
\end{tabular}
}
\-\\
\note{\emph{Note}: Reports results from regressions of actual input levels on optimal input levels and as described in Equation \ref{eq:wedge_predict}. Robust standard errors reported; $^{*} p<0.10, ^{**} p<0.05, ^{***} p<0.01$.
}
\end{table}
\begin{table}[htb]\centering \footnotesize
\caption{Input Wedges per Common Productivity Proxies}\label{tab_reg_productivityproxies}
\-\\
{
\def\sym#1{\ifmmode^{#1}\else\(^{#1}\)\fi}
\begin{tabular}{l*{4}{c}}
\hline\hline
                    &\multicolumn{1}{c}{(1)}         &\multicolumn{1}{c}{(2)}         &\multicolumn{1}{c}{(3)}         &\multicolumn{1}{c}{(4)}         \\
\hline
& \\ 
 \multicolumn{4}{l}{\emph{\underline{Panel (a): Actual research time}}} \\ 
 & \\ 
 Optimal level            &       0.508\sym{***}&       0.506\sym{***}&       0.509\sym{***}&       0.508\sym{***}\\
                    &    (0.0164)         &    (0.0166)         &    (0.0162)         &    (0.0163)         \\
[1em]
Recent research funding, own&       0.130\sym{***}&                     &                     &                     \\
                    &    (0.0311)         &                     &                     &                     \\
\hline\hline
\end{tabular}
}
\-\\
\note{\emph{Note}: Reports results from regressions of actual input levels on optimal input levels and common proxies for researchers' producitivities. All variables are standardized. All proxies are based on data from one to two years prior to the survey. Robust standard errors reported; $^{*} p<0.10, ^{**} p<0.05, ^{***} p<0.01$.
}
\end{table}

\subsection{Mechanical Composition Effect in the Counterfactual}\label{subsec_compeffect}
In Section \ref{subsec_countarfactuals_summary}, we estimate the growth in funding using current allocations that is necessary to achieve the same growth in output that we achieve using alternative allocations of the current funding. We choose guaranteed funding ($G$) as our policy lever for injecting funding into the market because it is exogenous in the model.\footnote{The other component of funding, which researchers obtain via fundraising, is the product of an exogenous component (researchers' fundraising productivity, $\phi_i$) and an endogenous component (researchers' time spent fundraising, $F_i$). Practically, it is much easier to engage with simulations that manipulate $G$, especially when behavioral responses are allowed.} In the main counterfactual, we find that a 210\% increase in guaranteed funding for all researchers---equivalent to a 210\% increase in \textit{aggregate} guaranteed funding ($G\equiv\sum_i G_i$)---translates into an 85\% increase in total aggregate funding ($B\equiv\sum_i B_i$), which then yields a 160\% increase in aggregate output ($Y\equiv\sum_i Y_i$). This may seem counterintuitive given that our specification of the scientific production function features decreasing returns to funding ($\gamma_i<1$ for all $i$). Below, we explain how this can occur. 

Recall the production functions is: $Y_i = \alpha_i B_i^{\gamma_i} R_i^{1-\gamma_i}$, where $B_i$ is researcher $i$'s total funding and $R_i$ is their time spent on research. A researcher's total budget is:$B_i =  G_i + B_{min} + \phi_i F_i$. Given these functional forms, the marginal change in a researchers' log-output ($d \ln(Y_i)$) given a change to their total budget, holding their time spent on research ($R_i$) fixed, is: $d \ln(Y_i) = \gamma_i d\ln(B_i)$. Next, define a researcher's share of total funding due to their guarantees as: $s_i \equiv G_i / B_i$.

This allows us to approximate the marginal change in a researcher's log-output given a change to their guaranteed funding: $d \ln(Y_i) \approx \gamma_i s_i d\ln(G_i)$, which holds as an approximation because it keeps $s_i$ fixed to its value before the change. Define a researcher's share of total budget as: $t_i\equiv  B_i/B$, and their share of total output under initial allocations as: $z_i \equiv Y_i / Y$. Thus, the relative change in aggregate budget and aggregate output given a change in each individual researcher's output are: $d \ln B \approx \sum_i t_i d  \ln(B_i)$ and $d \ln Y \approx \sum_i z_i d  \ln(Y_i)$. Combining these last two equations with previous derivations gives us two expressions for the relative change in aggregate budget and in aggregate output given an average relative change in funding guarantees: $d \ln (B) \approx \sum_i t_i s_i d \ln (G_i) \approx \sum_i t_i d \ln (B_i)$ and $d \ln (Y) \approx \sum_i z_i \gamma_i s_i d\ln(G_i) \approx \sum_i z_i \gamma_i d \ln(B_i)$.

If there is a positive covariance between the $d \ln (B_i)$ and $\gamma_i$ terms, \emph{then the relative growth in aggregate output ($Y$) can exceed the relative growth in the total budget ($B$)}. This is what we find in practice. Furthermore, note that the change in aggregate budget never exceeds the change in funding guarantees ($d \ln (G_i)$), because it is a convex combination of individual $d \ln (B_i)$, and the latter never exceed $d \ln (G_i)$ because the weights $s_i$ are smaller than one. In contrast, $d \ln (Y)$ is not a convex combination of variations in individual budget $d \ln (B_i)$, because the sum of $z_i \gamma_i$ weights may exceed one. Therefore, a positive covariance between $\gamma_i$ and $d \ln (B_i)$ inflates $d \ln (Y)$.

\end{document}